\documentclass[lettersize,journal]{IEEEtran}

\usepackage[utf8]{inputenc}
\usepackage{amsmath, amsfonts, amssymb, bbm}
\usepackage{algorithm}
\usepackage{algorithmic}
\usepackage{array}
\usepackage{textcomp}
\usepackage{stfloats}
\usepackage{url}
\usepackage{verbatim}
\usepackage{graphicx}
\usepackage{booktabs}
\usepackage{cite}
\usepackage{tabularx}
\usepackage{float}
\usepackage{amsthm}
\usepackage{enumitem}
\usepackage[dvipsnames]{xcolor}
\usepackage{dsfont}
\usepackage[makeroom]{cancel}
\usepackage{subcaption}
\usepackage{caption}
\usepackage{tikz}
\usetikzlibrary{calc}

\DeclareGraphicsExtensions{.pdf,.jpeg,.png}
\renewcommand{\eqref}[1]{Eq.~(\ref{#1})}

\begin{document}

\title{Modelling the Closed Loop Dynamics Between a Social Media Recommender System and Users' Opinions}

\author{Ella C. Davidson and Mengbin Ye
\thanks{E. C. Davidson and M. Ye are with the Adelaide Data Science Centre, Adelaide University, Adelaide, Australia (\texttt{\{connor.davidson;ben.ye\}@adelaide.edu.au}). E. C. Davidson is supported by the Australian Government with a Research Training Program scholarship. M. Ye is supported by the Australian Government through the Australian Research Council (DE250100199).}
}


\maketitle

\begin{abstract}
This paper proposes a mathematical model to study the coupled dynamics of a Recommender System (RS) algorithm and content consumers (users). The model posits that a large population of users, each with an opinion, consumes personalised content recommended by the RS. The RS can select from a range of content to recommend, based on users' past engagement, while users can engage with the content (like, watch), and in doing so, users' opinions evolve. This occurs repeatedly to capture the endless content available for user consumption on social media. We employ a campaign of Monte Carlo simulations to study how recommender systems influence users' opinions, and in turn how users' opinions shape the subsequent recommended content. Both the performance of the RS (e.g., how users engage with the content) and the polarisation and radicalisation of users' opinions are of interest. We find that different opinion distributions are more susceptible to becoming polarised than others, many content stances are ineffective in changing user opinions, and creating viral content is an effective measure in combating polarisation of opinions.
\end{abstract}


\section{Introduction}\label{sec:intro}

The internet has revolutionised how people generate, interact with, and consume information, and thus how people form opinions. Social media platforms (social media for short) are at the forefront of this revolution.
The proliferation of social media has brought about obvious benefits, but there are increasing concerns about their role in the polarisation and radicalisation of society~\cite{flaxman,whitepaper_onlineharm_2020}. 
Understanding how social media platforms shape users' opinions and behaviours has become a major focus among scientists and policymakers~\cite{whitepaper_onlineharm_2020, KGI_better_2025, stray2021designing}. The affordances of a social media platform, including its structure and design, as well as the algorithmic recommender systems (RSs) that determine which content users consume and engage with, define how users experience the platform. Platform affordances have steadily evolved over the years, from the text-centric  Facebook and Twitter to the image-focused Instagram. TikTok popularised the video-centric ``endless stream'' approach with its scrollable For You page~\cite{matsakis}, in which a user is presented with an endless stream of video content tailored for them. Instagram and YouTube have implemented their own Reels and Shorts, respectively, in response. 

Following other recent literature, we study social media platforms as a socio-technological system in which the platform and its users coexist in a closed loop, mutually influencing each other. Most platforms utilise a \textit{recommender system} (RS), an algorithm which determines the most relevant/appropriate content to recommend to users~\cite{saifudin} by taking into account user preferences and prior content consumption. Meanwhile, users are influenced by and engage with this content, thereby changing the information used by the RS in determining new content to recommend. Since data collection from social media APIs has become increasingly difficult~\cite{mimizuka2025postAPI}, an emerging approach is to study the socio-technological system using mathematical models~\cite{rossi,dean2025policy,ionescu2023CC,zhang2024misinformation, chen2025coevolution}. This approach exploits two previously disjoint streams of literature (RSs and opinion dynamics), each with extensive modelling foundations, by combining existing models into a dynamical system in which an RS and its users' opinions coevolve over time.

\subsection{Literature Review}

We now review the literature on the two key elements of our system, RSs and user opinion dynamics, and recent efforts to model the two as a socio-technological system. 

Much of the research into recommender systems has typically focused on the different algorithmic paradigms used to determine recommendations. These range from simple techniques like pattern and rule mining to intricate filtration techniques such as tensor factorisation models and deep learning algorithms~\cite{wang2023,saifudin,jannach}. 
One fundamental filtering technique is content-based filtering, which records users' previous interactions with content to predict future recommendations~\cite{jannach}, which follows a \textit{more of the same} philosophy. Collaborative filtering identifies clusters of users that behave similarly online and posits that users within a cluster have comparable wants and needs~\cite{jannach}. How a user engages with recommended content, such as ``liking'' the content (explicit engagement) or how long they may view the content (implicit engagement)~\cite{ricci}, may be used to help generate recommendations.
Recent literature has developed mathematical models for RSs, focusing on how to optimise the performance of an RS, i.e. maximise user engagement and minimise biases in users' ratings, while taking into account user preferences and biases~\cite{adomavicius,mansoury2020feedback}. However, such works assume opinions or preferences are static, and little attention has been paid to how, over time, the recommended content could change user opinions.

Opinion formation models (commonly referred to as opinion dynamics in the literature) are dynamical systems models that describe how opinions evolve over time due to social influence within interaction networks~\cite{anderson,proskurnikov2017tutorial}. The French--DeGroot model is a foundational model, in which agents in a network share opinions. Through an iterative weighted averaging process, a consensus of opinions is reached under mild network connectivity conditions. The classical Friedkin--Johnsen model extends the French--DeGroot model by positing that agents remain attached to their initial prejudices~\cite{friedkin}, and has extensive empirical studies and data supporting it~\cite{friedkin2011social_book}.
In contrast to the literature on RSs, opinion dynamics research has focused on how opinions evolve through social influence among peers, overlooking the fact that RSs are now central to how people obtain information and connect with other users on social media. 

Our work contributes to the growing efforts to unite the two streams of literature on RSs and opinion dynamics, modelling and studying a socio-technological closed-loop system evolving over time. One popular approach considers the RS acting to connect different users within a social network (as opposed to connecting users to content, as described below). More specifically, different RS algorithms have been proposed that create and remove links between users in a social network, allowing users to exchange opinions. These algorithms have considered mechanisms such as collaborative filtering, content-based filtering, virality/popularity-based filtering, and time-based mechanisms. However, crucially, user interactions or engagements are not fed back into the RS to tailor the next set of recommendations more effectively. Moreover, such models assume the RS is aware of \textit{the user opinions}, whereas real-world RSs typically are only aware of \textit{user interactions or engagements}. Ref.~\cite{perra} examined different sorting mechanisms, such as sorting chronologically and by opinion alignment, and identified that only some enabled status quo change. 
To examine the spread of misinformation, five recommendation algorithms were proposed and evaluated through numerical simulations using a real-world network structure from Reddit~\cite{zhang2024misinformation}. 
It was found that algorithmic interventions worsened polarisation, particularly when relying on content-based filtering. Analytical and computational approaches showed that an RS forming connections between users through collaborative filtering could facilitate the emergence of polarisation and filter bubbles~\cite{chen2025coevolution}.

Another approach focuses on the RS recommending content to users. One of the first models in this approach considered just one user and a simple RS~\cite{rossi}. The user had an opinion on a given topic, and at discrete timesteps, the RS presented the user with one of two possible articles (or content), representing a position at either extreme of some given topic. The user decided whether to ``click'' on the recommended article, whereby their opinion changed after receiving the recommendation, and their click decision was relayed back to the RS to better tailor future recommendations. This process occurred iteratively, such that the recommendations and the user's opinion coevolved over time. Through numerical simulations and analysis, it was found that the simple content-based filtering algorithm accelerated the radicalisation of user opinions, and in turn, extreme opinions contributed positively to user engagement~\cite{rossi}. The model in~\cite{rossi} was extended to consider an ensemble population of users and articles that held any position on the opinion spectrum~\cite{lanzetti}. Both the microscopic impact (on an arbitrary user's opinion) and macroscopic impact (on the distribution of opinions in the population) were studied~\cite{lanzetti}. A closed loop control paradigm was used to modify the recommendation algorithm to maximise user engagement in~\cite{sprenger}; here the presence of radicalised users contributed to the overall radicalisation of all users. 

\subsection{Contributions of This Work}
This paper provides two major contributions to extend the works discussed above. The first contribution is the development of a mathematical model inspired by~\cite{rossi} and ~\cite{lanzetti}, but with key additional features. We draw inspiration from the video-centric approach of the ``endless scroll'' feature, popularised by the For You Page, Shorts, and Reels. There is empirical evidence that ``liking'' a video on the For You page affects future recommendations~\cite{boeker}, and thus we introduce a more sophisticated stochastic decision-making mechanism for how users engage with content; users can like, dislike, or not engage with content, depending on their opinion. We also explicitly model how long users watch content, and this  determines how much the content changes their opinion. Extending the model of~\cite{rossi}, our proposed model considers a large number of users and an adjustable amount of content. The RS considers three factors when recommending a piece of content to a user: past engagement and watch rate from that user on said content, and whether the content is popular among other users (i.e., going viral~\cite{al}). The last two factors (watch rate and virality) are novel when compared to existing models by~\cite{rossi,lanzetti} and reflect key metrics incorporated into the algorithms of real-world RSs. In \cite{rossi}, the RS follows an $\epsilon$-greedy action selection, while \cite{lanzetti} further optimises over all previous timesteps. In our model, we propose an intuitive softmax-type selection process, often encountered in modern machine learning algorithms. 

Our second key contribution is the use of Monte Carlo numerical simulations to systematically explore the key factors that influence RS-opinion dynamics in a closed loop process. We investigate how the softmax parameter of the RS affects both the RS's performance and user opinions, as well as the impact of increasing the diversity of available content. Finally, we study how the virality in the RS impacts user opinions. The following statements summarise our findings: (i) after exposure to an RS, populations with neutrally distributed initial opinions are more susceptible to becoming polarised, (ii) a large number of content stances is ineffective in changing user opinions, and when a smaller number of stances is available, neutral content is effective in reducing polarisation and radicalisation, and (iii) virality-based filtering can decrease the levels of polarisation and radicalisation in user opinions.

The rest of the paper is organised as follows. In Section \ref{sec:background_model}, we present mathematical preliminaries and the model. Section \ref{sec:problem_form},  describes the Monte Carlo simulation setup and relevant metrics. Our main results are presented in Sections \ref{sec:rq1}--\ref{sec:rq3}. Section \ref{sec:conc} concludes the paper.

\section{Model}\label{sec:background_model}
\begin{figure}[h]
    \centering
    \includegraphics[width=0.99\linewidth]{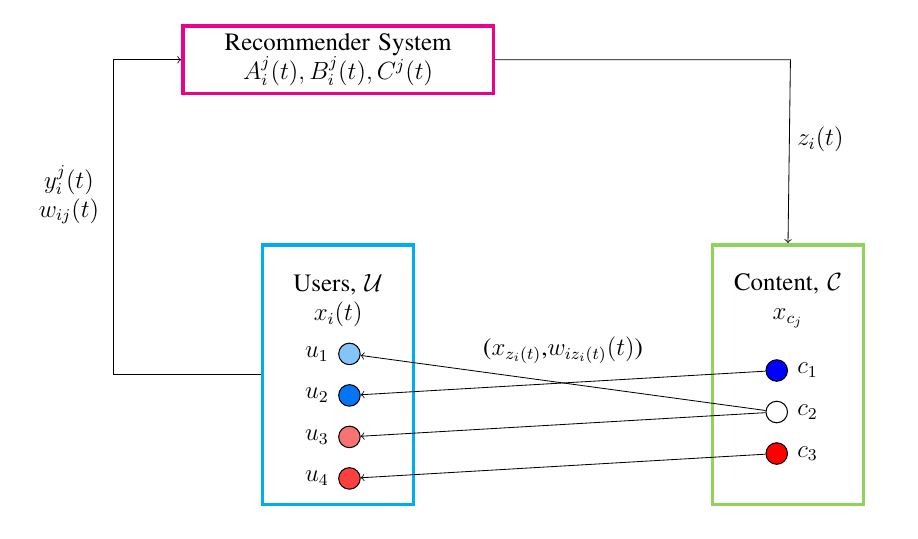}
    \caption{A schematic showing the components of the model (users, content, RS) as a closed-loop system. Here, the example consists of four users ($n=4$), and three pieces of content ($k=3$). At each timestep, user~$i$ receives a piece of content $z_i(t)$ from the RS, with stance $x_{z_i(t)}$. The user watches the content for a time $w_{iz_i(t)}(t)$, according to \eqref{eq:watch_rate}, and engages with the content, according to \eqref{eq:log_linear_like}. Watching this content causes user~$i$'s opinion $x_i(t)$ to evolve via \eqref{eq:user_opinion}. The user's engagement decision, $y_i^j(t)$, and their watch rate, $w_{ij}(t)$ are then relayed back the RS and used to update user-specific and content-specific variables $A_i^j(t)$, $B_i^j(t)$ and $C^j(t)$. These variables are used to better tailor the next recommendation, $z_i(t+1)$ according to \eqref{eq:softmax_func}, closing the loop.}
    \label{fig:rec_system}
\end{figure}

In this section, we present and discuss the proposed model, a schematic of which is shown in Fig.~\ref{fig:rec_system}. There are three key components, which are covered sequentially: the graph representing the social media environment, the user behaviour and opinion dynamics, and the algorithm governing the RS. 

\subsection{Social Media Environment Represented as a Graph}
We represent the system as a time-varying graph with two types of nodes. In particular, consider the directed graph $\mathcal{G}(t)=(\mathcal{U},\mathcal{C},\mathcal{E}(t))$. There are $n\geq 2$ users, indexed by the node set $\mathcal{U}=\{u_1,u_2,\ldots,u_n\}$, and they interact with content about a specific topic at discrete timesteps $t=0,1,2,\ldots$. For each user~$i$ we assume that they hold an opinion $x_i(t) \in [-1,1]$, which is a real valued scalar that can evolve over time as a result of consuming and interacting with the content. The content is indexed by the node set $\mathcal{C}=\{c_1,c_2,\ldots,c_k\}$. Each piece of content~$j$ has a particular opinion stance, 
\begin{equation}\label{eq:x_cj}
    x_{c_j} = \frac{2}{n-1}(j-1)-1
\end{equation}
that does not change over time. Thus, the $k$ pieces of content are equally spaced between $-1$ and $1$.

The quantities $x_i(t)$ and $x_{c_j}$ are abstract representations of the opinions of the users and the stance of the content, and can be understood as follows. Consider a specific topic, e.g. COVID-19 vaccinations are safe and effective. Then, $x_i(t)$ represents user~$i$'s orientation towards the statement on a continuous range, so that $x_i(t) = -1$ and $x_i(t) = 1$ represent maximal rejection and maximal support of the statement, while $x_i(t) = 0$ means they are neutral~\cite{anderson}. Meanwhile, $x_{c_j}$ represents the stance of the content on the same topic, so that $x_{c_1}$ and $x_{c_k}$ represent content which are fully against and fully supportive of COVID-19 vaccinations. Note that each content piece $j$ does not represent a specific instance of content (e.g. a specific news article), but rather, it represents any content which conveys a stance equal to $x_{c_j}$.

The RS is not part of the graph, but rather, can be considered as a separate entity that generates the edges which connect users with content on $\mathcal{G}(t)$, representing the delivery of said content to each user's social media feed. More specifically, at each timestep $t$ and for each user~$i$, the RS generates a directed edge $(c_j,u_i)$ from a single selected piece of content~$j$ to user~$i$. The specific process of how this content is selected is detailed in the sequel, but this process generates the edges of $\mathcal{G}(t)$, captured by the set $\mathcal{E}(t)$. Note that edges are always from content nodes to user nodes, and thus $\mathcal{G}(t)$ is a time-varying bipartite graph. We use the variable $z_i(t) \in \mathcal C$ to record the content recommended to user~$i$ at time $t$, and thus $x_{z_i(t)}$ is the stance of the content seen by user~$i$.

\begin{table}
\caption{Summary of key notation used throughout the paper.}
\label{tab:notation}
\centering
\renewcommand{\arraystretch}{1.15} 
\setlength{\tabcolsep}{4pt}        

\begin{tabularx}{\columnwidth}{|c||X|}
\hline
\textbf{Variable} & \textbf{Definition} \\
\hline\hline
$\tau$ & Total number of timesteps in the simulation \\
\hline
$\mathcal{U} = \{u_1, u_2, \ldots, u_n\}$ & Set of $n \geq 2$ users \\
\hline
$\mathcal{C} = \{c_1, c_2, \ldots, c_k\}$ & Set of $k \geq 2$ content \\
\hline
$x_i(t) \in [-1,1]$ & Opinion of user~$i$ at timestep~$t$ \\
\hline
$x_{c_j} \in [-1,1]$ & Fixed stance of content~$j$ \\
\hline
$x_{z_i(t)} \in [-1,1]$ & Stance of content $z_i(t) \in \mathcal{C}$ recommended to user~$i$ at timestep~$t$ \\
\hline
$w_{ij}(t) \in [0,1]$ & How long user~$i$ watched content~$j$ \\
\hline
$y_i^j(t) \in \{1, -1, 0\}$ & User’s engagement decision \\
\hline
$\beta_i \in [0, \infty)$ & User’s rationality for engagement decision \\
\hline
$\lambda_i \in [0,1]$ & User’s susceptibility to social influence \\
\hline
$\overline{w}_{ij} \in [0,1]$ & Average watch rate of user~$i$ for content~$j$ \\
\hline
$\delta \in [1, \tau]$ & Memory of the recommender system \\
\hline
$\omega \in [0,1]$ & Weight of virality-based filtering \\
\hline
$\alpha \in [0, \infty)$ & Softmax parameter \\
\hline
\end{tabularx}
\end{table}

\subsection{User Behaviour and Dynamics}

We model three key processes for a user when they are presented with a piece of content: the user's engagement decision (whether they like the content), the user's watch rate (how long they watch the content), and the user's opinion dynamics. These processes are modelled using TikTok's For You page as an inspiration, but are intended to be general and representative of many different social media experiences.

\subsubsection*{User Engagement Decision} \label{subsubsec:user_engagement}

The first process is how a user engages with content that they consume. In many platforms, there is an ability to click a ``like'' button or ``dislike'' button as a way of engaging with content (and a user can also choose to do neither). At each timestep $t$, user~$i$ is recommended content~$j$, and they must choose how to engage with the content. This engagement is captured by the variable $y_i^j(t)$, with $y_i^j(t) = 1$ representing liking the content, $y_i^j(t) = -1$ disliking the content, and $y_i^j(t) = 0$ indicating the user remained neutral and did not engage. The decision model is inspired by game theory and utility theory; three payoff functions ($\pi_i(1), \pi_i(-1)$ and $\pi_i(0)$) dictate the payoff or utility a user obtains from the engagement decision. This payoff is dependent on the difference $\Bar{x}_i(t) = x_i(t) - x_{z_i(t)}$. We propose
\begin{subequations}\label{eq:payoff_engage}
    \begin{align}
        \pi_i(y_i^j(t) = 1) & = 0.5\cos(1.5\Bar{x}_i(t))+0.5 \label{eq:payoff_engage_like} \\
        \pi_i(y_i^j(t) = - 1) & = -0.5\cos(1.5\Bar{x}_i(t))+0.5 \label{eq:payoff_engage_dislike} \\
        \pi_i(y_i^j(t) = 0) & = 0.5\cos(3\Bar{x}_i(t) -\pi)+0.5. \label{eq:payoff_engage_neutral} 
    \end{align}
\end{subequations}
In opinion formation models, it is standard to use distance-based functions that depend on the similarity of the opinions $x_i(t)$ and $x_{z_i}(t)$ to capture the influence between users~\cite{mas2013cultural}. Here, we adopt the same principle for the payoff functions in~\eqref{eq:payoff_engage_like} and \eqref{eq:payoff_engage_dislike}. These are even functions that are monotonically decreasing with respect to $\vert\Bar{x}_i(t)\vert$. The function in~\eqref{eq:payoff_engage_neutral} captures the relationship between the two functions $\pi_i(y_i^j(t) = 1)$ and $\pi_i(y_i^j(t) = - 1)$, such that it is maximal when the two are equal and zero when either takes a value of $1$. Graphical representations of the functions are given in the supplementary material for the interested reader.

User~$i$'s engagement decision, $y_i^j(t)$, is determined by a stochastic process. Specifically, the probability of choosing $y_i^j(t) =  s$ is
\begin{equation}\label{eq:log_linear_like}
\mathbb{P}[y_i^j(t) = s] = 
\frac{\exp\{\beta_i\pi_i(s)\}}{D_i},
\end{equation}
where
\begin{equation*}
D_i = 
\exp\{\beta_i\pi_i(-1)\} +
\exp\{\beta_i\pi_i(0)\} +
\exp\{\beta_i\pi_i(1)\}.
\end{equation*}
where $\beta_i \in [0, \infty)$ represents user~$i$'s rationality. This specific process has been widely used to model human decision-making with bounded rationality, including in stochastic games~\cite{blume}, and is used widely in Machine Learning as the softmax function.

\subsubsection*{User Watch Rate}

Users are immediately presented with content by design on TikTok and many other platforms like Shorts and Reels. Therefore, when the platform is opened, users immediately watch content until they swipe to the next video. We model this type of consumption as with the engagement decision above; this process is based on the user's opinion, $x_i(t)$, and the stance of the recommended content, $x_{z_i(t)}$. The user's watch rate at timestep $t$ is represented by $w_{iz_i(t)}(t) \in (0,1)$ and affects future recommendations and the user's opinion at the next timestep. Note that $w_{iz_i(t)}(t)$ represents the fraction of the content they consume, e.g. $w_{iz_i(t)}(t) \approx 0$ and $w_{iz_i(t)}(t) \approx 1$ could represent the user watching a video briefly before moving on, and watching the entire video, respectively.

First, let us define an intermediate variable $v(t) = x_i(t) x_{z_i(t)}$. We posit that each user's watch rate is determined by the following function
\begin{equation}\label{eq:watch_rate}
    w_{iz_i(t)}(t) = \dfrac{1}{1+e^{-\gamma v(t)}},
\end{equation}
where $\gamma>0$ is a tuning parameter. The function is plotted with various $\gamma$ values in the supplementary material. This function indicates that user~$i$ consumes more content if the content and their own opinion are aligned (have the same sign), and the higher the magnitude of their opinion, the more they watch. This modelling aligns with findings from~\cite{yamaguchi2023extreme} that extreme opinions are positively associated with user engagement and selective exposure theory, which refers to the tendency of individuals to prefer information that is consistent with their own pre-existing attitudes and beliefs~\cite{knobloch2014choice}.

\subsubsection*{User Opinion Dynamics}

The final user-specific process we model is the evolution of their opinion over time; this important process plays a major part in the user's engagement decision and watch rate as modelled above, and critically, allows us to understand how the recommended content shapes user opinions over time. We employ the classical Friedkin--Johnsen model~\cite{friedkin}, consistent with other works modelling the RS-user opinion dynamics problem~\cite{rossi,lanzetti}. Note that the Friedkin--Johnsen model is one of the few opinion dynamics models with extensive empirical studies supporting its applicability in a broad range of scenarios~\cite{friedkin2011social_book,friedkin2017_truthwins_pnas}.

The evolution of user~$i$'s opinion is:
\begin{align}\label{eq:user_opinion}
    x_i(t+1) &= \lambda_i\Big[w_{iz_i(t)}(t)x_{z_i(t)}+\big(1-w_{iz_i(t)}(t)\big)x_i(t)\Big] \notag \\
    &\quad + (1 - \lambda_i)x_i(0),
\end{align}
where the constant $\lambda_i \in [0,1]$ captures their susceptibility to influence. In other words, user~$i$'s updated opinion is a convex combination of their initial opinion $x_i(0)$, their current opinion $x_i(t)$, and the stance of content they consumed, $x_{z_i(t)}$. Content influence is minimised and maximised as $\lambda_i \to 0$ and $\lambda_i \to 1$, respectively, and content influence increases as user~$i$ consumes the content for longer, as $w_{iz_i}(t)\to 1$.

\subsection{Recommender System}
Having introduced the model of the user-specific processes, we now introduce how the RS assigns personalised recommendations to its users. To that end, we define the variables $A_i^j(t)$, $B_i^j(t)$ and $C^j(t)$, which relate to user~$i$'s engagement decision, user~$i$'s watch rate and the viral element, respectively. The RS uses these three variables to calculate a payoff or utility $P_i^j(t)$, which it then uses to determine the recommended content piece at the next timestep via a stochastic process.

\subsubsection*{User Likes}
Here, we describe how the RS keeps track of each user's engagement with content in the past. First, we define the column vector $R_i(t) = [r_i^1(t), r_i^2(t), \dots, r_i^k(t)]^\top \in \mathbb{N}_0^k$. The entry $r_i^j(t) \in \mathbb N_0$ captures the number of times user~$i$ has been recommended content piece $j$ before timestep $t$. Thus, the RS is able to track its recommendations to each user, and it is updated as $R_i(t+1) = R_i(t) + \mathbf{e}_{z_i(t)}$, with $\mathbf{e}_j\in \mathbb{R}^k$ being the $j^{th}$ canonical basis vector of $\mathbb{R}^k$. 

Recall from Section \ref{subsubsec:user_engagement} that $y_i^j(t)$ represents user~$i$'s engagement decision regarding the recommended content piece $j$ at timestep $t$. We propose that the RS keeps track of what each user has liked in the past. Using an indicator function $\textbf{1}$, we define $a_i^j(t) = \sum_{s=1}^t \textbf{1}_{\{y_i^j(s)=1\}}$
as the number of times that user~$i$ has liked a particular content recommendation $j$.

Next, we define the variable 
\begin{equation}\label{A_value}
    A_i^j(t) = \dfrac{a_i^j(t)}{r_i^j(t)}.
\end{equation}
as the ratio between the number of times user~$i$ has liked a piece of content~$j$, and the total number of times content~$j$ has been recommended user~$i$. Conveniently, $A_i^j(t) \in [0,1]$. If $r_i^j(t) = 0$ (user~$i$ has not yet been recommended content piece $j$), then we set $A_i^j(t) = 0$, to avoid domain errors.

\subsubsection*{User Watch Rate}
Empirical evidence from~\cite{boeker} found that watching a TikTok video for at least 25\% of its duration significantly affected subsequent content recommendations on the For You page. Motivated by this, we define the variable $\overline{w}_{ij}(t) = \sum_{s=1}^tw_{ij}(s)/r_i^j(t)$ as the average watch rate of user~$i$ for content~$j$. We define the variable,
\begin{align}\label{eq:B_value}
    B_i^j(t) = \frac{1}{2}\tanh(\mu(\overline{w}_{ij}(t)-0.25))+0.5,
\end{align}
where $\mu>0$ is a tuning parameter, that affects future recommendations; the function for different $\mu$ can be found in the supplementary material where it can be seen that the function illustrates the aforementioned observation from~\cite{boeker}. By design, $B_i^j(t) \in (0,1)$ takes on a small strictly positive value when $\overline{w}_{ij}(t)=0$, and approaches $1$ if user~$i$ is consuming content~$j$ heavily when it is recommended to them.

\subsubsection*{Virality-Based Filtering}
It is important that the RS recognises which content is popular in the system. Let $L^j(t)$ denote the total number of likes content~$j$ has received at timestep $t$ from all users. We define, for some memory size $\delta \in \mathbb N_+$:
\begin{equation}\label{eq:C_value}
    C^j(t) = \dfrac{\sum_{s=1}^{\min(t,\delta)} L^j(t-s)}{\sum_{s=1}^{\min(t,\delta)} \sum_{\ell=1}^kL^\ell(t-s)}.
\end{equation}
as how much positive engagement, or likes, content~$j$ is receiving in comparison to everything else. This enables us to compare the popularity of different content, i.e. determine if a particular piece of content has gone viral.

Notice that only the previous $\delta$ timesteps are accounted for, to model the freshness of viral content and the memory of the RS. By design, $C^j(t) \in [0,1]$, and also represents the collaborative filtering element in the RS, as it is recording what content is performing well (likes for all users) in the system. We henceforth refer to this variable as that associated with virality-based filtering.

\subsubsection*{Generating Recommendations}
By design, $A_i^j(t)$, $B_i^j(t)$ and $C^j(t) \in [0,1]$ which enables a relative comparison of the different terms. In particular, $A_i^j(t)$ and $B_i^j(t)$ are user-specific and form the content-based filtering part of the RS, whereas $C^j(t)$ is universal across all users, and captures collaborative filtering with a focus on viral content. 

A payoff for each user~$i$ and each content~$j$ is computed:
\begin{equation}\label{eq:P_value}
    P_i^j(t) = \dfrac{(1 - \omega)(A_i^j(t)+B_i^j(t))}{2} + \omega C^j(t).
\end{equation}
The parameter $\omega \in [0,1]$ represents the weight placed on the virality-based filtering, relative to the content-based filtering. The RS then uses the payoff $P_i^j(t)$, $j = 1,2, \hdots, k$ to determine the content to recommend to user~$i$ at timestep $t+1$. Specifically, the RS uses the softmax function, a standard Machine Learning approach to transforms real values (scores, utilities) into a probability distribution~\cite{bishop}: 
\begin{equation}\label{eq:softmax_func}
    \mathbb{P}[z_i(t+1) = c_j] = \dfrac{\exp\{\alpha P_i^j(t)\}}{\sum_{j=1}^k\exp\{\alpha P_i^j(t)\}},
\end{equation}
where $\alpha \in [0, \infty)$ represents the softmax parameter~\cite{franke2023softmax}. We define the reciprocal of the softmax parameter as the temperature of the RS, i.e. $\alpha = \frac{1}{T}$. As the softmax parameter increases, the RS moves along a scale from exploration to exploitation. That is, for small $\alpha$ values, the output of the probability function approaches a uniform distribution and recommendations are assigned randomly. Conversely, as $\alpha \longrightarrow \infty$, there holds $\mathbb{P}[z_i(t+1) = c_\ell] \longrightarrow 1$ where $\ell = \mathrm{argmax}_{j\in \{1,\hdots, k\}}\{P_i^j(t)\}$. That is, the RS is more exploitative, as it will only recommend for user~$i$ the content $c_\ell$ that has the highest utility/payoff among all the content.

\begin{figure}
    \centering
    \begin{subfigure}[b]{0.35\textwidth}
        \centering
        \includegraphics[width=\textwidth]{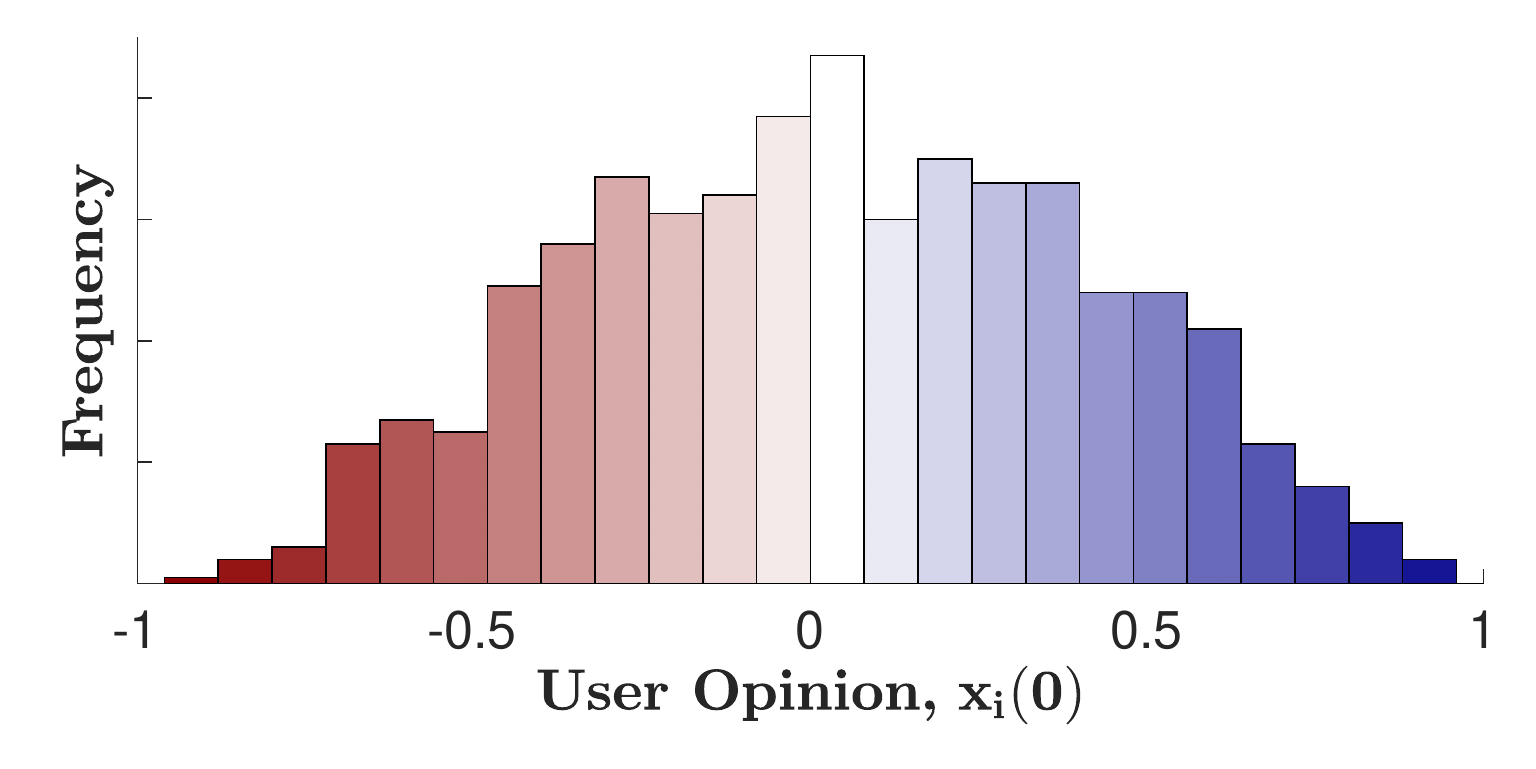}
        \caption{Neutrally distributed initial condition, NDIC.}
        \label{subfig:normal_initial}
    \end{subfigure}
    \hfill
    \begin{subfigure}[b]{0.35\textwidth}
        \centering
        \includegraphics[width=\textwidth]{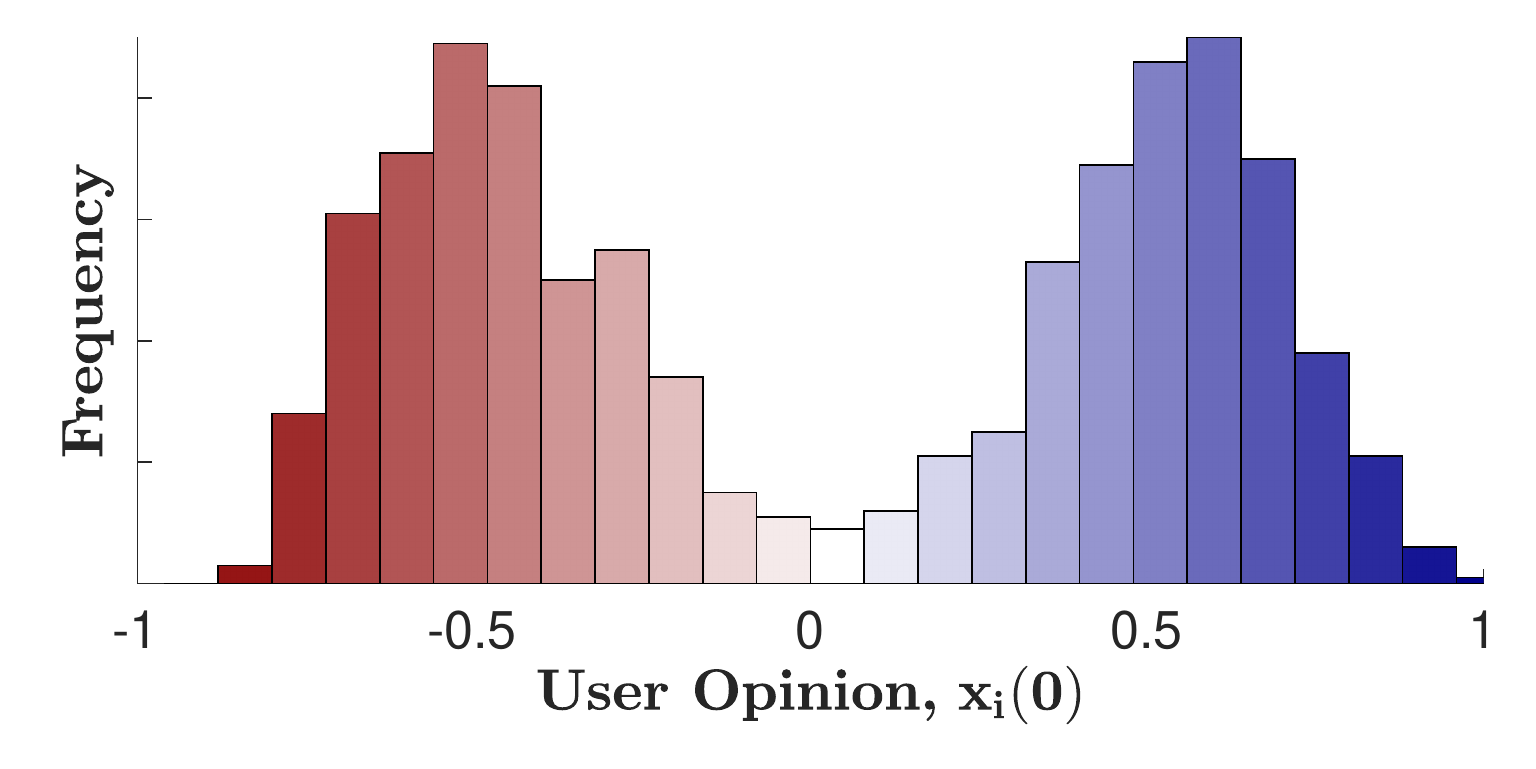}
        \caption{Bimodally distributed initial condition, BDIC.}
        \label{subfig:bimodal_initial}
    \end{subfigure}
    \caption{The two initial opinion distributions, $x_i(0)$, used throughout simulations.}
    \label{fig:initial_op_dists}
\end{figure}

\section{Methodology}\label{sec:problem_form}

\begin{figure*}[ht]
    \centering
    \begin{subfigure}[b]{0.32\textwidth}
        \centering
        \includegraphics[width=\textwidth]{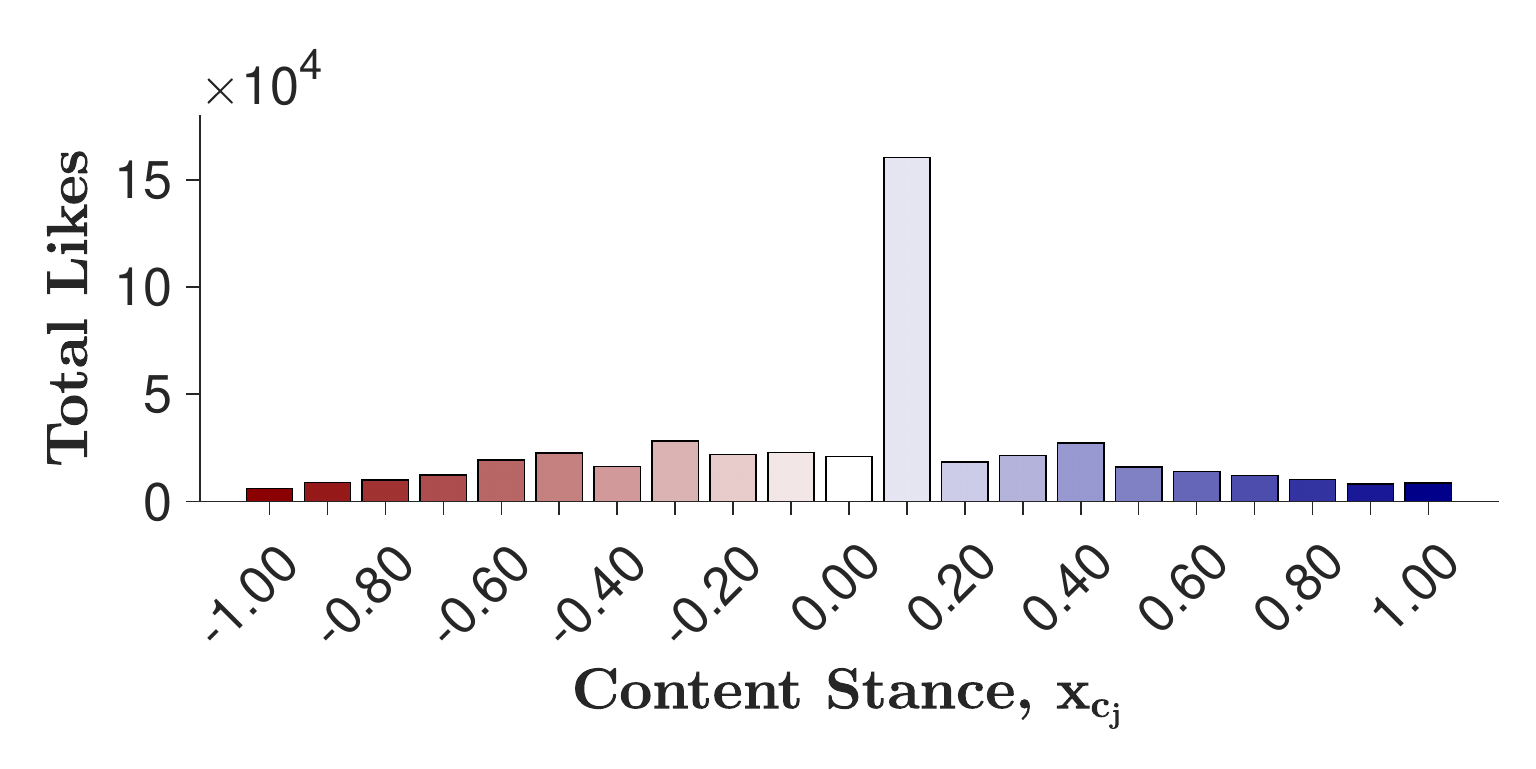}
        \caption{Total likes per piece of content at $\tau=1000$.}
        \label{subfig:rq1_beta14_likes}
    \end{subfigure}
    \begin{subfigure}[b]{0.32\textwidth}
        \includegraphics[width=\textwidth]{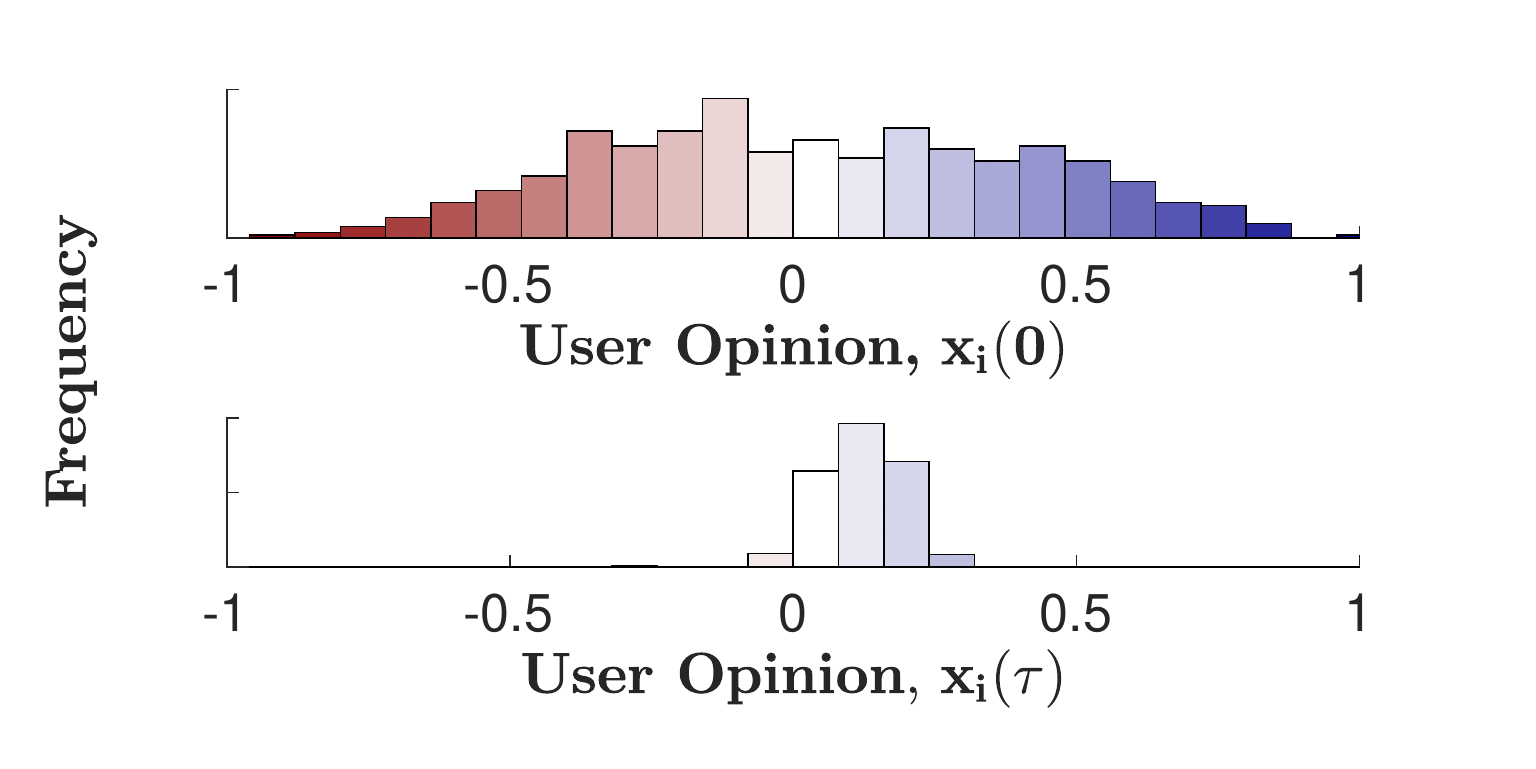}
        \caption{Changes in opinion distributions over time.}
        \label{subfig:rq1_beta14_opinions}
    \end{subfigure}
    \begin{subfigure}[b]{0.32\textwidth}
        \includegraphics[width=\textwidth]{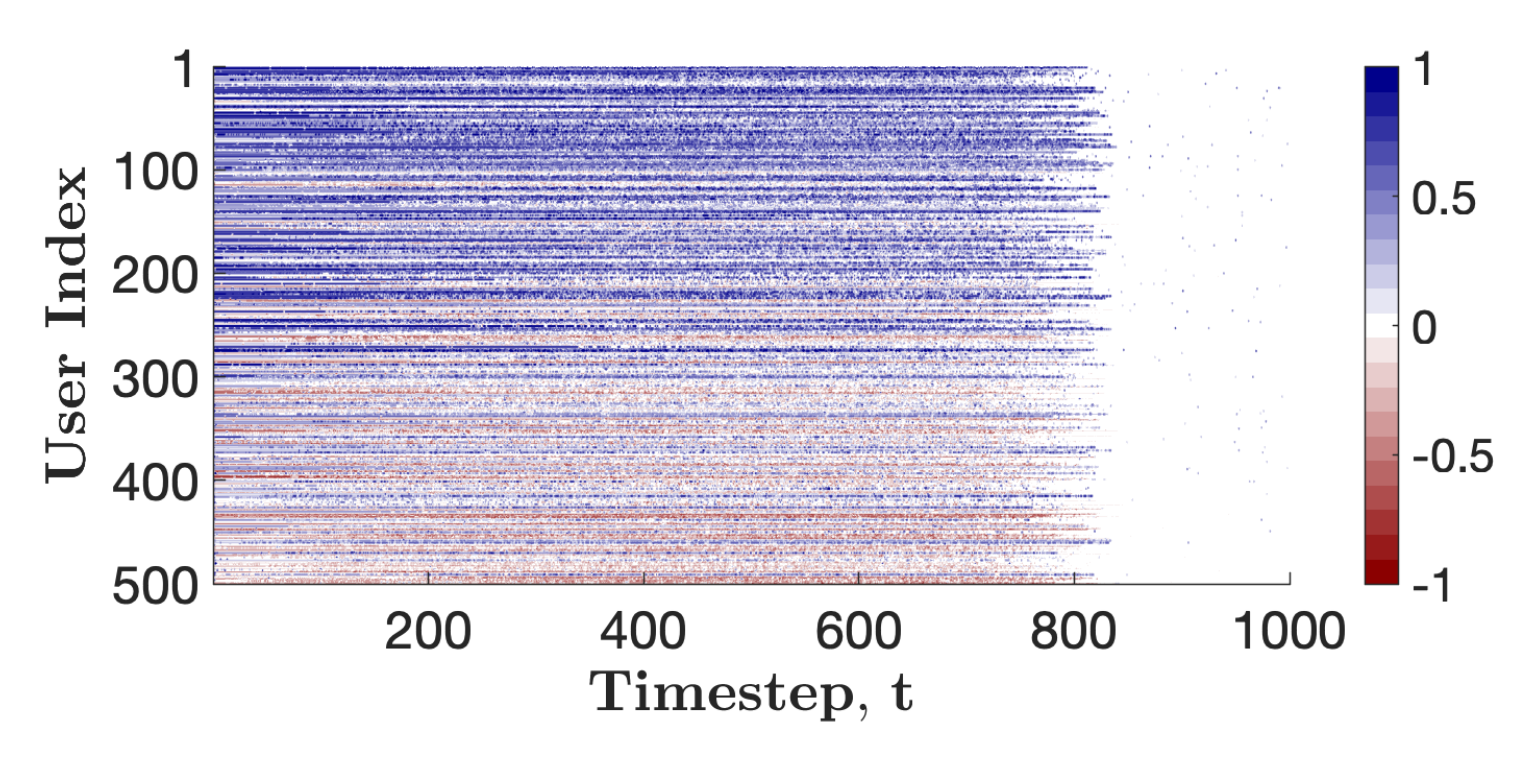}
        \caption{Content recommendations over time, indexed by $x_i(\tau)$.}
        \label{subfig:rq1_beta14_content}
    \end{subfigure}
    \vfill
    \begin{subfigure}[b]{0.32\textwidth}
        \centering
        \includegraphics[width=\textwidth]{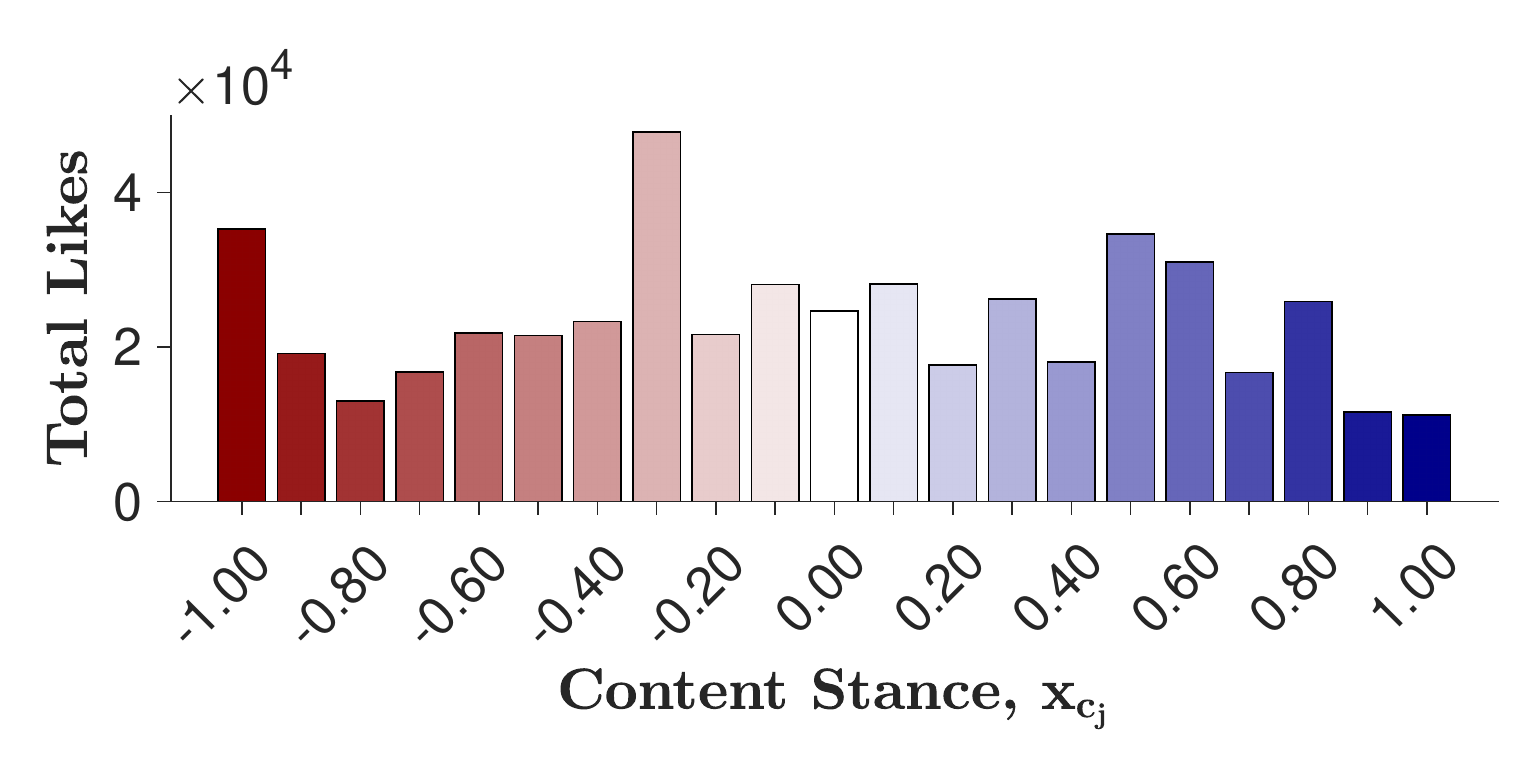}
        \caption{Total likes per piece of content at $\tau=1000$.}
        \label{subfig:rq1_beta18_likes}
    \end{subfigure}
    \begin{subfigure}[b]{0.32\textwidth}
        \includegraphics[width=\textwidth]{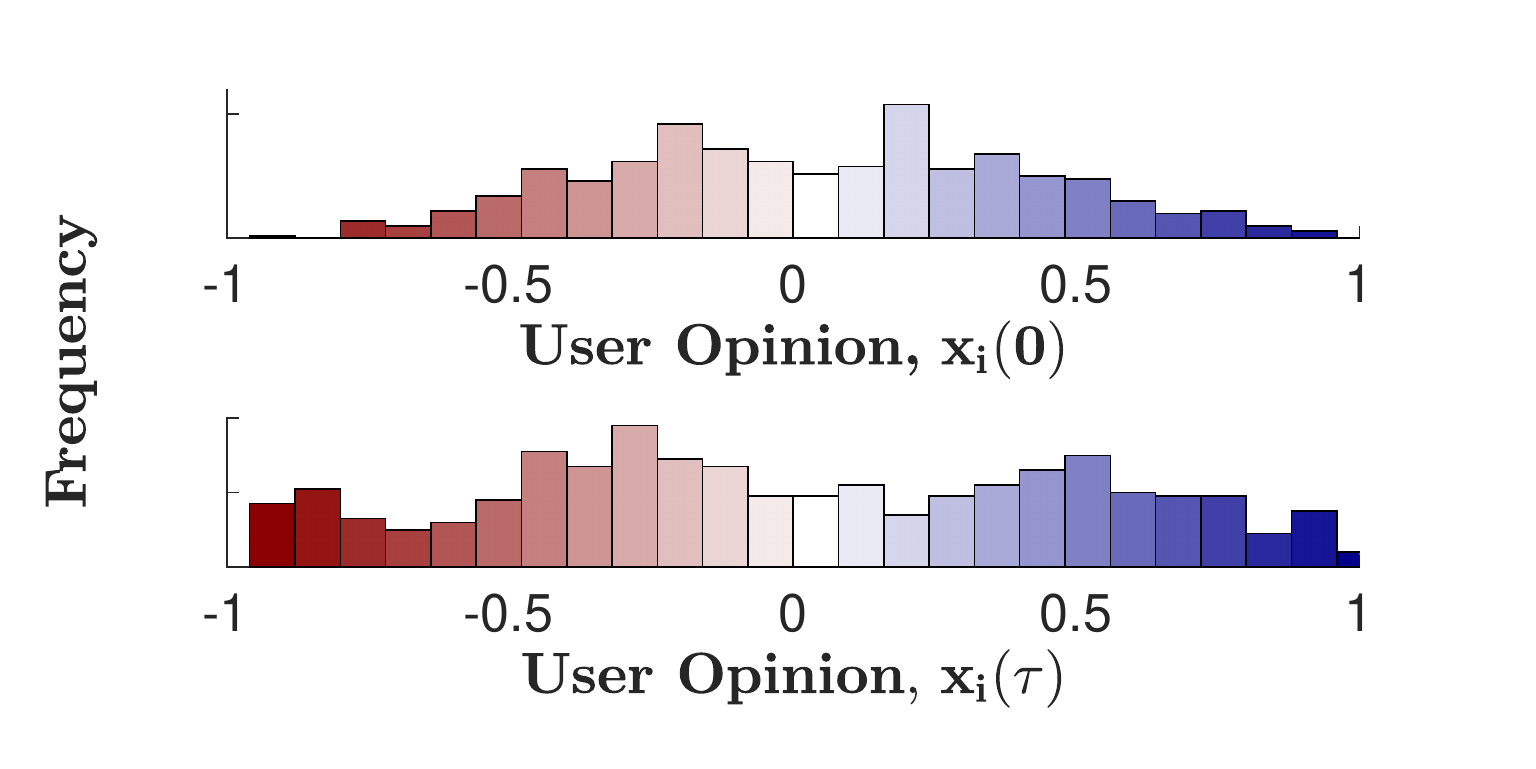}
        \caption{Changes in opinion distributions over time.}
        \label{subfig:rq1_beta18_opinions}
    \end{subfigure}
    \begin{subfigure}[b]{0.32\textwidth}
        \includegraphics[width=\textwidth]{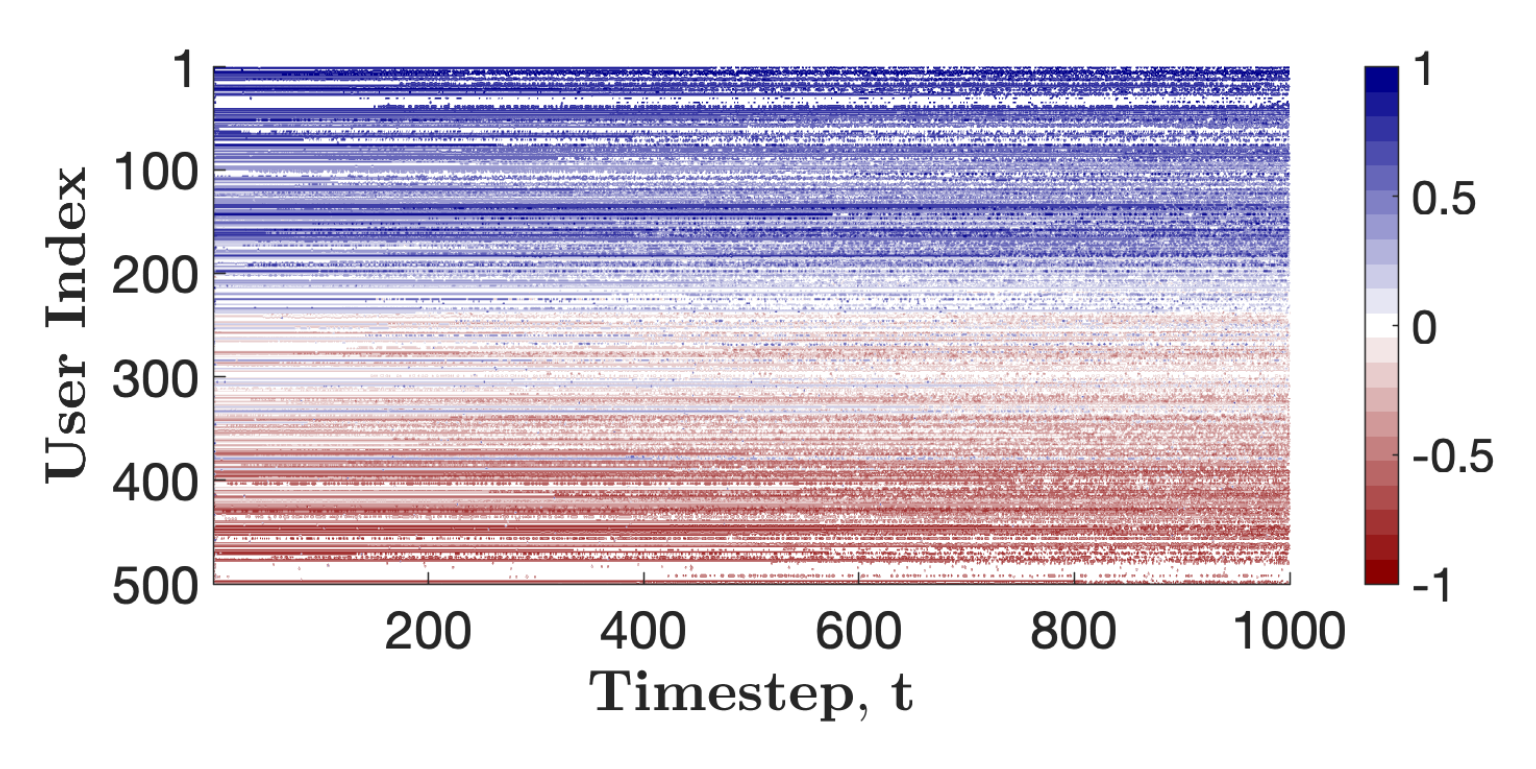}
        \caption{Content recommendations over time, indexed by $x_i(\tau)$.}
        \label{subfig:rq1_beta18_content}
    \end{subfigure}
    \caption{Example simulations showing the total likes per piece of content, the opinion distributions, and the content recommendations over time for $\alpha=14$ (Panels (a), (b), and (c)) and for $\alpha = 18$ (Panels (d), (e), and (f). In more detail, (a) and (d) show the total likes from the entire population, per piece of content. Panels (b) and (e) show the initial opinion distribution compared to the final opinion distribution, highlighting the effect of the RS on the population opinions. Finally, (c) and (f) show a contour plot representing all content recommended to each user (indexed on the $y$-axis) over time.}
    \label{fig:rq1_beta14}
\end{figure*}

We are now in a position to describe how we will use the mathematical model, detailed in Section \ref{sec:background_model}, to study its effects on users' opinions. Due to the model's stochastic nature and highly non-linear dynamics, we employ a Monte Carlo simulation approach, breaking our study into three specific research questions that we will describe in detail in the relevant sections below. These questions aim to unveil the impact of various parameters associated with the RS. As such, for each set of parameters, we run 50 replication simulations and use four metrics to quantify the outcomes, two for the RS and two for the opinion distribution of the population. For each parameter setting and each metric, we take the mean and standard deviation of the 50 replications to construct a 95\% confidence interval around our results. Broadly speaking, our first question will examine the softmax parameter $\alpha$ of the RS, our second question will examine the diversity of content stances through $k$, and our final question will focus on the impact of virality filtering by studying $\omega$ and $\delta$.

\subsection{Simulation Setup}\label{subsec:sim_setup}

Here, we outline parameters and settings that are universal across all of our simulations, while the specific adjustments are detailed in the relevant sections below. We run each simulation for $\tau = 1000$ timesteps, as preliminary simulations show that this is a sufficient amount of time for the system dynamics to stop fluctuating substantially. For each simulation, there are $n = 500$ users, with a homogeneous rationality $\beta_i = 9$ for all $u_i\in \mathcal U$. The setting $\beta_i = 9$ provides a ``sweet spot'' whereby individuals are almost always rational but have some room for spontaneity, see~\cite{ye2021collective}. 

For the two tuning parameters in \eqref{eq:watch_rate} and \eqref{eq:B_value}, we set $\gamma = 5$ and $\mu = 5$. First, $\gamma = 5$ ensures the watch rate takes on a small non-zero value when $v(t)=-1$, and rapidly increases as $v(t)$ increases, approaching $w_{iz_i(t)}(t) \approx 1$ when $v(t)=1$. This means there is some small influence even when a user watches only a small amount of the content, and near-maximal influence when their watch rate is high. Second, $\mu = 5$ ensures a small positive $B_i^j(t)$ value when $\overline{w}_{ij}(t)\approx0$, allowing for a small payoff value even when a user has not consumed a lot of that content. We set a homogenous susceptibility of $\lambda_i=0.9$, for all $u_i\in \mathcal U$, for the opinion evolution in \eqref{eq:user_opinion}. This means the content assigned by the RS has a significant but not complete influence on each user, as users will have some mild attachment to their initial opinion, $x_i(0)$. In the supplementary material, we repeat all simulations reported with $\lambda_i = 0.2$ for all users. Intuitively, the impact of the RS is much smaller, but we observe similar qualitative phenomena.

We now introduce heterogeneity into our user dynamics with their initial opinions. For the initial opinions of the users, $x_i(0)$, we randomly sample $x_i(0)$ from a distribution for each simulation. In fact, we utilise two distributions: a neutral distribution, see Fig.~\ref{subfig:normal_initial}, and a bimodal distribution, see Fig.~\ref{subfig:bimodal_initial}. We henceforth refer to these initial opinion distributions as NDIC (neutrally distributed initial condition) and BDIC (bimodally distributed initial condition), respectively. The NDIC approximates normal distribution centred at $0$ and truncated at $-1$ and $1$, representing a balanced population. The BDIC instead reflects a society that is already polarised. To create the initial distributions, we draw from a beta distribution that has codomain $[0,1]$, the shape of which is controlled by parameters $\alpha$ and $\beta$~\cite{mcdonald}. Therefore, we have for the NDIC a mean of $\mu=0$ and standard deviation of $\sigma=0.378$ and for the BDIC $\mu=0$ and $\sigma=0.524$.

\subsection{Metrics for the System}\label{subsec:metrics}

\begin{figure*}[ht]
    \centering
    \begin{subfigure}[b]{0.49\textwidth}
        \centering
        \includegraphics[width=\textwidth]{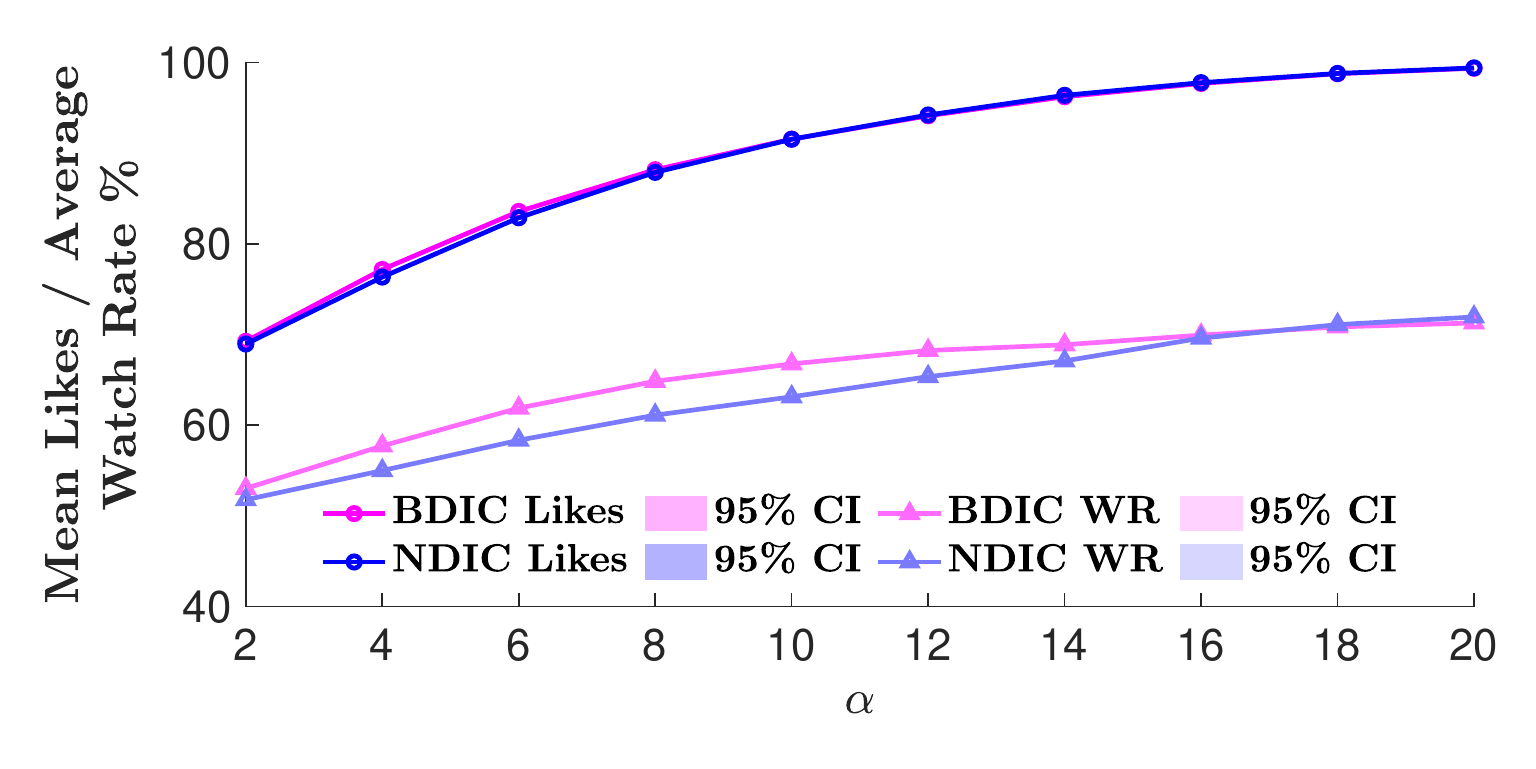}
        \caption{RS performance metrics.}
        \label{fig:RQ1_performance}
    \end{subfigure}
        \hfill
        \begin{subfigure}[b]{0.49\textwidth}
        \centering
        \includegraphics[width=\textwidth]{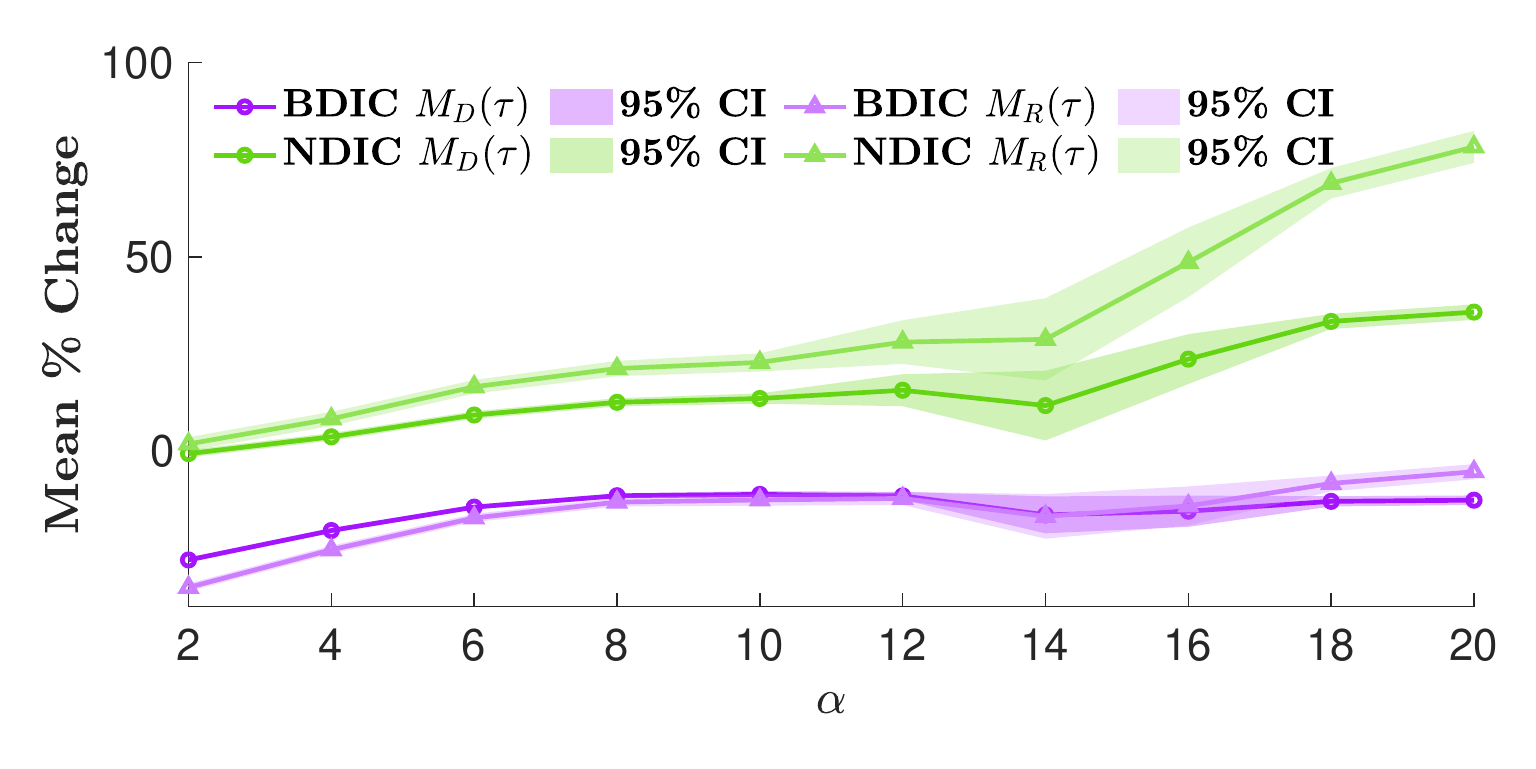}
        \caption{Opinion distribution metrics.}
        \label{fig:RQ1_polarisation}
    \end{subfigure}
    \caption{\textit{Research Question One}, in (a) the performance metrics of the RS; $\mathrm{Likes\,\%}$, \eqref{eq:likes}, and $\mathrm{Average\,Watch\,Rate\,\%}$ (WR), \eqref{eq:avg_watch_rate}, are shown for different values of the softmax parameter, $\alpha$. In (b), the final percentage changes of the polarisation metric $M_D(\tau)$, \eqref{eq:dispersion}, and radicalisation metric $M_R(\tau)$, \eqref{eq:radicalisation}, are plotted for different values of~$\alpha$.}\label{fig:RQ1}
\end{figure*}

We record the state of the system at the end of the simulation window of $\tau = 1000$ timesteps, and compute four metrics that are used to address the research questions. Two metrics examine the RS's performance, and two metrics describe the final opinion distribution to ascertain the impact of the RS on the users. 

To measure the performance of the RS, we record the user-specific processes of liking and watching recommended content. In each simulation, we record the total amount of likes received as a percentage of total amount of likes possible, i.e
\begin{equation}\label{eq:likes}
    \mathrm{Likes}\, \% = \dfrac{\sum_{j=1}^k\sum_{s=1}^\tau L^j(s)}{n\tau} \times 100,
\end{equation}
where $n\tau$ represents the total number of recommendations made by the RS to the $n$ users over $\tau$ timesteps. Similarly, we define the average watch rate of all users across that simulation as a percentage:
\begin{equation}\label{eq:avg_watch_rate}
    \mathrm{Average \, Watch \, Rate}\, \% = \dfrac{\sum_{s=1}^\tau \sum_{i=1}^n w_{iz_i(s)}(s)}{n\tau} \times 100.
\end{equation}

For the population, we are especially interested in polarisation and radicalisation. To capture the former, we use dispersion $M_D$ as introduced in \cite{bramson}, which  measures the shape of the whole opinion distribution and how far each user opinion is away from the mean. Formally, we have:
\begin{equation}\label{eq:dispersion}
    M_D(t)=\dfrac{2}{n}\sum_{i=1}^n|x_i(t)-\Bar{X}(t)|,
\end{equation}
where $\Bar{X}(t)$ represents the average user opinion at timestep $t$. Dispersion can increase as subgroups move further apart, when user opinions move further away from the middle ground toward the fringes of the distribution, and when the distribution flattens~\cite{bramson}. The second metric $M_R$ focuses on measuring the presence of extremist opinions in population:
\begin{equation}\label{eq:radicalisation}
    M_R(t) = \dfrac{1}{n}\sum_{i=1}^n(x_i(t))^2.
\end{equation}
This metric increases as more users shift towards the fringes of the population, i.e. become radicalised.

Note that the metrics $M_D$ and $M_R$ serve two distinct purposes. The dispersion metric, $M_D$, measures the shape of the opinion distribution as a whole and how that shape changes, including whether the two subgroups are moving further away or closer together. This allows for an insight into how polarised user opinions are. Meanwhile, the radicalisation metric, $M_R$, measures the prevalence of extremist opinions in the distribution. Both metrics allow insight into how the opinion distribution is moving away from the neutral. 

\section{Research Question One: Softmax Parameter}\label{sec:rq1}

Our first research question focuses on the impact of the softmax parameter $\alpha$, used by the RS to probabilistically determine what content to recommend to each user, as in \eqref{eq:softmax_func}. Intuitively, smaller and larger values of $\alpha$ correspond to an RS that aims to be more explorative (recommending a range of content to a user) and more exploitative (recommending only content that is likely to receive user engagement), respectively. To that end, we fix $k = 21$, $\omega = 0.5$ and $\delta = 5$, and then vary $\alpha$ from $\alpha = 2$ to $\alpha = 20$ with a step size of $2$.  

Before we present the main results of this section, we showcase two exemplar simulations to help visualise certain mechanics of the mathematical model, including the content recommendation and opinion evolution over time. Fig.~\ref{fig:rq1_beta14} shows a single simulation with $\alpha = 14$ (panels (a), (b) and (c)) and with $\alpha = 18$ (panels (d), (e) and (f)). With $\alpha = 14$, a single piece of content goes \textit{viral}. As seen in Fig.~\ref{subfig:rq1_beta14_likes}, the content with stance $x_{c_j} = 0.1$ has over 100,000 more likes than the next most liked piece of content. Meanwhile, Fig.~\ref{subfig:rq1_beta14_opinions} indicates that from an NDIC, the opinions of the users coalesce around $0.1$ to $0.2$, which shows the viral content $x_{c_j} = 0.1$ has a substantial effect on the opinion evolution of all users. In Fig.~\ref{subfig:rq1_beta14_content}, we see the recommended content is broadly distributed for a long time (red and blue in roughly equal amount at timesteps below $t = 800$) but then the viral content emerges at $t\approx 800$ and is recommended to all users thereafter. 

With a larger softmax parameter at $\alpha = 18$, the RS is less explorative, preferring to recommend highly personalised content it thinks will get consistent engagement from the specific user. In Fig.~\ref{subfig:rq1_beta18_likes}, the number of likes per piece of content is more broadly distributed, and no content has gone viral. Rather, users receive more consistent and extreme recommendations (compare this to Fig.~\ref{subfig:rq1_beta14_likes}). The effect of more extreme content being recommended and liked by users can be seen in Fig.~\ref{subfig:rq1_beta18_opinions} with more users at the fringes of the opinion range; for this simulation at $\alpha=18$, there was a $88.34\%$ increase in $M_R$ and $39.77\%$ increase in $M_D$. These large increases are due to users consistently receiving more extreme recommendations, as shown in Fig.~\ref{subfig:rq1_beta18_content}.

\begin{figure*}[ht]
    \centering
    \begin{subfigure}[b]{0.49\textwidth}
        \centering
        \includegraphics[width=\textwidth]{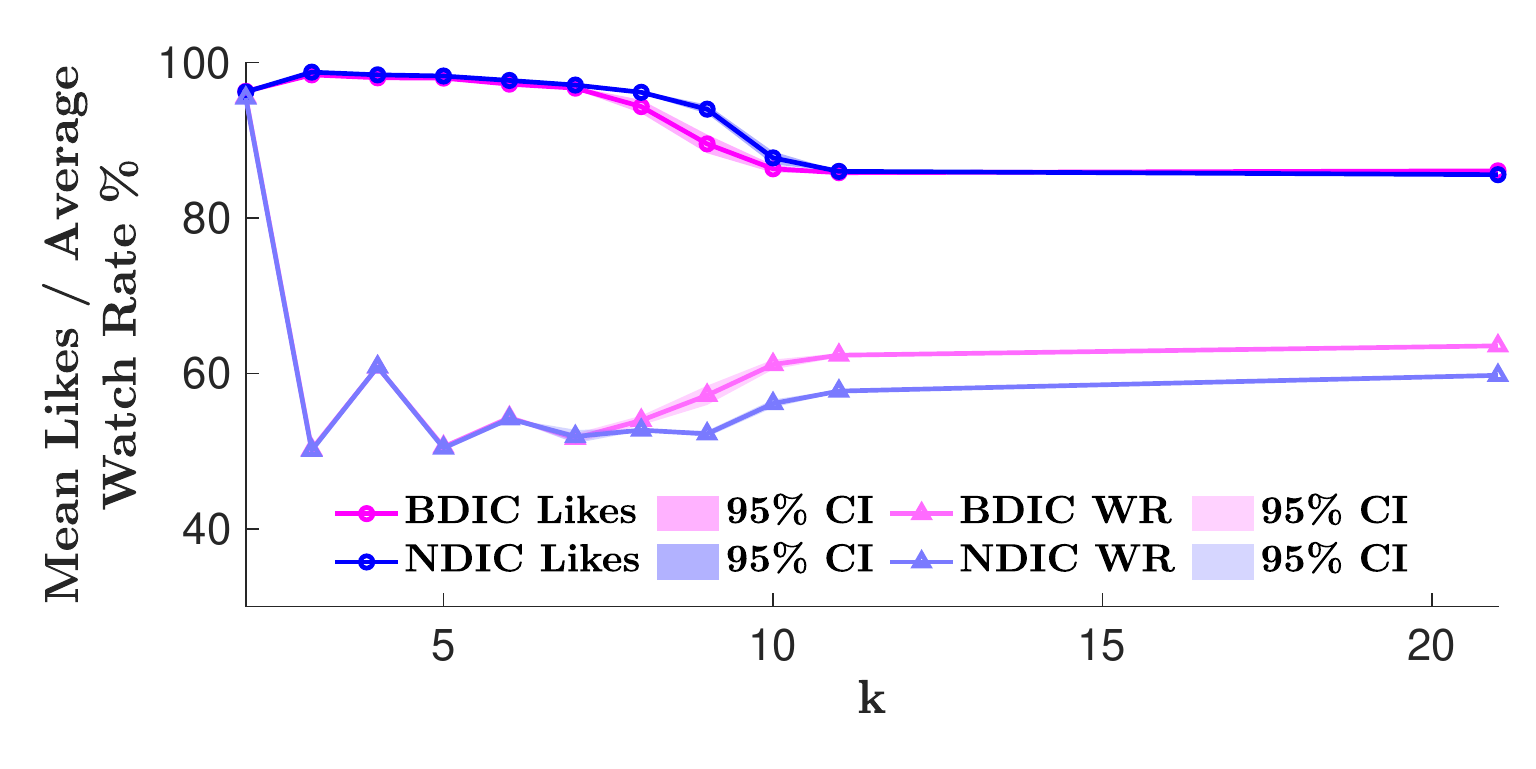}
        \caption{RS performance metrics.}
        \label{subfig:RQ2_performance_21}
    \end{subfigure}
    \hfill
    \begin{subfigure}[b]{0.49\textwidth}
        \centering
        \includegraphics[width=\textwidth]{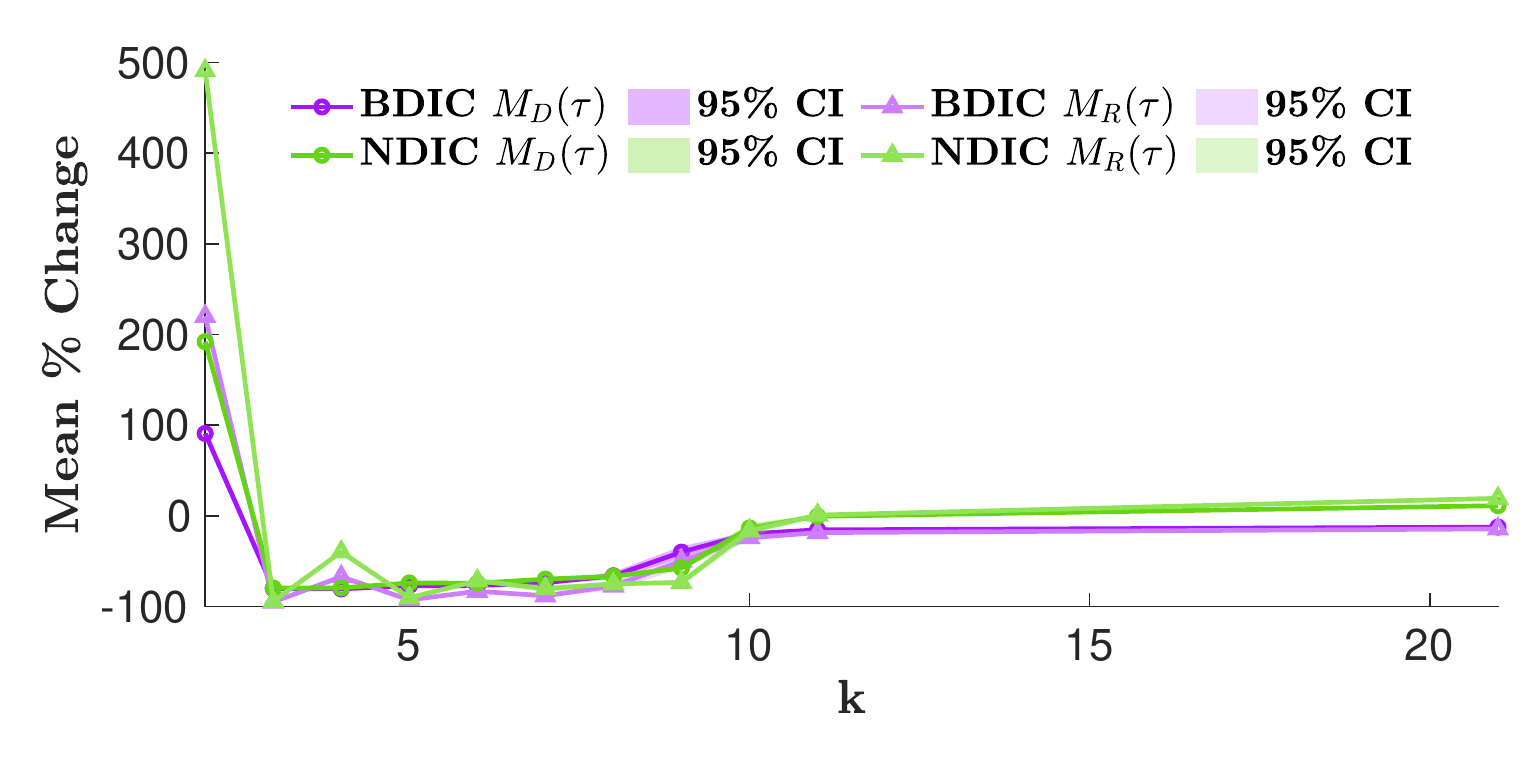}
        \caption{Opinion distribution metrics.}
        \label{subfig:RQ2_polarisation_21}
    \end{subfigure}
    \caption{\textit{Research Question Two}, in (a) the performance metrics of the RS; $\mathrm{Likes\,\%}$, \eqref{eq:likes}, and $\mathrm{Average\,Watch\,Rate\,\%}$ (WR), \eqref{eq:avg_watch_rate}, are shown for different number different of content, $k$. In (b), the final percentage changes of the polarisation metric $M_D(\tau)$, \eqref{eq:dispersion}, and radicalisation metric $M_R(\tau)$, \eqref{eq:radicalisation}, are plotted for different values of~$k$.}
    \label{fig:RQ2_performance}
\end{figure*}

\subsection{Monte Carlo Simulations}

We now present the full results for \textit{Research Question~One}. As noted in Section~\ref{sec:problem_form}, for each value of $\alpha$, we run $50$ independent simulations. We compute the mean and $95\%$ confidence intervals for our four metrics described in Section~\ref{subsec:metrics}, and present the results in Fig.~\ref{fig:RQ1}. 

Fig.~\ref{fig:RQ1_performance} shows the two RS metrics (mean and $95\%$ confidence interval) at the end of the simulation, i.e. $t = \tau$, as a function of the softmax parameter $\alpha$. Clearly, as $\alpha$ increases, user engagement also increases. Interestingly, the average watch rate of the NDIC is strictly lower than that of the BDIC until $\alpha=18$, when the NDIC and BDIC average watch rates converge. From the formulation of the watch rate function, \eqref{eq:watch_rate}, it is not surprising that a population with an NDIC would yield a lower average watch rate than a population with BDIC. The term $v(t)$ increases as the magnitude of the opinion increases, and the recommended content aligns with the opinion. User opinions typically have a larger magnitude in a BDIC than in an NDIC, explaining the higher average watch rate in the former. 

Fig.~\ref{fig:RQ1_polarisation} showcases the opinion distribution metrics as a function of $\alpha$. Here, we plot the percentage change in $M_D$ and $M_R$ from $t=0$ to $t=\tau$, e.g. $100\%\times(M_R(\tau)-M_R(0))/M_R(0)$. This enables us to make a fair comparison between the NDIC and BDIC, in terms of how the RS changes the initial opinion distribution to the final opinion distribution. We observe that a population with an NDIC will always become more polarised and more radicalised, as $M_R$ and $M_D$ both increase as $\alpha$ increases. There is substantial change beyond $\alpha = 16$,  with a 68.89\% increase in $M_R$ and 34.35\% increase in $M_D$. This, paired with the previous observation in Fig.~\ref{fig:RQ1_performance} of the NDIC's average watch rate converging with that of the BDIC when $\alpha \geq 16$, suggests the RS shifts a population with an NDIC towards a polarised distribution like the BDIC. As for the BDIC, consistently, the change in both opinion distribution metrics is below zero, meaning the population is becoming less polarised and less radicalised as $\alpha$ increases. That is, for a population with a BDIC, the RS is bringing people's opinions closer together, but having an opposite effect for populations with an NDIC.

While the trend is largely monotonically increasing for all metrics, notice that for the NDIC in Fig.~\ref{fig:RQ1_polarisation}, there is a decreasing trend at $\alpha=14$. This discrepancy at $\alpha=14$ is due to one piece of content going viral, as showcased in the exemplar simulation in Fig.~\ref{fig:rq1_beta14}. This suggests that when the softmax parameter of the RS is $\alpha=14$, there is a balance between exploration and exploitation, such that it is likely for a piece of content to go viral and be recommended to all users. The large confidence interval around the mean of $M_R$ and $M_D$ at $\alpha=14$ suggests that a piece of content going viral is not a consistently occurring phenomenon, and does not happen in every simulation. When the softmax parameter is lower, one has a more explorative RS that recommends content randomly. In contrast, higher $\alpha$ values lead to a more exploitative RS that exploits its early success by consistently recommending content that users have previously explicitly liked. In both cases, the likelihood of one content going viral appears to be substantially lower. We conclude by noting that $\alpha$ is not the only parameter that can generate viral content. In the sequel, we will observe that other RS parameters, especially $\omega$ and $\delta$, can also greatly impact the likelihood of viral content. 

\begin{figure*}[ht]
    \centering
        \begin{subfigure}[b]{0.24\textwidth}
        \centering
        \includegraphics[width=\textwidth]{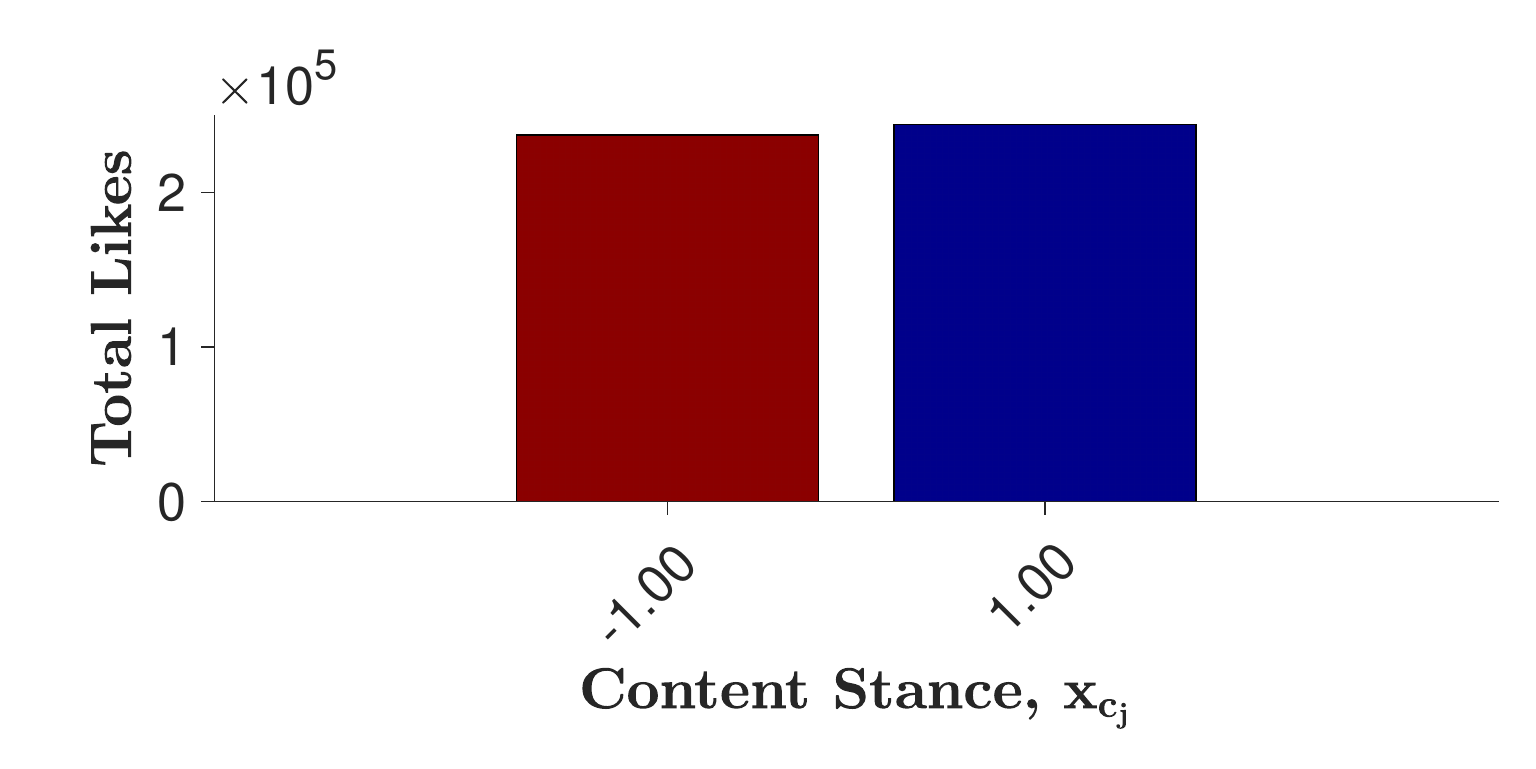}
        \caption{$k=2$}
        \label{subfig:RQ2_k2_likes}
    \end{subfigure}
    \begin{subfigure}[b]{0.24\textwidth}
        \centering
        \includegraphics[width=\textwidth]{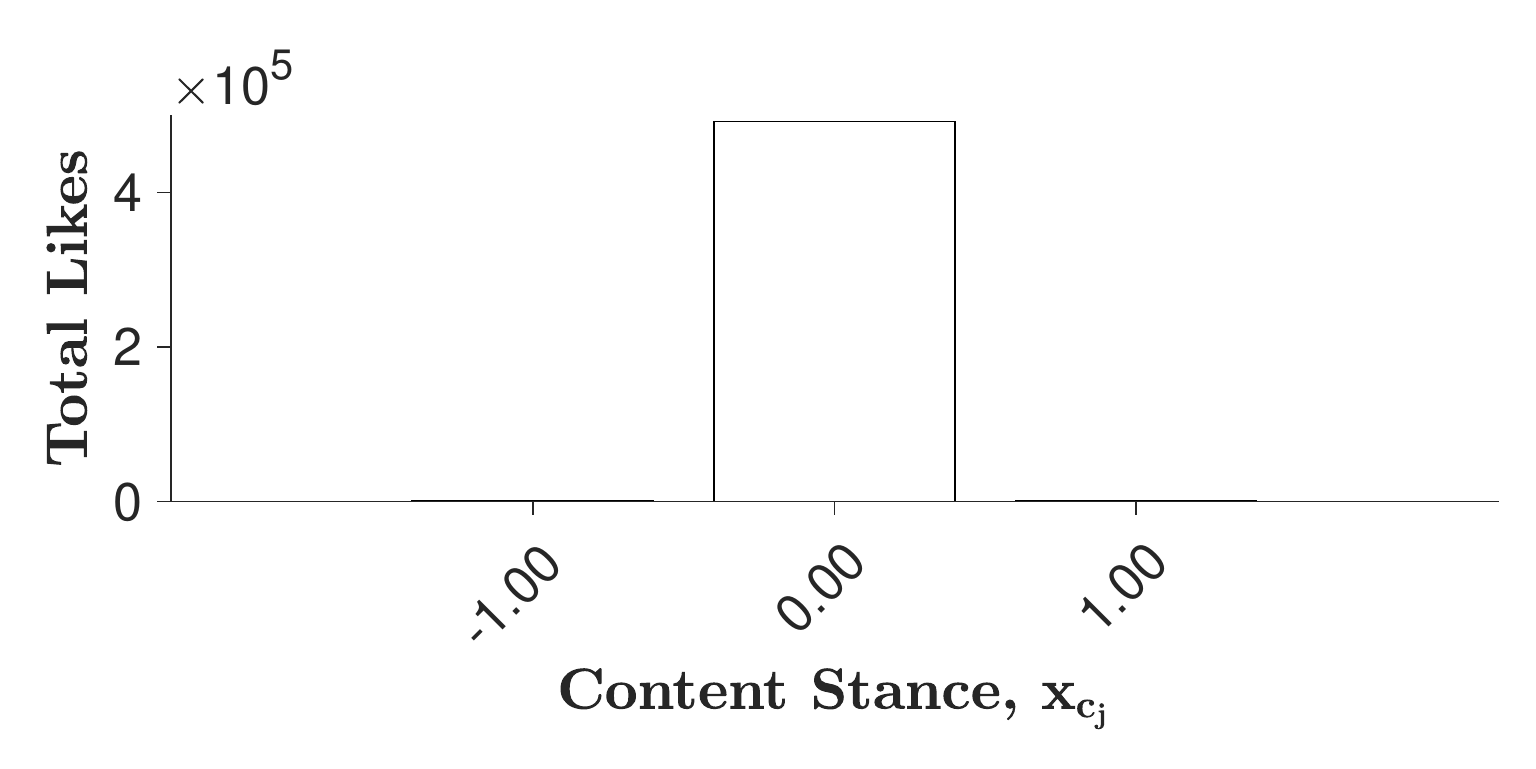}
        \caption{$k=3$}
        \label{subfig:RQ2_k3_likes}
    \end{subfigure}
    \begin{subfigure}[b]{0.24\textwidth}
        \centering
        \includegraphics[width=\textwidth]{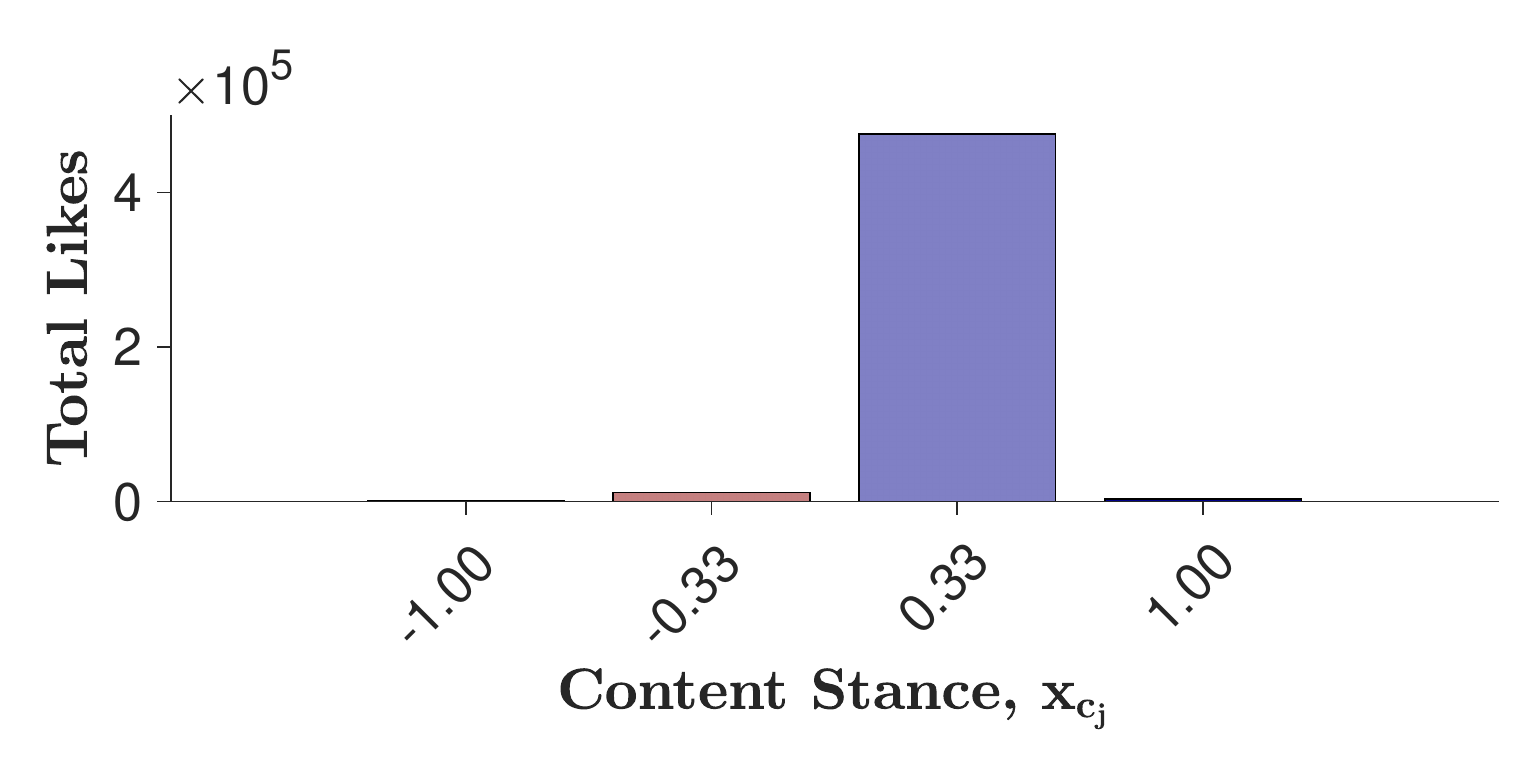}
        \caption{$k=4$}
        \label{subfig:RQ2_k4_likes}
    \end{subfigure}
    \begin{subfigure}[b]{0.24\textwidth}
        \centering
        \includegraphics[width=\textwidth]{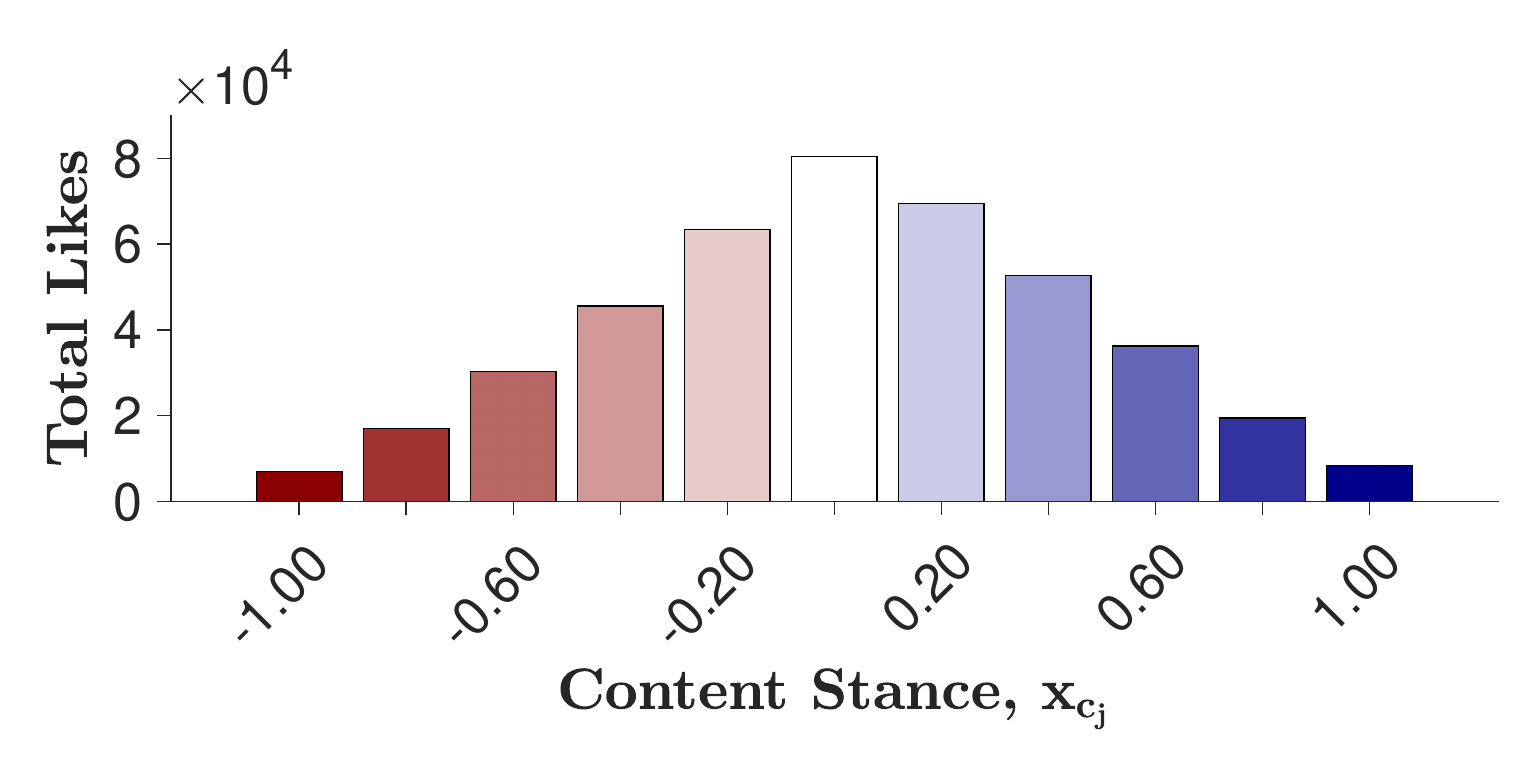}
        \caption{$k=11$}
        \label{subfig:RQ2_k11_likes}
    \end{subfigure}
    \vfill
    \begin{subfigure}[b]{0.24\textwidth}
        \centering
        \includegraphics[width=\textwidth]{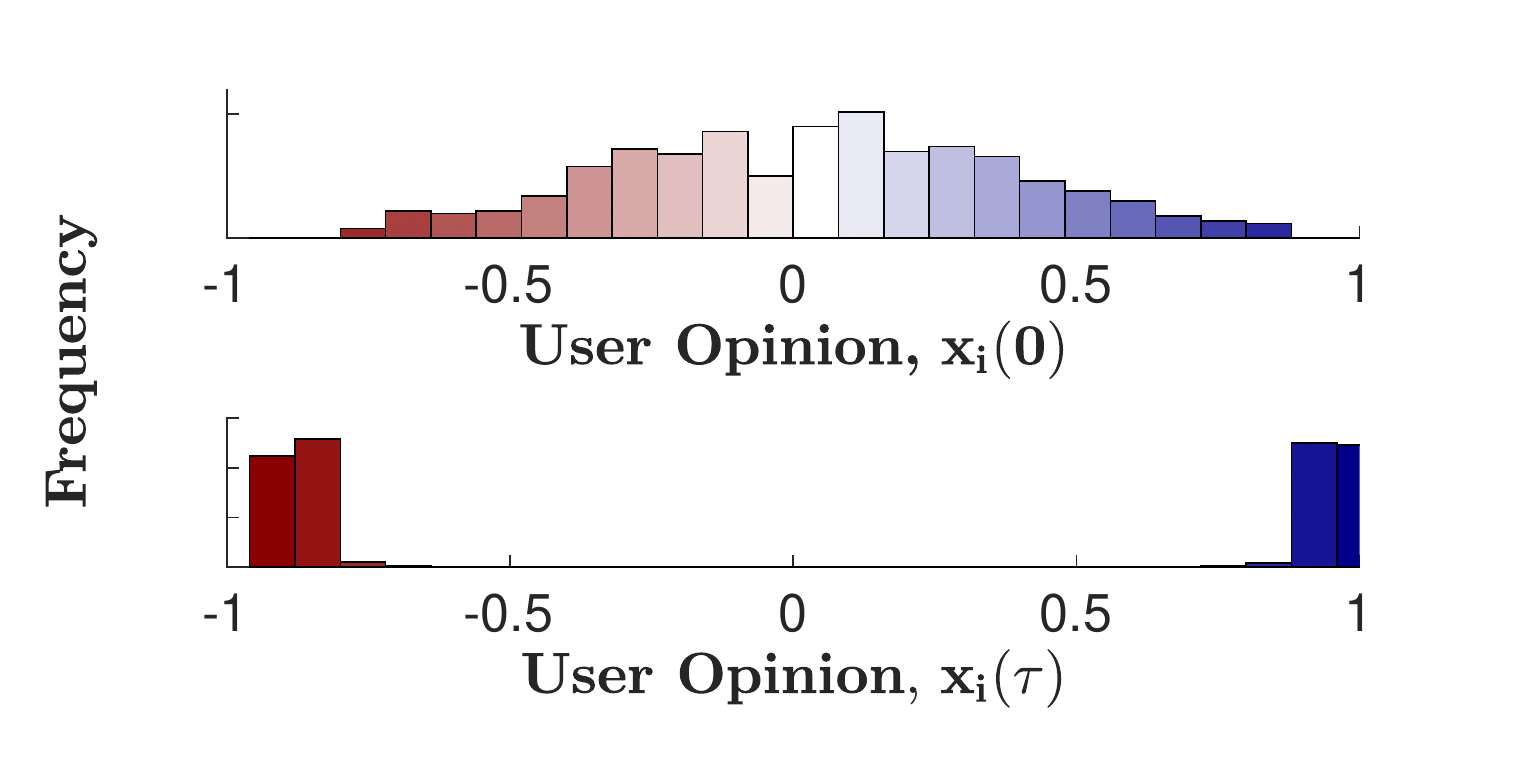}
        \caption{$k=2$}
        \label{subfig:RQ2_k2_ops}
    \end{subfigure}
    \begin{subfigure}[b]{0.24\textwidth}
        \centering
        \includegraphics[width=\textwidth]{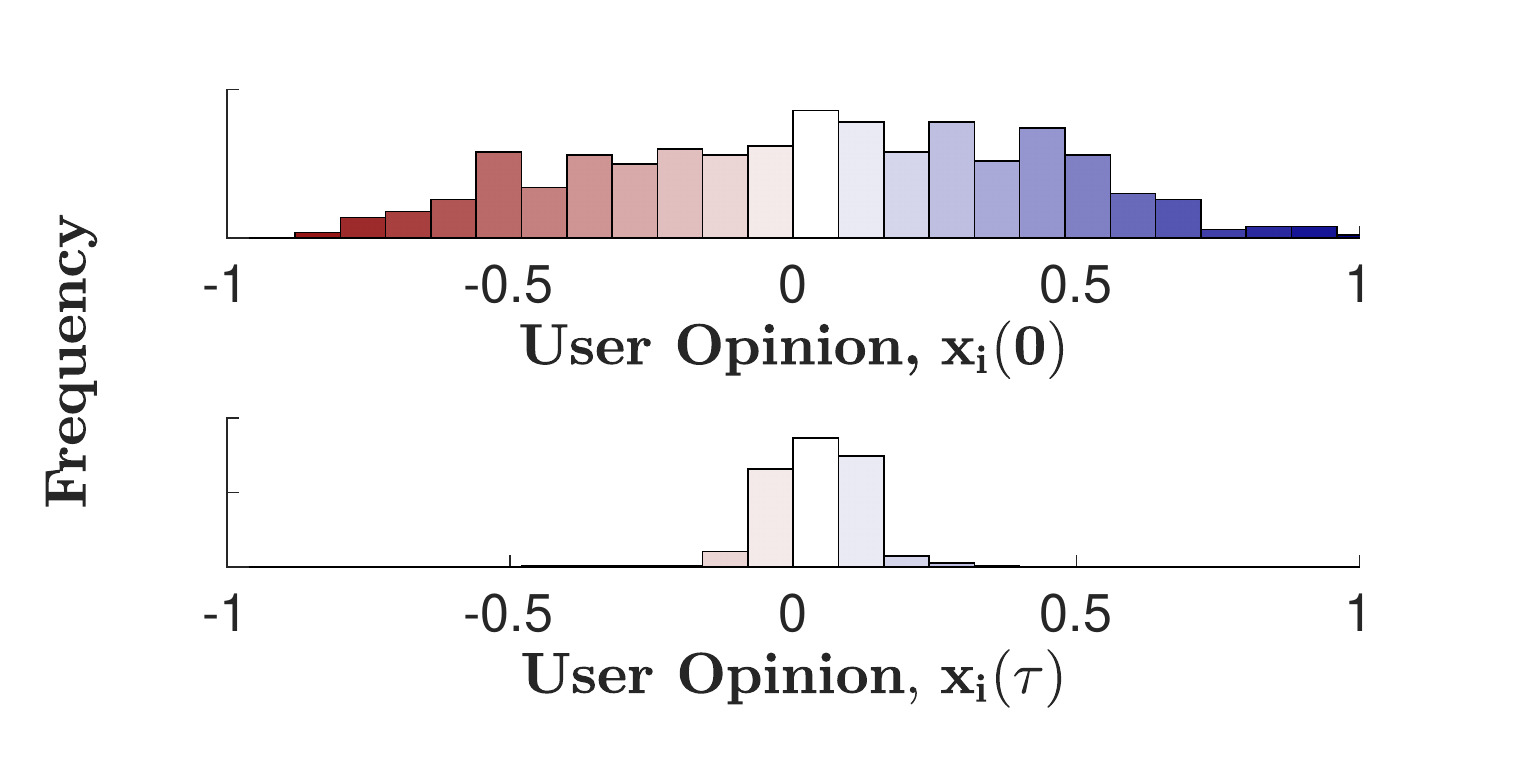}
        \caption{$k=3$}
        \label{subfig:RQ2_k3_ops}
    \end{subfigure}
    \begin{subfigure}[b]{0.24\textwidth}
        \centering
        \includegraphics[width=\textwidth]{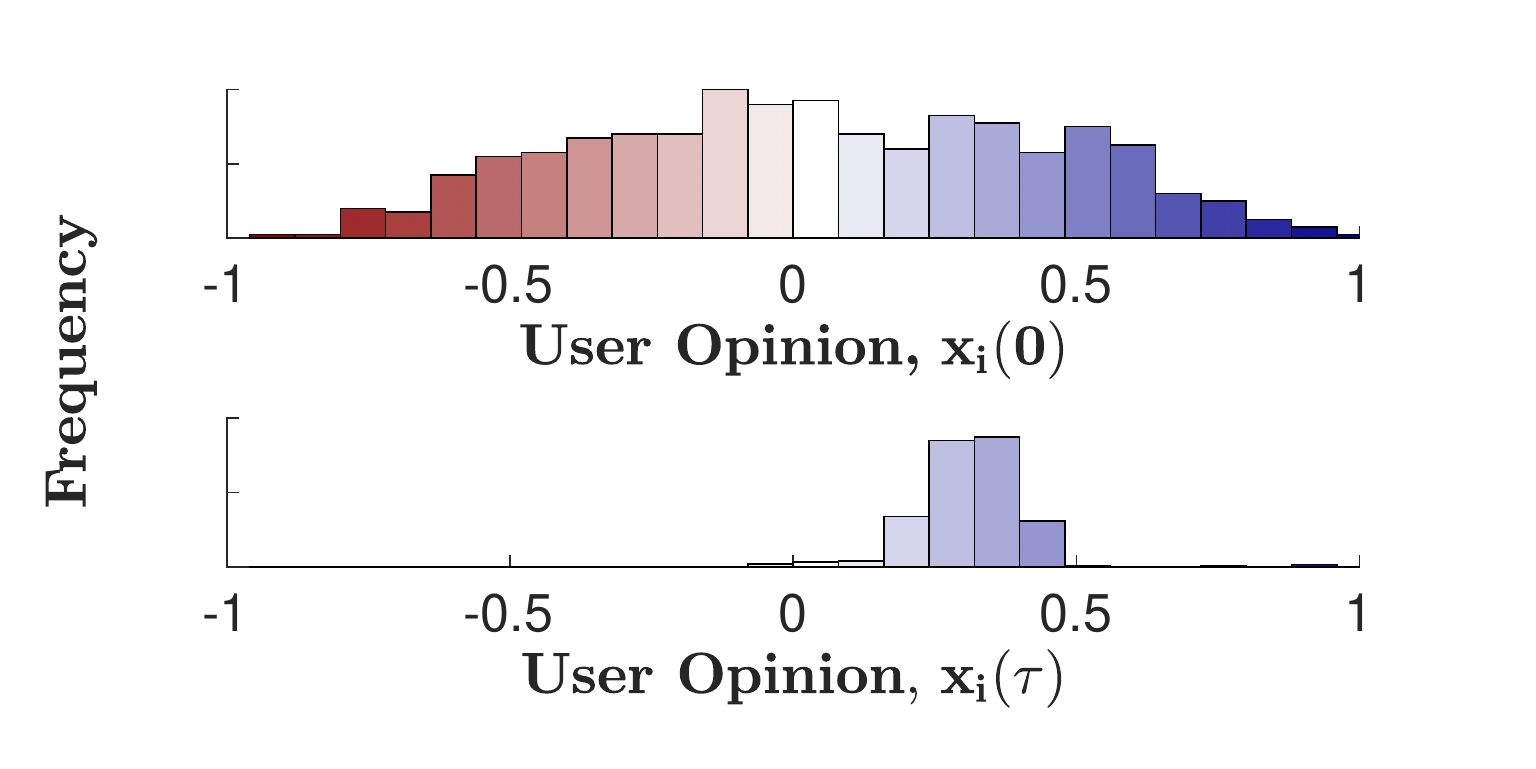}
        \caption{$k=4$}
        \label{subfig:RQ2_k4_ops}
    \end{subfigure}
    \begin{subfigure}[b]{0.24\textwidth}
        \centering
        \includegraphics[width=\textwidth]{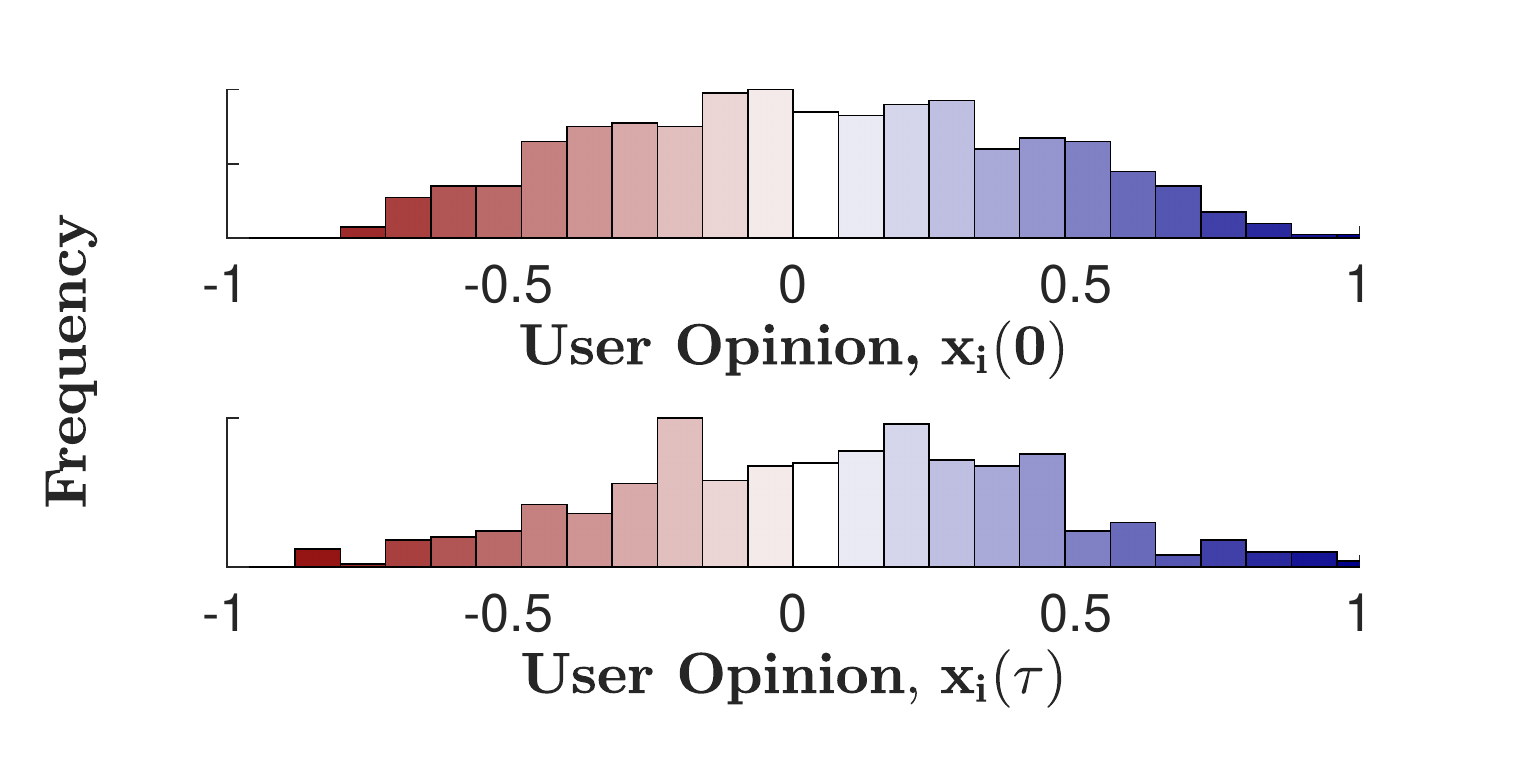}
        \caption{$k=11$}
        \label{subfig:RQ2_k11_ops}
    \end{subfigure}
    \caption{Example simulations with the population starting with an NDIC and $\alpha= 7$, for different number of content $k$. For $k = 2$, (a) shows the total number of likes for each piece of content, and (e) the initial and final opinion distributions showing the effects on the final opinion distributions. Panels (b) and (f) show the simulation for $k = 3$, panels (c) and (g) show the outcome for $k = 4$, and panels (d) and (h) show the outcome for $k=11$.}
    \label{fig:RQ2_opinions_effect_neutral}
\end{figure*}

\section{Research Question Two: Diversity of Content}\label{sec:rq2}

Our second research question concerns the integer parameter $k \geq 2$, which as detailed in \eqref{eq:x_cj}, helps to define the content stance. Intuitively, $k$ represents the diversity of the content/information available for the RS to recommend to users. With $k = 2$, there are precisely two pieces of content taking opposing extreme stances in the opinion spectrum $[-1,1]$, with $x_{c_1} = -1$ and $x_{c_2} = 1$. As $k$ increases, there is more content available (evenly spaced across the opinion spectrum, see \eqref{eq:x_cj}). When $k$ is odd, there is always one piece of content at the neutral stance of $0$, whereas for even $k$, there is not; the impact of this becomes salient in our results. Previous research by \cite{rossi} and \cite{perra} set $k=2$, i.e., $x_{c_j} \in \{-1,1\}$, and it was found that the RS led user(s) to become polarised. In contrast, \cite{lanzetti} considered an RS that could recommend content of any stance to a user, meaning opinions and content existed on the interval $(-\infty,\infty)$, and found the opinion distribution of the population was unchanged by the RS. For our model, this can be effectively captured by allowing $k\to\infty$ on the interval $[-1,1]$. Given $k$ is a parameter within our model, it becomes of great interest to understand the impact that $k$ has on  the RS's performance and user opinions.

Similar to Section~\ref{sec:rq1}, we set $\omega = 0.5$ and $\delta = 5$ and we now vary $k$ in an unevenly spaced grid of $\{2,3,4,...,10,11,21\}$ to place a higher emphasis on low diversity scenarios. We set $\alpha = 7$ based on our findings in Section~\ref{sec:rq1}. Namely, when $\alpha = 7$, the RS receives more than $80\%$ of possible likes and an average watch rate of more than $60\%$, guaranteeing high engagement, see Fig.~\ref{fig:RQ1_performance}. Simultaneously, $\alpha = 7$ allows for some variety in the content users receive, as the RS will occasionally recommend to a user content not tailored specifically for them. This ``variety injection'' is a noted feature of a real-world RS, such as TikTok's For You page, which regularly performs a ``similarity check" to ensure consecutive recommendations are different from one another and that users see a greater variety of content~\cite{tiktok_nd}.

\subsection{Monte Carlo Simulations}

The results in Fig.~\ref{fig:RQ2_performance} show the four metrics for varying $k$, obtained from $50$ Monte Carlo repeats with $95\%$ confidence intervals, similar to Section~\ref{sec:rq1}. The metrics converge asymptotically for $k > 21$; hence, we focus on $k \leq 21$. Results for $k \geq 21$ are provided in the supplementary material for completeness.

In Fig.~\ref{subfig:RQ2_performance_21}, the percentage of total likes and the average watch rate are plotted for different $k$. First, notice that when $k=2$, i.e., $x_{c_j} \in \{-1, 1 \}$, both the NDIC and BDIC populations exhibit remarkably high engagement, with the average watch rate and total percentage of likes exceeding 90\%. As $k$ increases, the broad trend is that the average watch rate increases, while the total percentage of likes decreases, with peaks at odd $k$. Fig.~\ref{subfig:RQ2_polarisation_21} shows the two opinion distribution metrics as $k$ varies. When $k = 2$, we observe substantial increases in polarisation and radicalisation of the final opinion distribution compared to the initial, for both NDIC and BDIC. In contrast, polarisation and radicalisation are substantially reduced when $3 \leq k \leq 11$. In Fig.~\ref{fig:RQ2_opinions_effect_neutral}, we show four example simulations at $k = 2$ (panels (a) and (e)), $k=3$ (panels (b) and (f)), $k = 4$ (panels (c) and (g)), and $k=11$ (panels ((d) and (h)), tracking the total number of likes each content has received, as well as the initial and final opinion distributions. 

\begin{figure*}[h]
    \centering
    \begin{subfigure}[b]{0.49\textwidth}
        \centering
        \includegraphics[width=\textwidth]{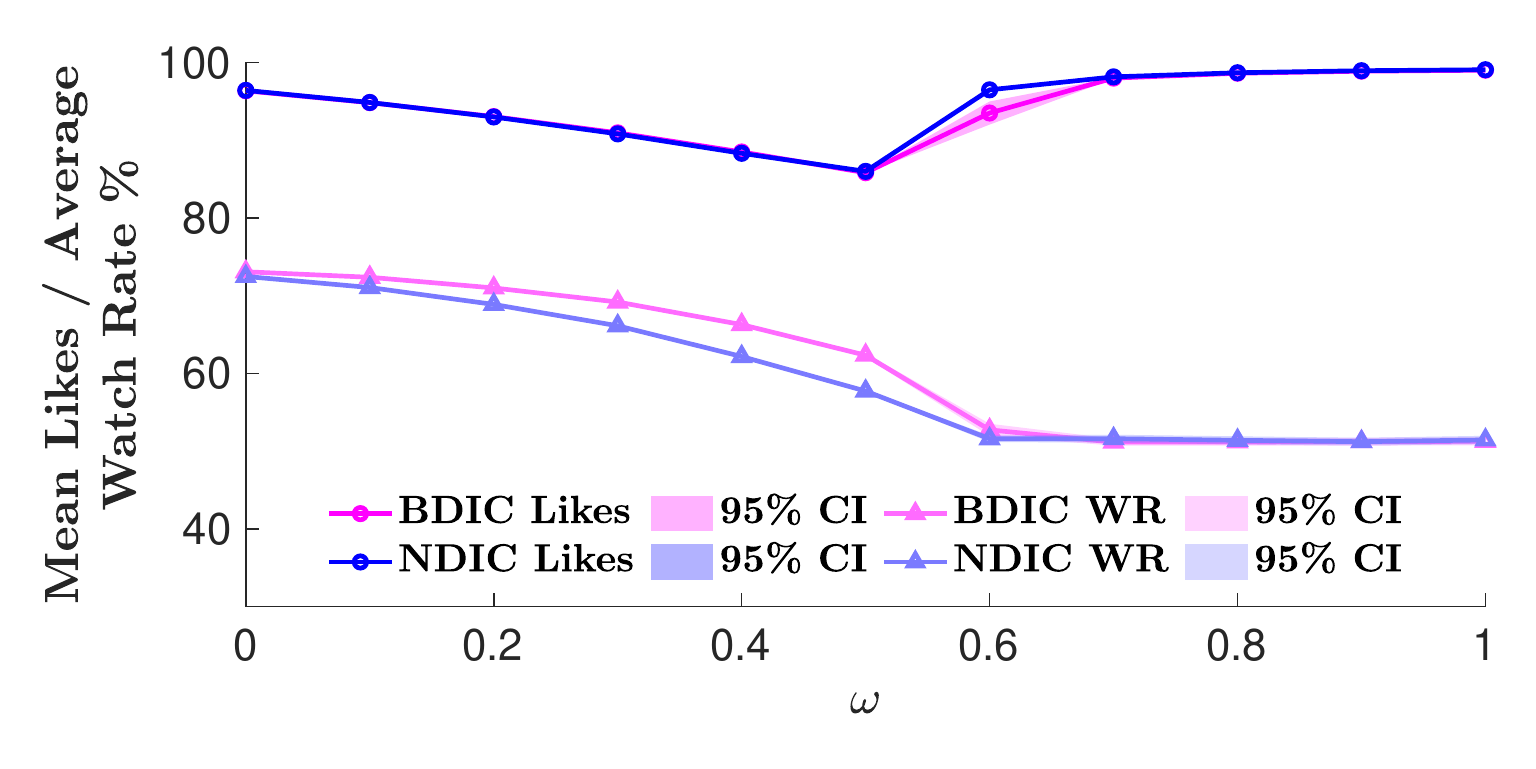}
        \caption{RS performance metrics.}
        \label{fig:RQ3_performance}
    \end{subfigure}
    \hfill
    \begin{subfigure}[b]{0.49\textwidth}
        \centering
        \includegraphics[width=\textwidth]{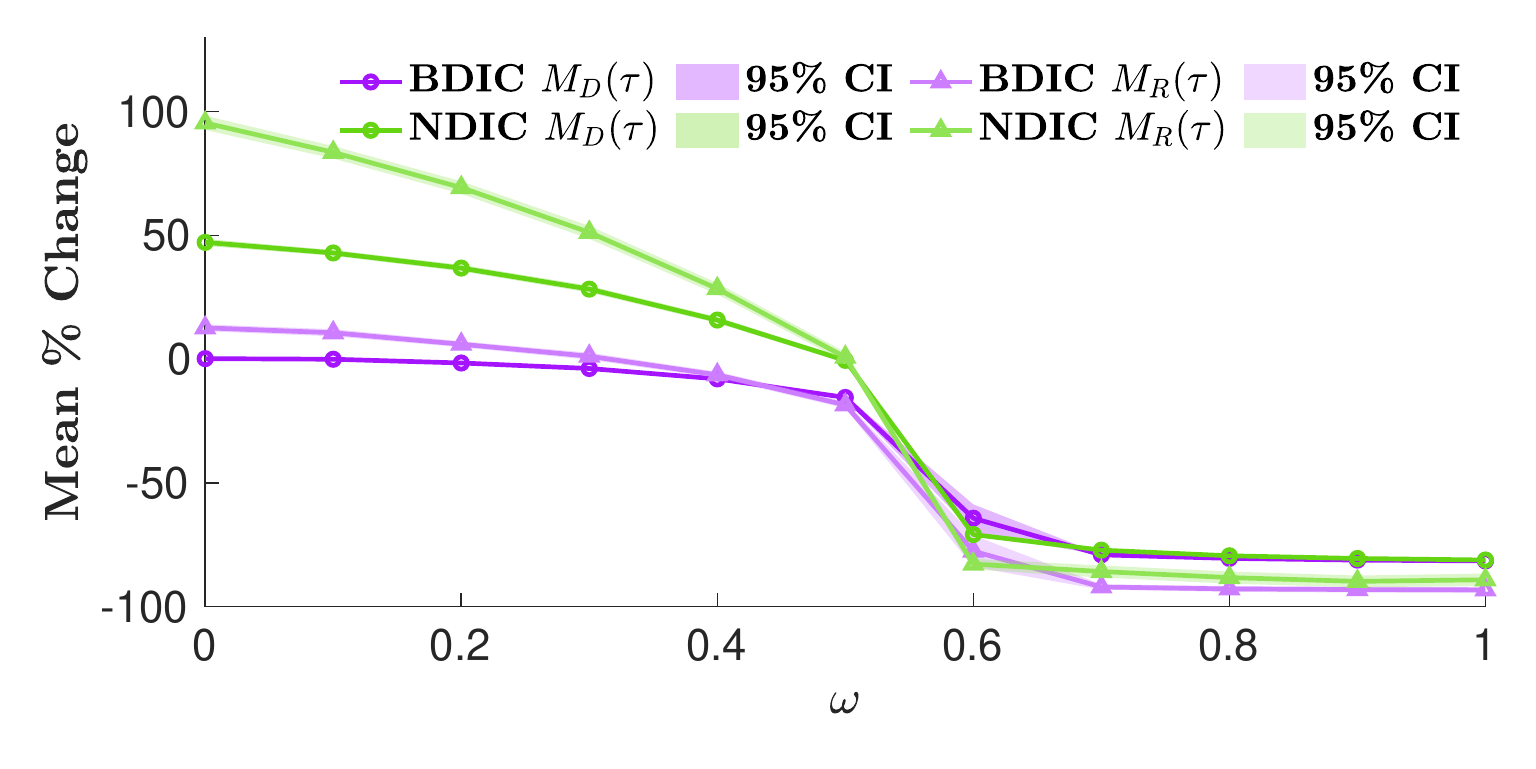}
        \caption{Opinion distribution metrics.}
        \label{fig:RQ3_polarisation}
    \end{subfigure}
    \caption{\textit{Research Question Three}, in (a) the performance metrics of the RS; $\mathrm{Likes\,\%}$, \eqref{eq:likes}, and $\mathrm{Average\,Watch\,Rate\,\%}$ (WR), \eqref{eq:avg_watch_rate}, are shown for different values of $\omega$, the virality-based filtering weighting parameter. In (b), the final percentage changes of the polarisation metric $M_D(\tau)$, \eqref{eq:dispersion}, and radicalisation metric $M_R(\tau)$, \eqref{eq:radicalisation}, are plotted for different values of~$\omega$. Note that $\omega = 0$ and $\omega = 1$ correspond to entirely content-based filtering and entirely virality-based filtering, respectively.}
\end{figure*}

The case of $k = 2$ differs from all other $k$ values. In this case, $x_{c_j} \in \{-1,1\}$, and thus the RS is recommending extreme content to all users, resulting in a highly polarised and radicalised population with the opinions of the users shifting strongly towards the opposite ends of the opinion spectrum, see Fig.~\ref{subfig:RQ2_k2_likes} and \ref{subfig:RQ2_k2_ops}. This is reflected in a large positive increase in both opinion distribution metrics. In other words, for a population of highly susceptible users (we set $\lambda_i = 0.9$ in our simulations), on a platform where only extremely polarised content is recommended, an extremely polarised and radicalised population emerges. 

For $3\leq k \leq 11$, we first comment that whenever $k$ is odd, the neutral content with stance $0$ goes viral, causing opinions to be centred around the viral content (see Fig.~\ref{subfig:RQ2_k3_likes} and \ref{subfig:RQ2_k3_ops}). In the RS metrics, we see the average watch rate drop to $50\%$. From \eqref{eq:watch_rate}, it follows that $x_{z_i(t)}=0$ implies $w_{iz_i(t)} = 50\%$, suggesting the neutral piece of content consistently went viral in all the simulations, for both the NDIC and BDIC. As user opinions are skewed toward the stance of the viral content, the term $\overline{x}_i(t) = x_i(t) - x_{z_i(t)}$ approaches zero, indicating that the probability of the user liking the content increases. When $k$ is odd, the opinion distribution metrics record a significant decrease for both the NDIC and BDIC. When $k$ is even, such as in Fig.~\ref{subfig:RQ2_k4_likes} and \ref{subfig:RQ2_k4_ops} with $k=4$, one piece of content can still go viral and cause the population's opinions to coalesce around that respective stance. However, because this content has a non-zero stance, the decrease in $M_D$ and $M_R$ are likely less than if the viral content had a stance of $0$. When $k=11$, see Fig.~\ref{subfig:RQ2_k11_likes} and \ref{subfig:RQ2_k11_ops}, we can see that there is sufficient variety in content to cater to the entire NDIC, so the likes per piece of content begins to mirror the NDIC itself. In this example, when $k=11$, $x_{c_j}=0$ remains the most popular piece of content; however, it receives only 21.48\% of all the likes, which is significantly less viral than when $k=3$ or $k=4$.

We can summarise the findings as follows. When $k = 2$, the RS performs the best, maximising engagement and watch rate, but this leads to a hyper-polarised and radicalised population. When there is limited diversity ($ 3 \leq k \leq 9$), the RS, in fact, reduces polarisation and radicalisation in the population, as opinions coalesce around the stance of the viral content. This is most effective when $k$ is odd, i.e., when there is a neutral content, $x_{c_j}=0$. When $k \geq 11$, there is sufficient diversity that each user receives recommendations close to their initial opinion, and content does not typically go viral. The opinion distributions are largely unchanged over time (although there is a small reduction in polarisation and radicalisation for BDIC).

\section{Research Question Three: Virality}\label{sec:rq3}

Our third and final research question focuses on how the RS integrates information about viral content into its recommendation process. In particular, we examine $\omega \in [0,1]$ and $\delta \in \mathbb N_+$. The former is the relative weight given to the payoff associated with the virality-based filtering in \eqref{eq:P_value} (relative to the payoffs associated with content-based filtering), so that as $\omega$ increases, the RS puts more weight on viral content over trying to maximise user-specific engagement. The latter captures the ``memory'' of the RS, i.e. as $\delta$ increases, the RS considers past engagement input from users over a longer window when deciding what is a viral piece of content.

The virality-based filtering mechanism introduced in our model is substantially different from existing models~\cite{lanzetti,rossi,sprenger,perra}. Thus, we are motivated to understand how this mechanism shapes the RS and user opinions. We vary $\omega$ from $0$ to $1$ in step sizes of $0.1$, and vary $\delta$ from $1$ (meaning the memory of the RS resets every timestep) to $\tau$ (meaning the RS remembers every like recorded up to the given timestep when determining $C^j(t)$) in steps of $10$. Based on the results of the previous sections, we fix $\alpha=7$ and $k=11$. We select $k=11$, given that it provides a sufficient variety of content but does not guarantee a piece of viral content at the parameter settings already explored.

\subsection{Monte Carlo Simulations - Varying $\omega$}

We now present the results of the Monte Carlo simulations as we vary $\omega$, while fixing $\delta = 5$. Fig.~\ref{fig:RQ3_performance} shows the RS metrics for different values of $\omega$, while the opinion distribution metrics are shown in Fig.~\ref{fig:RQ3_polarisation}.

In Fig.~\ref{fig:RQ3_performance}, we can see that the RS performs well when content-based filtering has a greater weight than virality-based filtering. As $\omega$ increases, we see user engagement decrease between $\omega = 0$ and $ \omega = 0.5$. This slight decrease suggests that some users are beginning to receive content recommendations that differ from their current preferences, which in turn reduces their engagement with the content. At $\omega =0.5$, we observe a shift; the average watch rate asymptotically approaches $50\%$, while the total number of likes increases. We have previously observed in Fig.~\ref{subfig:RQ2_performance_21} that when the average watch rate approaches $50\%$, a neutral piece of content in the system is going viral. This is more likely to happen when the RS has a larger emphasis on virality-based filtering ($\omega \geq 0.6$). However, we note that while the average watch rate converges to 50\%, the RS performs better in receiving likes, suggesting that users will engage with the content by liking it but not watching for long. 

Fig.~\ref{fig:RQ3_polarisation} shows the opinion distribution metrics. We see that when $\omega < 0.5$ (majority content-based filtering), there is an increase in polarisation and radicalisation of user opinions, and this is substantially more pronounced with an NDIC opinion distribution. However, as $\omega$ increases, the level of polarisation and radicalisation decreases, with a phase transition-type behaviour at $\omega = 0.5$. Indeed, as $\omega$ increases above $0.5$, polarisation and radicalisation are reduced substantially (negative percentage change in all metrics). Below the threshold value of $\omega = 0.5$, the content-based filtering provides users with content that aligns with their opinions, creating a reinforcing effect that results in increased polarisation and radicalisation, consistent with existing literature~\cite{rossi}. However, increasing $\omega$ means that all users are more likely to be recommended the same content, which reduces polarisation and radicalisation. At a sufficiently high $\omega$ (above $0.5$), we conjecture that one piece of content is always going viral, and it is often content with a mild or neutral stance, since the average watch rate in Fig.~\ref{fig:RQ3_performance} converges to $50\%$. The effect of neutral content going viral on the population is evident; opinions become more neutral, and both opinion distribution metrics decrease, resulting in negative percentage changes.

\subsection{Virality with $\omega$ and $\delta$}

\begin{figure}
    \centering
    \includegraphics[width=0.7\linewidth]{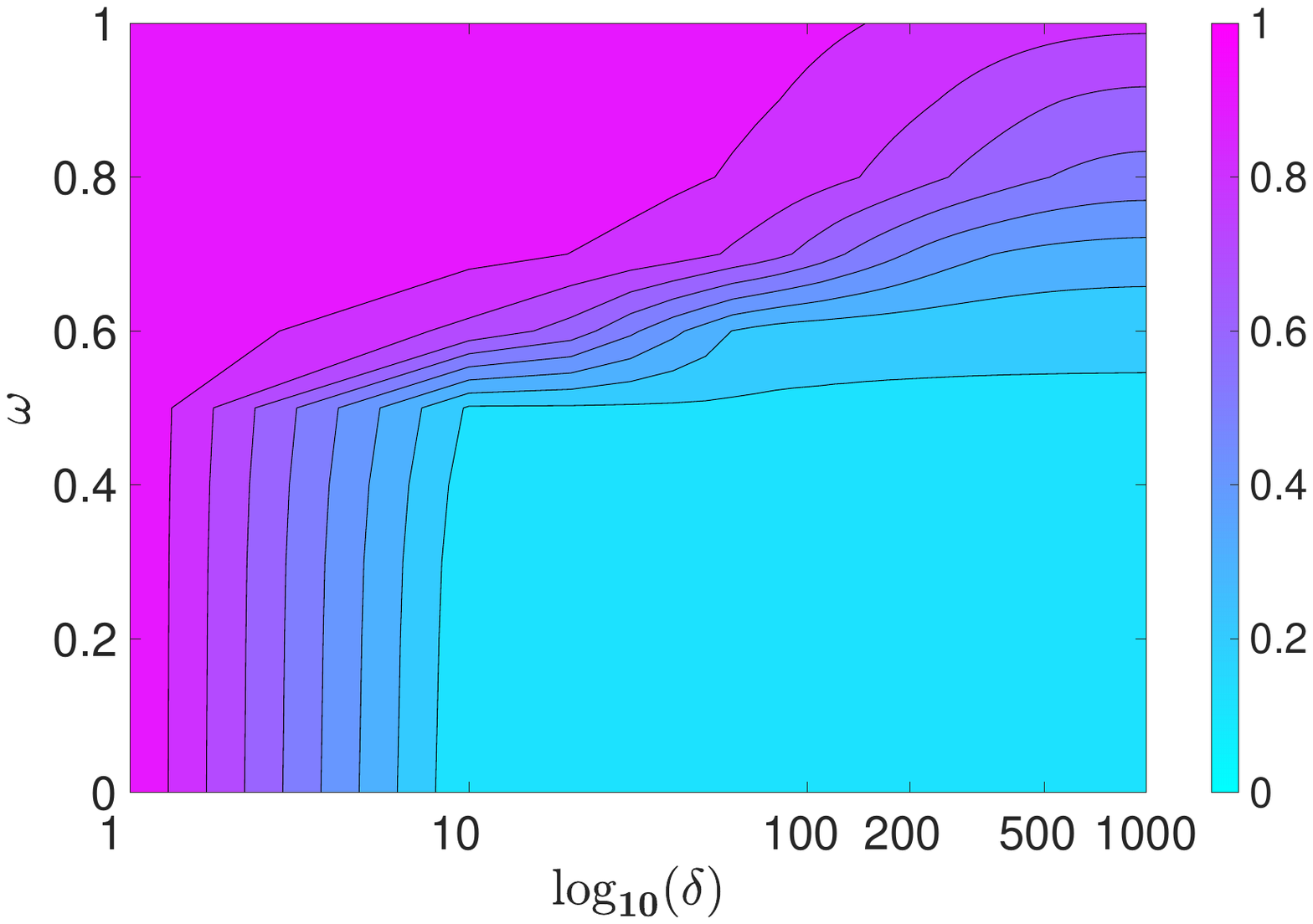}
    \caption{A contour plot exploring the relationship between $\omega$, $\delta$ and when content goes viral for an NDIC. Here, $\omega$ is varied from 0 to 1, $\delta$ is varied from 1 to 1000 (note the log scale), and the colour bar refers to the proportion of the amount of likes the most liked piece of content received out of the total amount of likes recorded in the system for that simulation.}
    \label{fig:RQ3_viral}
\end{figure}

We now introduce the second part of our results for \textit{Research Question Three}, as shown in the heatmap in Fig.~\ref{fig:RQ3_viral}. Now, we vary not only $\omega$, but also $\delta$. Rather than focus on the RS and opinion distribution metrics, we examine when viral content emerges. To do so, in each simulation, we identify the most liked piece of content, and then compute the proportion of likes it received from all users relative to the total number of likes across all content. In other words, this proportion represents how ``dominant'' this one piece of content was over the other content. The heatmap showcases this proportion as a function of $\omega$ and $\delta$. The results between an NDIC and a BDIC do not differ substantially; therefore, the results for the BDIC are reported in the supplementary material.

In Fig.~\ref{fig:RQ3_viral}, we can see that when $\delta = 1$, one piece of content will go viral due to the RS locking onto the most liked piece of content after one timestep, see \eqref{eq:C_value}. After $\delta=1$, there is a sharp decrease in the level of virality when $\delta>1$. Beyond $\delta=1$, we can see that when $\omega<0.5$, one piece of content is unable to go viral, meaning there must be a majority of virality-based filtering to ensure a viral piece of content. We also observe that as $\delta$ increases, $\omega$ must be greater to ensure a piece of content goes viral, meaning that as the memory of the RS increases, there must be more virality-based filtering to guarantee a piece of viral content. 

\section{Discussion and Conclusions}\label{sec:conc}

This paper proposed a mathematical model to describe the closed loop dynamics between an RS and users' opinions. 
We have extended the current literature by allowing the model to facilitate many users and pieces of content interacting through an RS. We have also introduced additional aspects to the user dynamics, such as liking content and how long users watch content, and further enhanced the RS dynamics with the introduction of virality-based filtering. 

Through a campaign of Monte Carlo simulations, we introduced three research questions to investigate important and novel parameters in our model---the softmax parameter of the RS ($\alpha$), diversity of content ($k$), and parameters associated with the virality-based filtering mechanism of the model ($\omega$ and $\delta$). 
When investigating $\alpha$, the softmax parameter of the RS in \eqref{eq:softmax_func}, we found that there was a balance between exploration (low $\alpha$) and exploitation (high $\alpha$) that resulted in a piece of content going viral. Furthermore, one could substantially increase the likelihood of a piece of content going viral by making the RS's virality-based filtering mechanism dominant relative to its content-based filtering elements and reducing the RS's memory---achieved by adjusting $\omega$ and $\delta$. In fact, viral content emerged as a consistent phenomenon in the proposed model, its effects highlight the closed loop nature of the system; user interactions make one piece of content go viral in the first place, and then the viral content shapes user opinions, and consequently, their interactions. Whenever a piece of content went viral, there was a major impact on all metrics; specifically for the opinion distribution metrics, we saw a sudden and substantial decrease in polarisation and radicalisation. This decrease was because the viral content was almost always of a (close to) neutral stance. In many cases, real-world viral content consists of inoffensive, funny memes and jokes that are not extreme or of a fringe origin \cite{ling_meme_2021}.

\textit{Research Question Two} revealed a key insight on how content diversity, $k$, had a non-trivial interplay with the opinion dynamics. Namely, when $k=2$ (there are only two pieces of content), we saw maximal polarisation and radicalisation, while as $k$ becomes large, there is minimal change to the opinions over time. The former is consistent with~\cite{rossi}, and the latter is consistent with~\cite{lanzetti}. 
Both scenarios allow the RS to optimise for performance as measured by user engagement. However, neither is ideal for reducing polarisation in user opinions. 
Our results suggest there is a ``sweet spot'' of moderate content diversity (moderate $k$) that decreases polarisation and radicalisation, and this effect is amplified i) if there are content with a neutral stance, and ii) if said content goes viral.
These findings reinforce similar real-world observations. In the bipartisan political system in America~\cite{iyengar2019} with only two choices, a polarised population emerges. 
In contrast, multiparty political systems offer people the ability to only hear and support opinions they already agree with, sometimes producing legislative deadlock that hinders the effectiveness of a democracy~\cite{mainwaring}. In contrast, when content is recommended to users that they may disagree with, it can broaden their understanding beyond their current ideological position~\cite{bruns}. 

Regarding our two selected initial opinion distributions, the NDIC and BDIC, we found that the NDIC was much more susceptible to polarisation and radicalisation than the BDIC. This effect is exacerbated whenever the RS aims for high performance (i.e. increasing $\alpha$ when $k=2$ and focusing on content-based filtering by setting a small $\omega$). Our results show that the BDIC's opinion distribution metrics decreased after interaction with the RS (except for the aforementioned scenarios). This suggests that an RS impacts an initially polarised population less, as each user will likely already favour a particular content stance and thus be recommended similar content. In contrast, it also highlights the dangers of neutral populations being influenced by RSs, as we observe an increase in polarisation and radicalisation over time.

While this paper has provided a much-needed look into RSs' impact on users' opinion dynamics, analysing the results revealed some limitations of the proposed mathematical modelling framework and highlighted problems that remain open for further exploration. Our current watch rate function results in a user always having a watch rate of approximately $50\%$ when recommended neutral or close-to-neutral content. In the future, defining a more sophisticated watch rate function when interacting with neutral content would benefit the modelling process. Another limitation of the model is that
our virality-based filtering mechanism considers engagement from all users. Implementing a form of collaborative filtering, to identify groups of similar users and serve them similar recommendations, in combination with virality-based filtering, could allow different content to go viral within different subgroups at the same time. Extending our framework to consider multiple topics, could be useful for studying the rabbit-hole/spillover effect, especially in how users might be recommended increasingly extremist content, until they are eventually immersed by conspiracy content~\cite{sutton_rabbit_2022}. The proposed model and planned future work will help broaden our understanding of the influence that RSs and the content they recommend have on society.

\bstctlcite{IEEEexample:BSTcontrol}
\bibliographystyle{IEEEtran}
\bibliography{Proposal1}

\newpage
\onecolumn
\begin{center}
    {\LARGE \textsc{Supplementary Material}}
\end{center}

\section{Functions in the Model}\label{sec: appendix functions}

\begin{figure}[ht]
    \centering
    \begin{subfigure}[b]{0.49\textwidth}
        \centering
        \includegraphics[width=\textwidth]{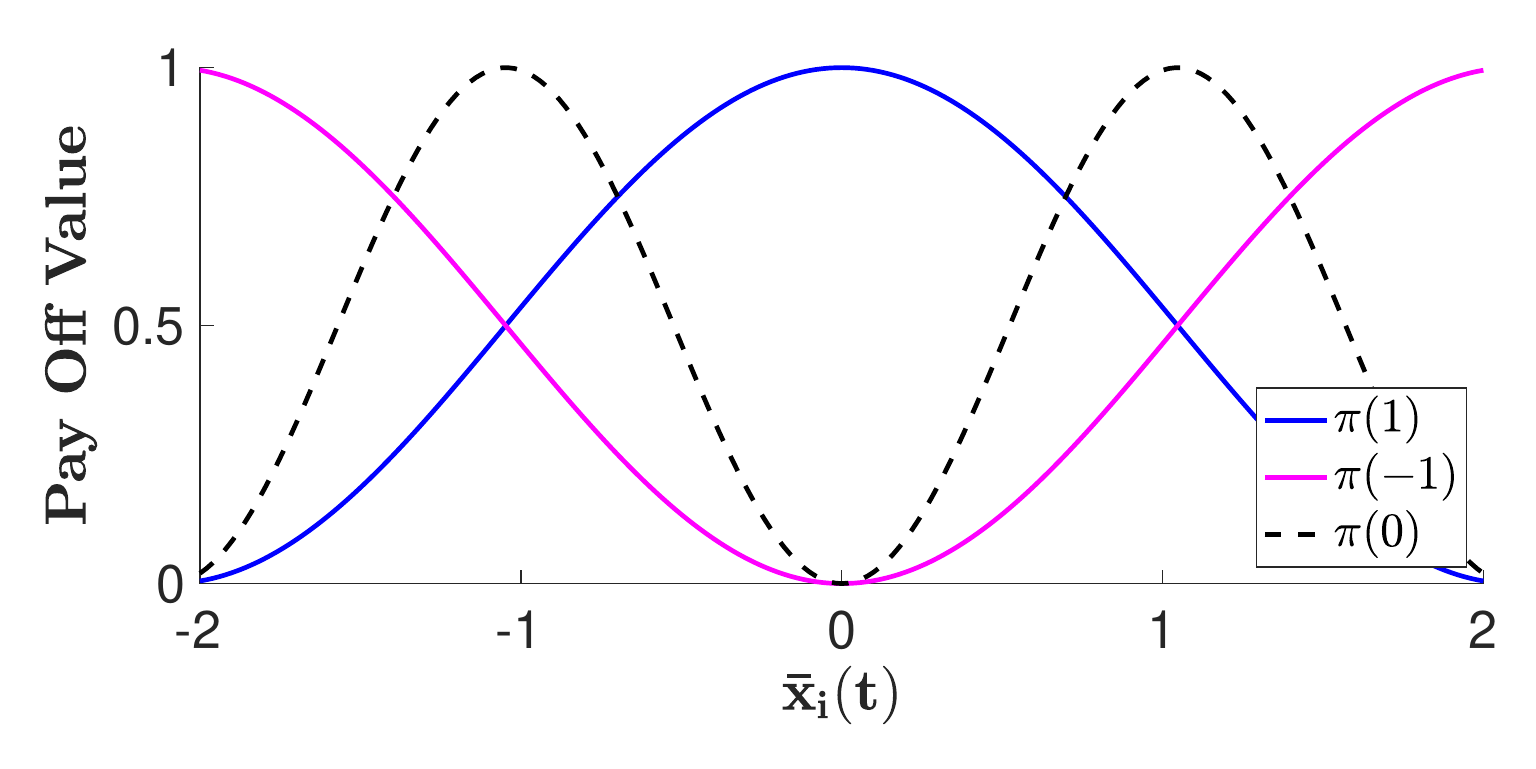}
        \caption{Payoff functions for the engagement decision, dependent on user opinion and content recommendation, where $\Bar{x}_i(t) = x_i(t) - x_{z_i(t)}$.}
        \label{fig: payoff like}
    \end{subfigure}
    \hfill
    \begin{subfigure}[b]{0.49\textwidth}
        \centering
        \includegraphics[width=\textwidth]{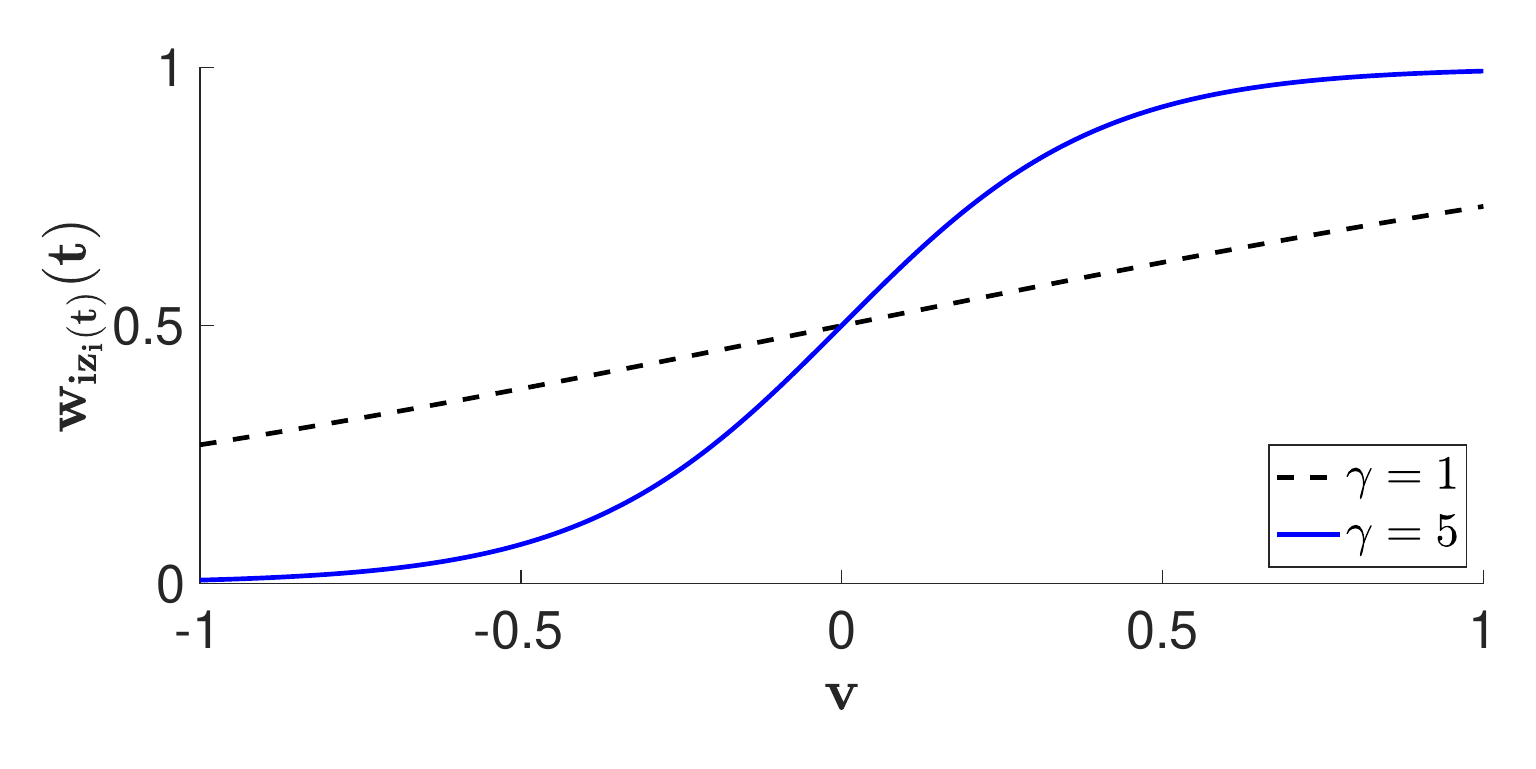}
        \caption{Function used to determine user watch rate, $w_{iz_i(t)}(t)$, where $v(t) = x_i(t) \times x_{z_i(t)}$.\\ }
        \label{fig: watch rate func}
    \end{subfigure}
    \hfill
    \begin{subfigure}[b]{0.49\textwidth}
        \centering
        \includegraphics[width=\textwidth]{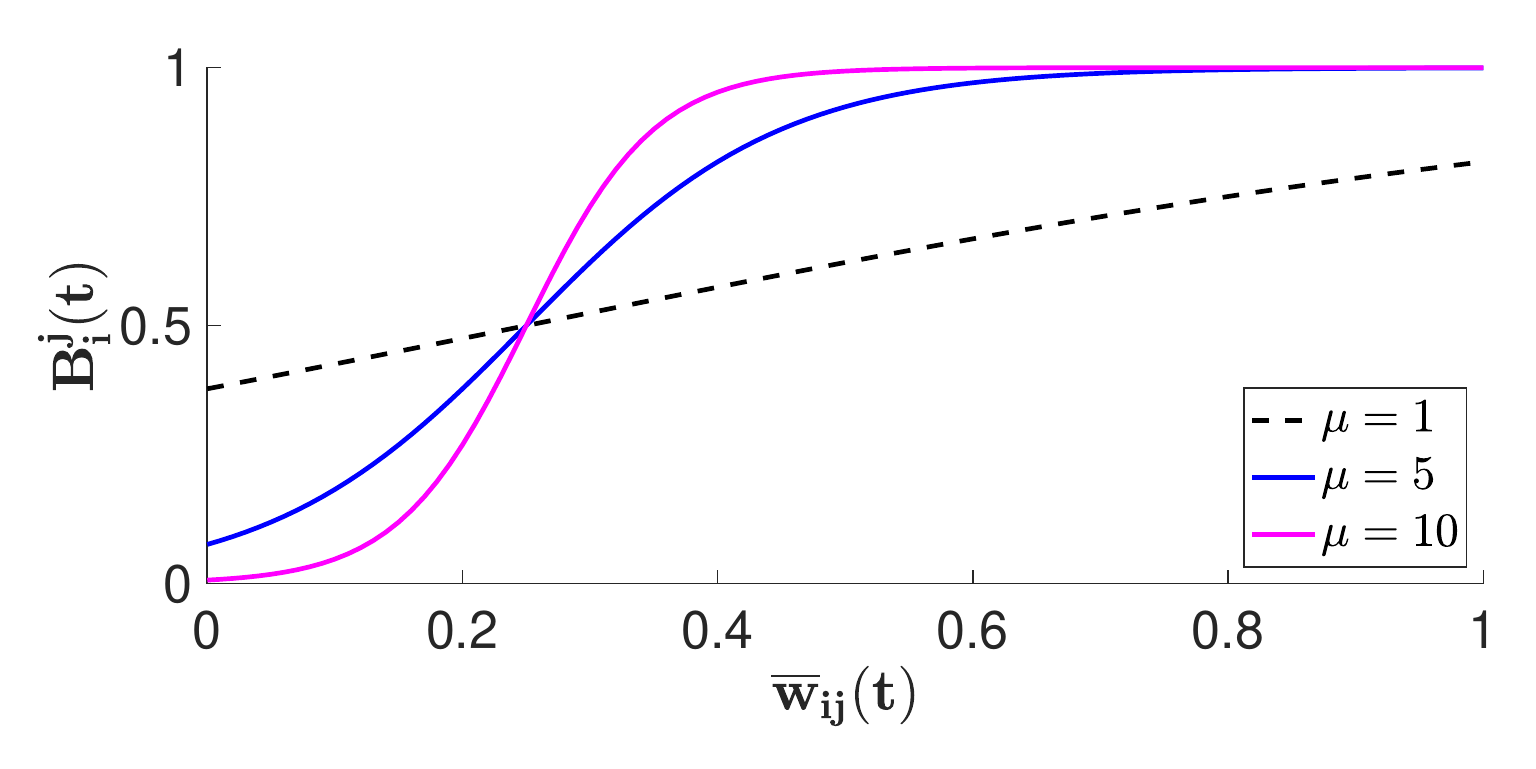}
        \caption{Effect of watch rate on future recommendations, generation of the value $B_i^j(t)$.}
        \label{fig: watchrate effect}
    \end{subfigure}
    \caption{Plots of the functions defined in Section II.}
\end{figure}

Here we provide graphical representations of the functions which underlie the user-specific processes of the engagement decision (Fig.~\ref{fig: payoff like}), the users' watch rate (Fig.~\ref{fig: watch rate func}) and the function for $B_i^j(t)$, the effect of the users' watch rate on future recommendations (Fig.~\ref{fig: watchrate effect}).

\section{Extended Results Research Question Two}

\begin{figure}[ht]
    \centering
    \begin{subfigure}[b]{0.49\textwidth}
        \centering
        \includegraphics[width=\textwidth]{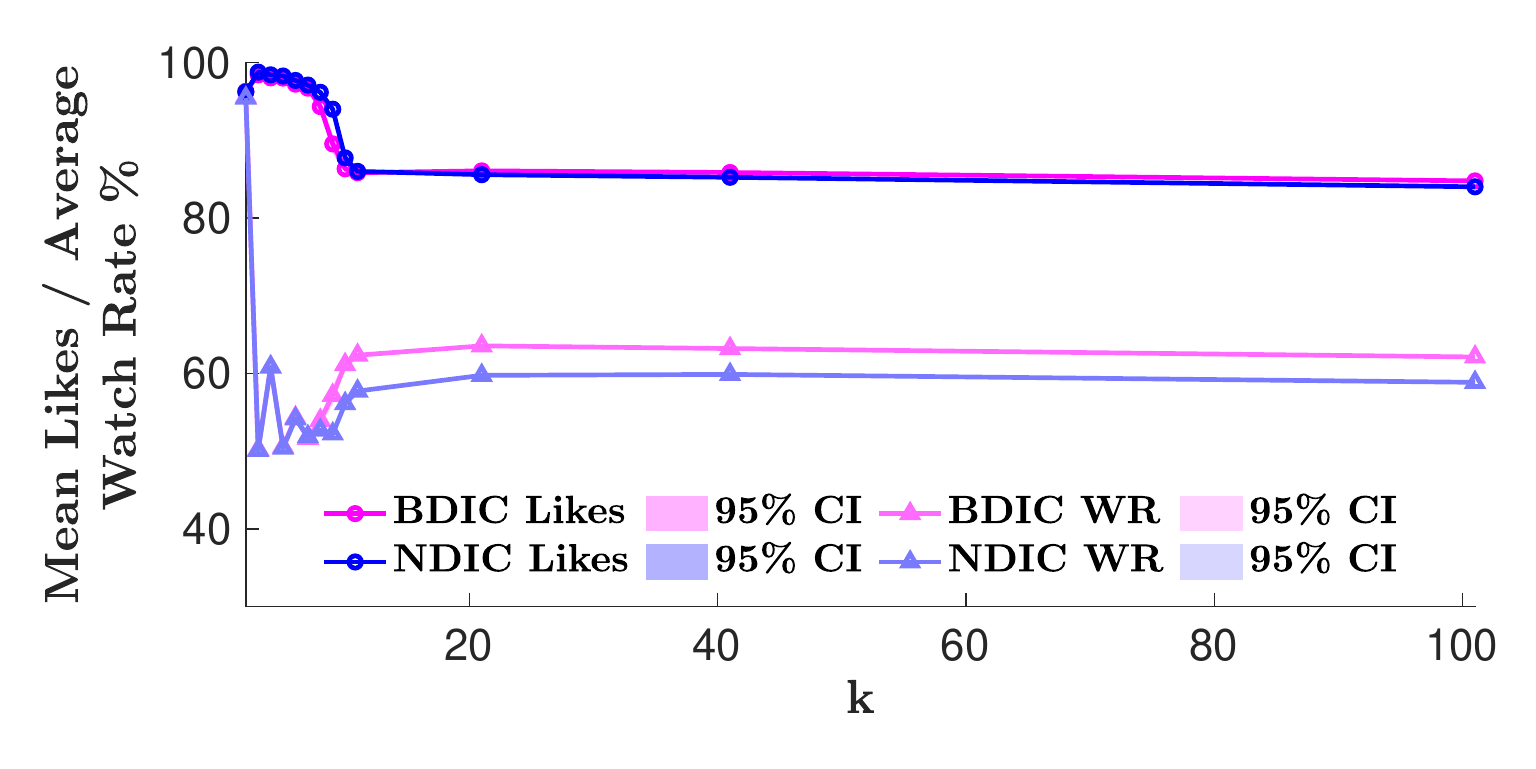}
        \caption{RS performance metrics}
        \label{subfig:RQ2_performance_101_extended}
    \end{subfigure}
    \hfill
    \begin{subfigure}[b]{0.49\textwidth}
        \centering
        \includegraphics[width=\textwidth]{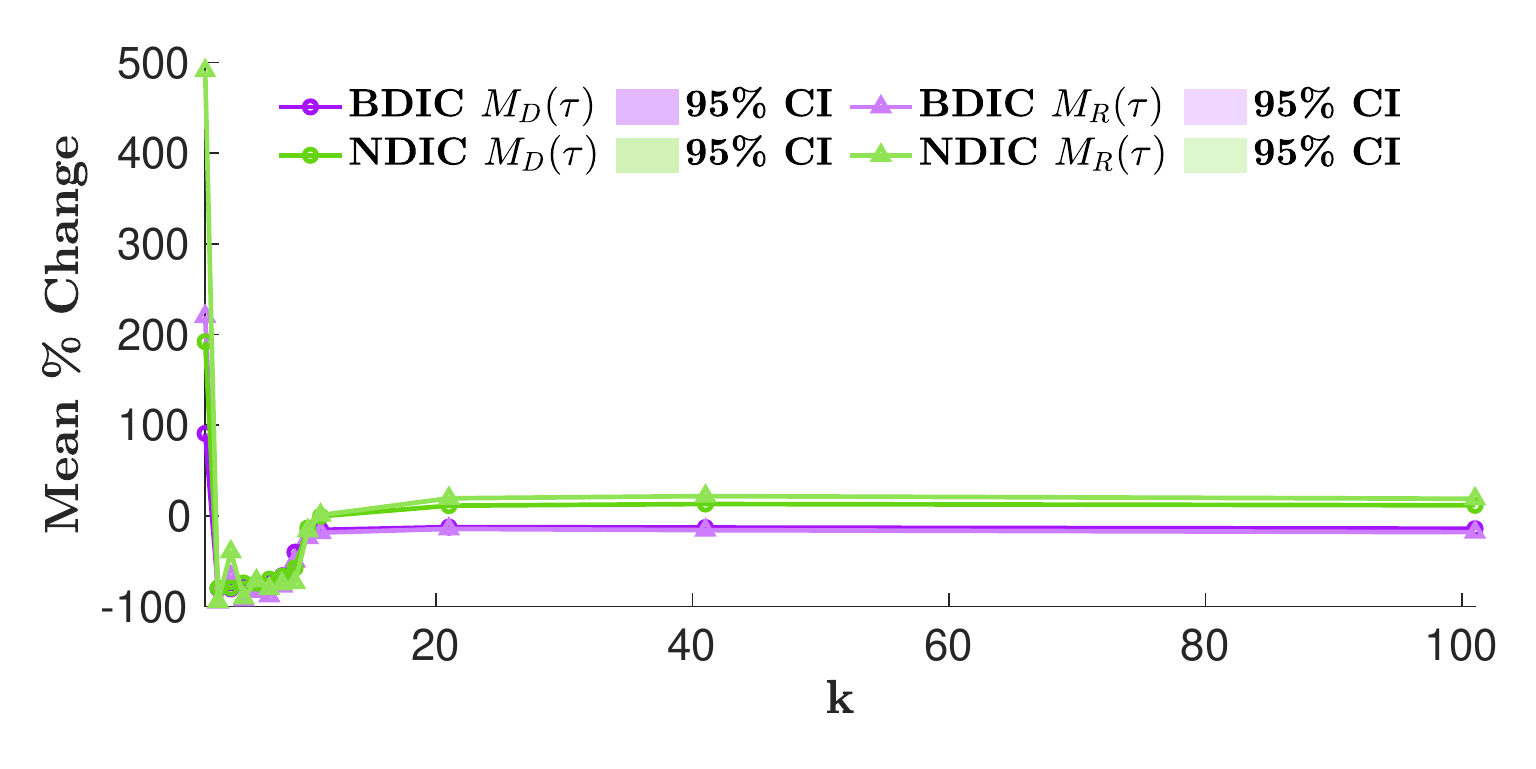}
        \caption{Opinion distribution metrics}
        \label{subfig:RQ2_radicalisation_101_extended}
    \end{subfigure}
    \caption{Extended results to $k=101$. In (a) the performance metrics of the RS; $\mathrm{Likes\,\%}$, Eq.(11), and $\mathrm{Average\,Watch\,Rate\,\%}$ (WR), Eq.(12), are shown for different number different of content, $k$. In (b), the final percentage changes of the polarisation metric $M_D(\tau)$, Eq.(13), and radicalisation metric $M_R(\tau)$, Eq.(14), are plotted for different values of~$k$.}
    \label{fig:RQ2_performance_extended}
\end{figure}

Here we provide the extended results for \textit{Research Question Two} (for $\lambda_i=0.9$). The full extent of $k$ values investigated can be seen in Fig.~\ref{fig:RQ2_performance_extended}, the notable difference from Fig.5 is the inclusion of $k=41$ and $k=101$.

In Fig.~\ref{fig:RQ2_performance_extended}, when $k \in [11,101]$ and the change in metrics has asymptotically diminished, the NDIC records a greater increase in $M_D$ and $M_R$ than the BDIC. This suggests that when $k \geq 10$, diversity of content does not have a great effect on the BDIC and that distribution remains polarised. However, for an NDIC the distribution becomes more polarised.

\section{Extended Results Research Question Three}\label{sec: supp results}

\begin{figure}[ht]
    \centering
    \includegraphics[width=0.49\linewidth]{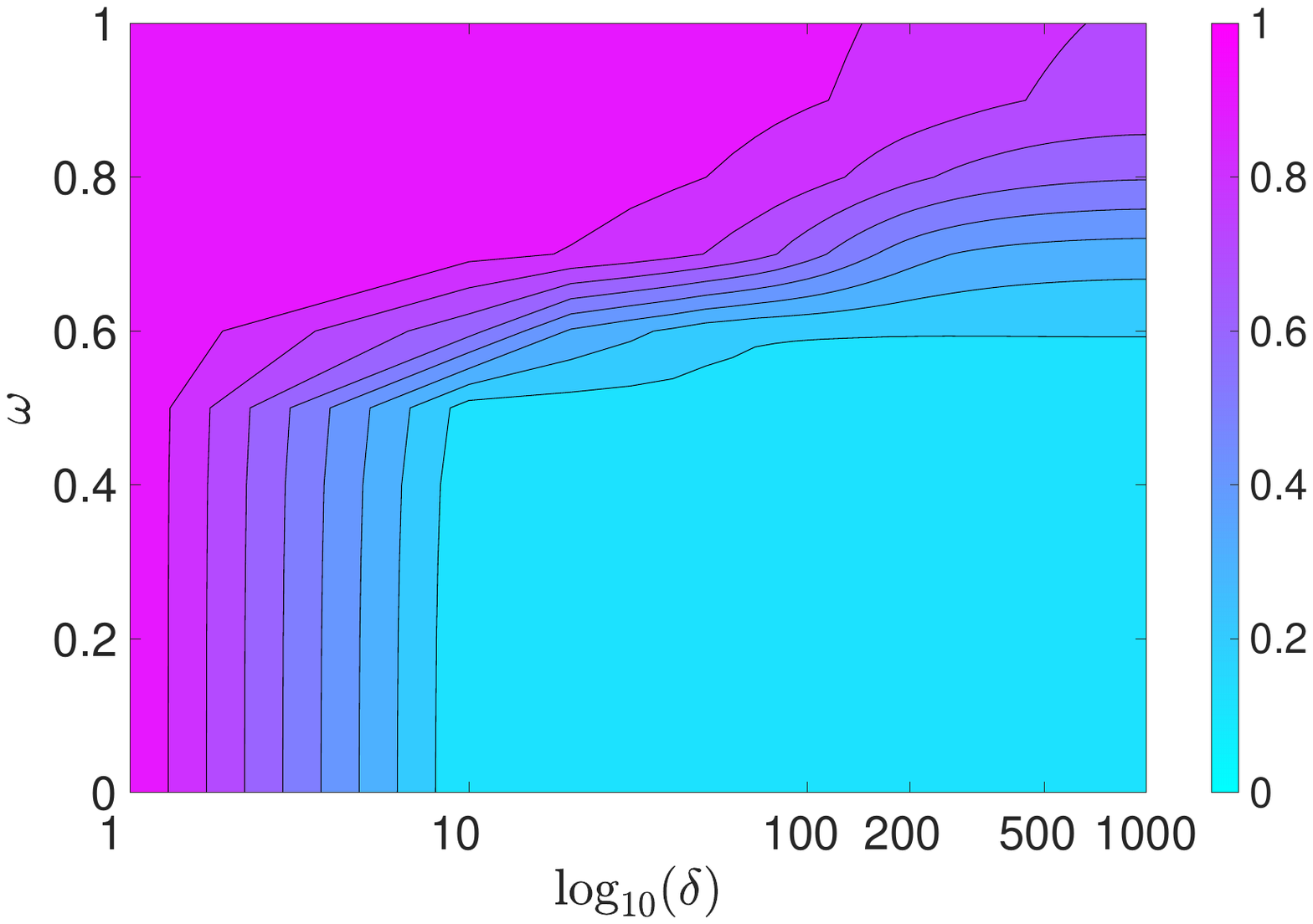}
    \caption{A contour plot exploring the relationship between $\omega$, $\delta$ and when content goes viral within simulations for a BDIC. Here, $\omega$ is varied from 0 to 1, $\delta$ is varied from 1 to 1000 (note the log scale), and the colour bar refers to the proportion of the amount of likes the most liked piece of content received out of the total amount of likes recorded in the system for that simulation.}
    \label{fig:RQ3_viral_BDIC}
\end{figure}

Here we present the result for measuring virality by varying $\omega$ and $\delta$ for a BDIC. The results seen in Fig.~\ref{fig:RQ3_viral_BDIC} do not substantially differ from those seen in Fig.~8.

\section{Supplementary Material Results for $\lambda_i=0.2$}\label{sec: supp results}

\begin{figure}[ht]
    \centering
    \begin{subfigure}[b]{0.49\textwidth}
        \centering
        \includegraphics[width=\textwidth]{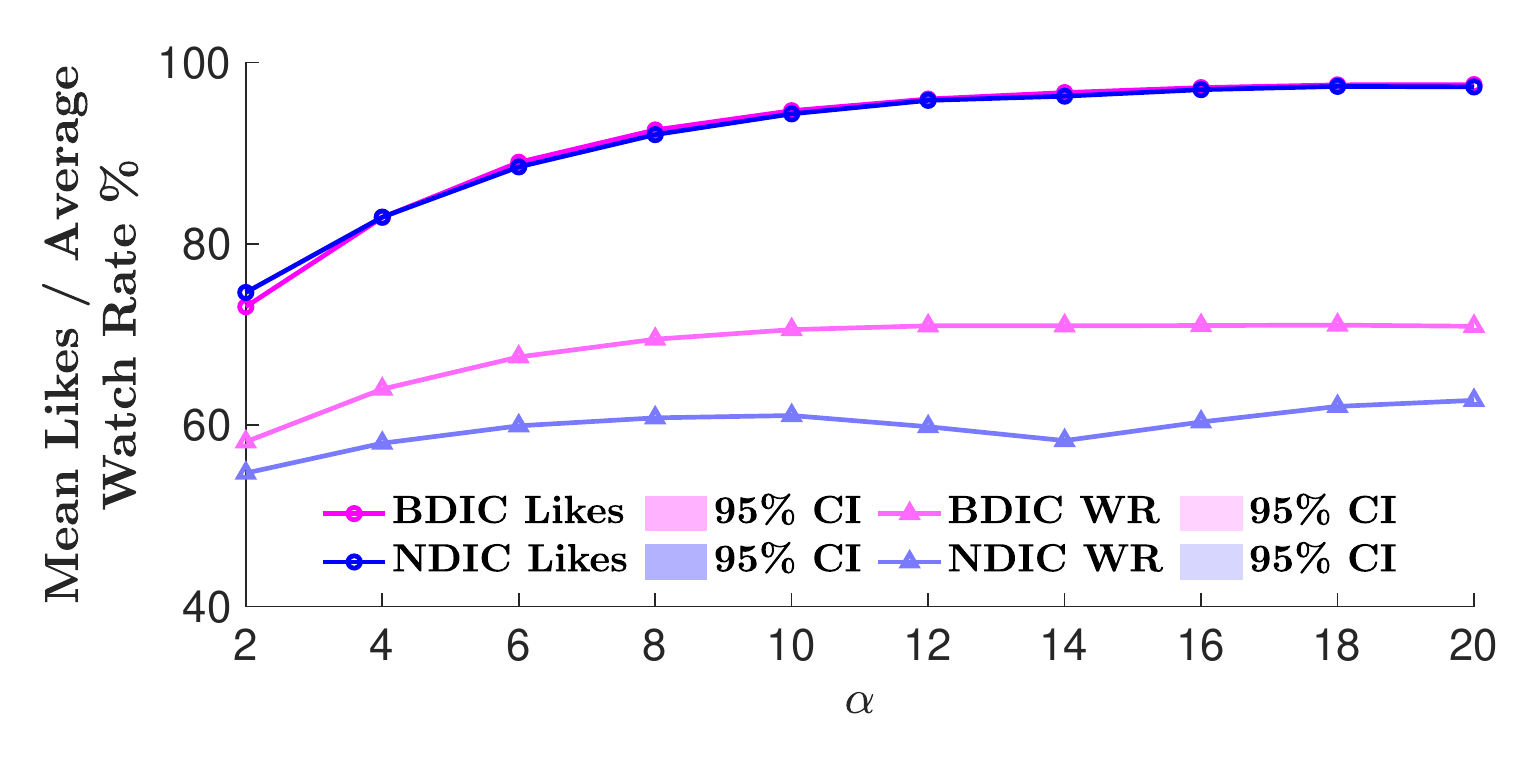}
        \caption{RS performance metrics}
        \label{fig:RQ1_performance_02}
    \end{subfigure}
    \hfill
    \begin{subfigure}[b]{0.49\textwidth}
        \centering
        \includegraphics[width=\textwidth]{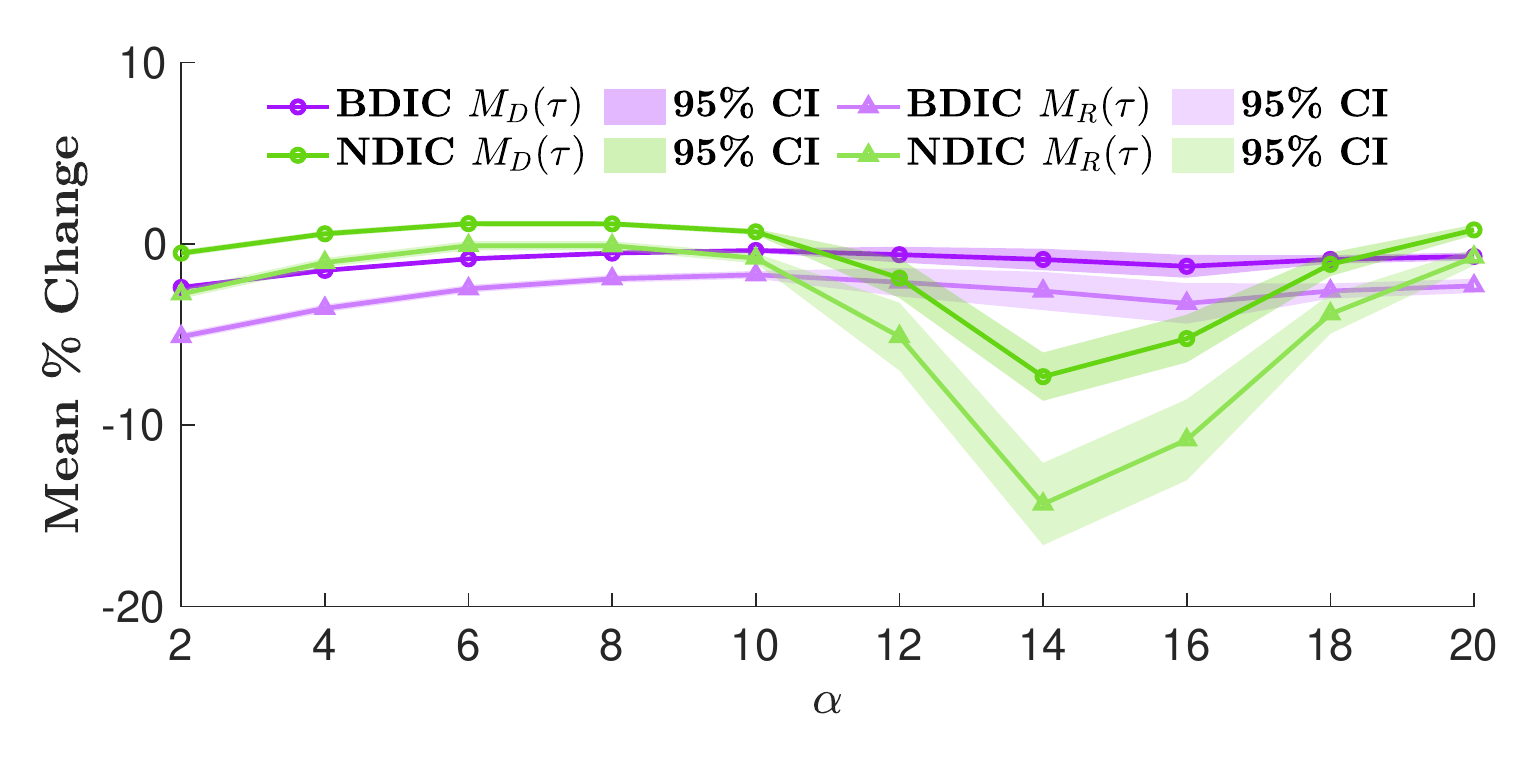}
        \caption{Opinion distribution metrics}
        \label{fig:RQ1_polarisation_02}
    \end{subfigure}
    \caption{\textit{Research Question One}: In (a) the performance metrics of the RS; $\mathrm{Likes\,\%}$, Eq.(11), and $\mathrm{Average\,Watch\,Rate\,\%}$ (WR), Eq.(12), are shown for different values of the softmax parameter, $\alpha$. In (b), the final percentage changes of the polarisation metric $M_D(\tau)$, Eq.(13), and radicalisation metric $M_R(\tau)$, Eq.(14), are plotted for different values of~$\alpha$.}
    \label{RQ1_together_02}
\end{figure}

\begin{figure}[ht]
    \centering
    \begin{subfigure}[b]{0.49\textwidth}
        \centering
        \includegraphics[width=\textwidth]{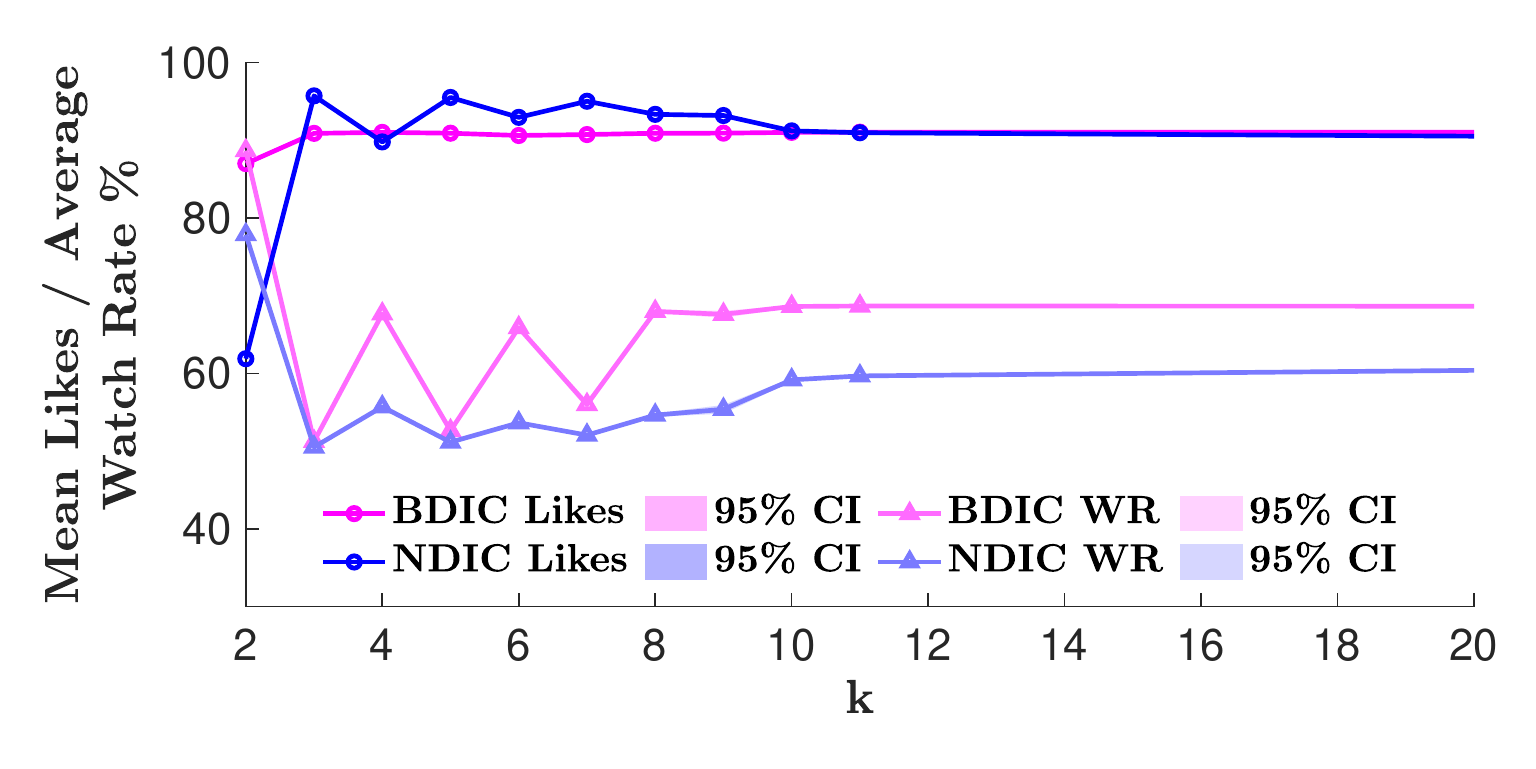}
        \caption{RS performance metrics}
        \label{subfig:RQ2_performance_21_02}
    \end{subfigure}
    \hfill
    \begin{subfigure}[b]{0.49\textwidth}
        \centering
        \includegraphics[width=\textwidth]{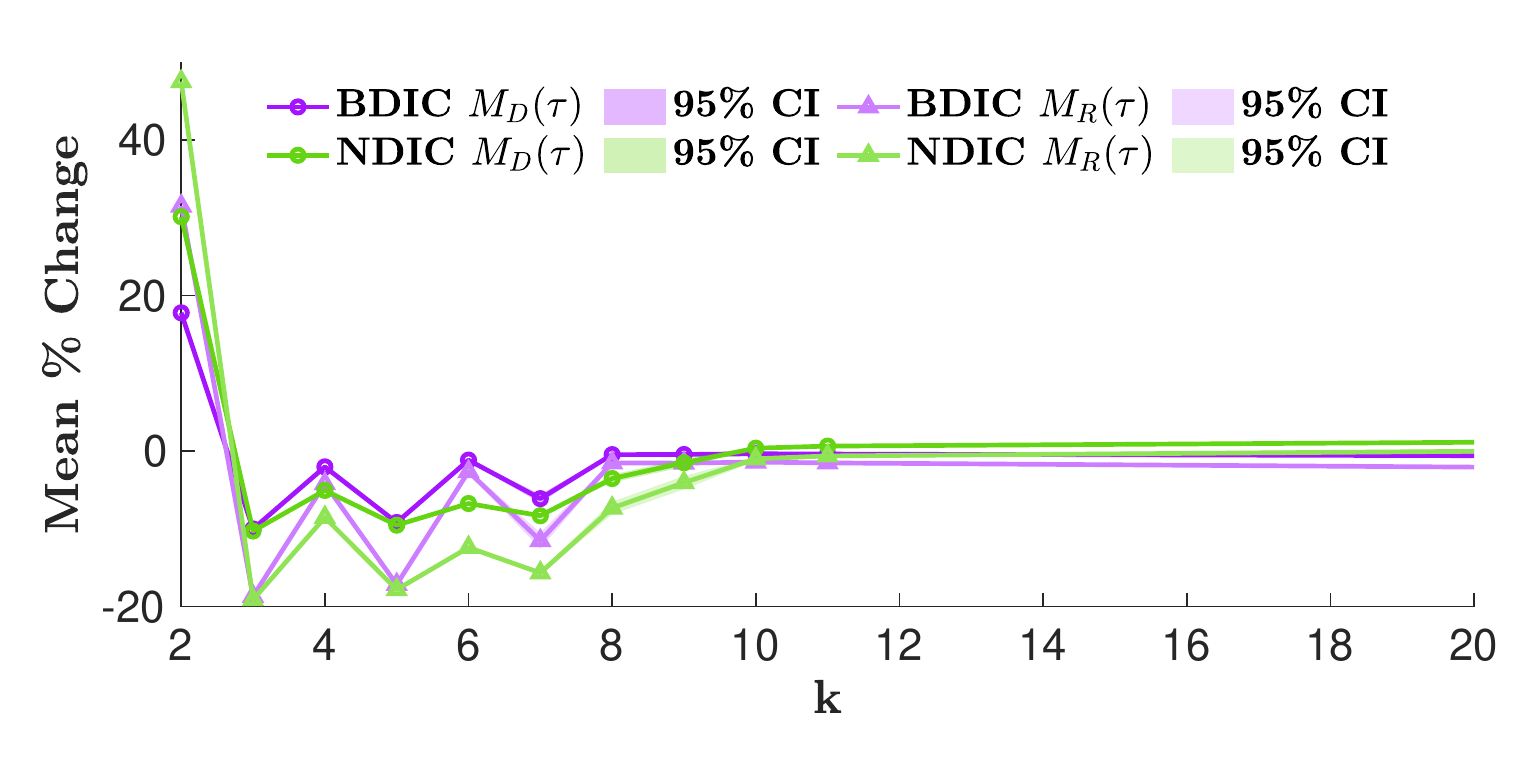}
        \caption{Opinion distribution metrics}
        \label{subfig:RQ2_polarisation_21_02}
    \end{subfigure}
    \caption{\textit{Research Question Two}: In (a) the performance metrics of the RS; $\mathrm{Likes\,\%}$, Eq.(11), and $\mathrm{Average\,Watch\,Rate\,\%}$ (WR), Eq.(12), are shown for different number different of content, $k$. In (b), the final percentage changes of the polarisation metric $M_D(\tau)$, Eq.(13), and radicalisation metric $M_R(\tau)$, Eq.(14), are plotted for different values of~$k$.}
    \label{fig:RQ2_performance_21}
\end{figure}

\begin{figure}[ht]
    \centering
    \begin{subfigure}[b]{0.49\textwidth}
        \centering
        \includegraphics[width=\textwidth]{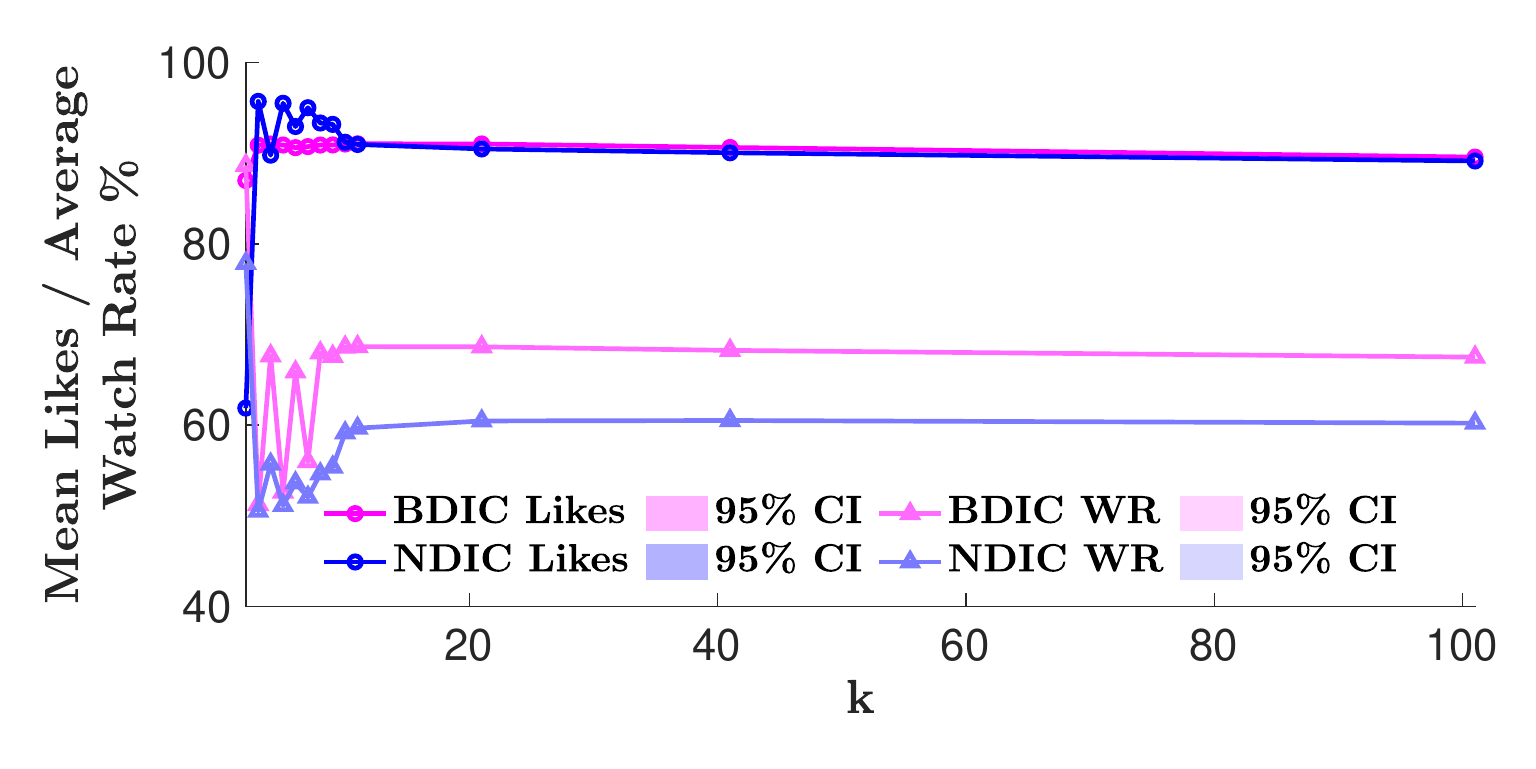}
        \caption{RS performance metrics}
        \label{subfig:RQ2_performance_101_02}
    \end{subfigure}
    \hfill
    \begin{subfigure}[b]{0.49\textwidth}
        \centering
        \includegraphics[width=\textwidth]{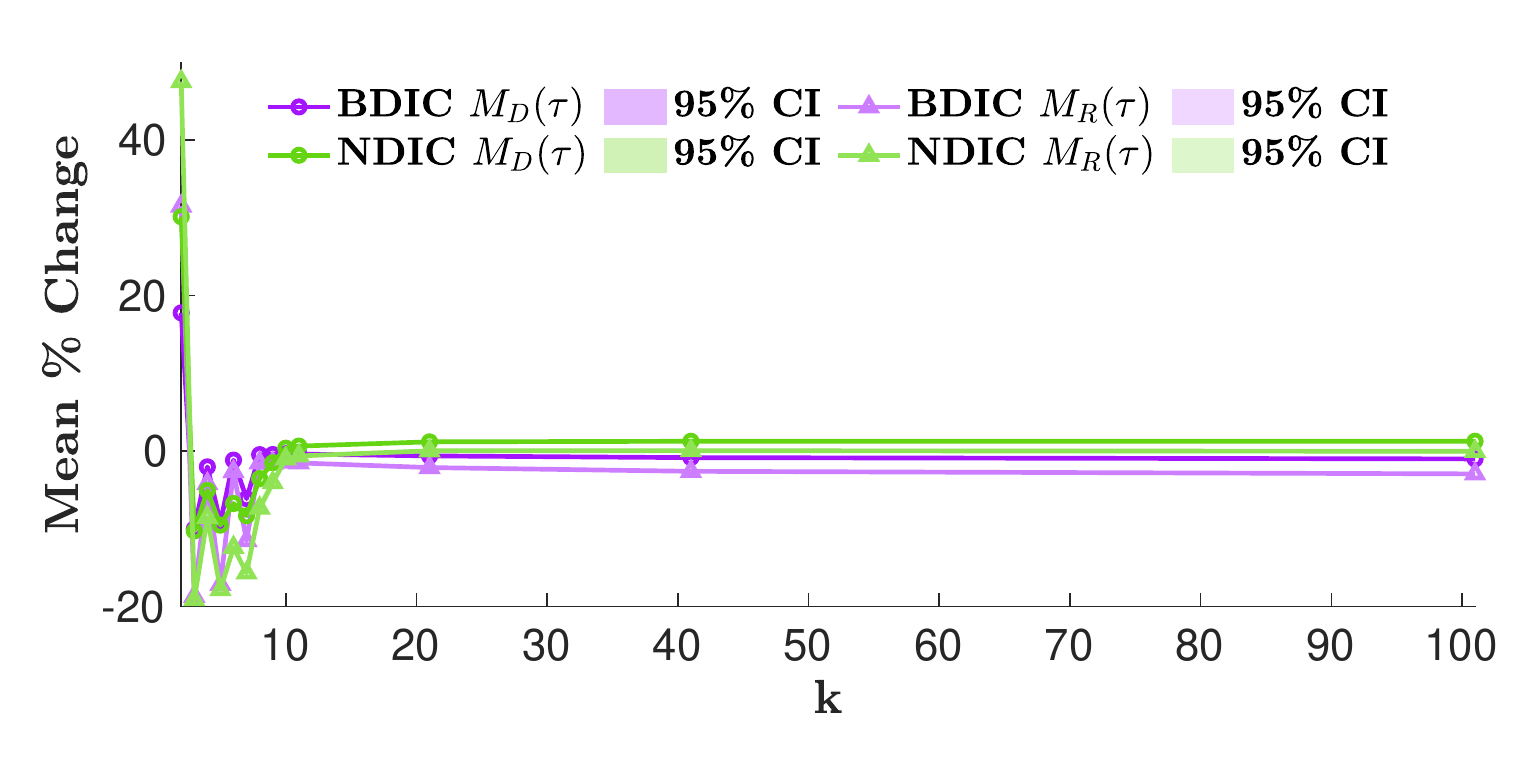}
        \caption{Opinion distribution metrics}
        \label{subfig:RQ2_radicalisation_101_02}
    \end{subfigure}
    \caption{\textit{Research Question Two}: Extended results to $k=101$. In (a) the performance metrics of the RS; $\mathrm{Likes\,\%}$, Eq.(11), and $\mathrm{Average\,Watch\,Rate\,\%}$ (WR), Eq.(12), are shown for different number different of content, $k$. In (b), the final percentage changes of the polarisation metric $M_D(\tau)$, Eq.(13), and radicalisation metric $M_R(\tau)$, Eq.(14), are plotted for different values of~$k$.}
    \label{fig:RQ2_polarisation_101}
\end{figure}

\begin{figure}[ht]
    \centering
    \begin{subfigure}[b]{0.49\textwidth}
        \centering
        \includegraphics[width=\textwidth]{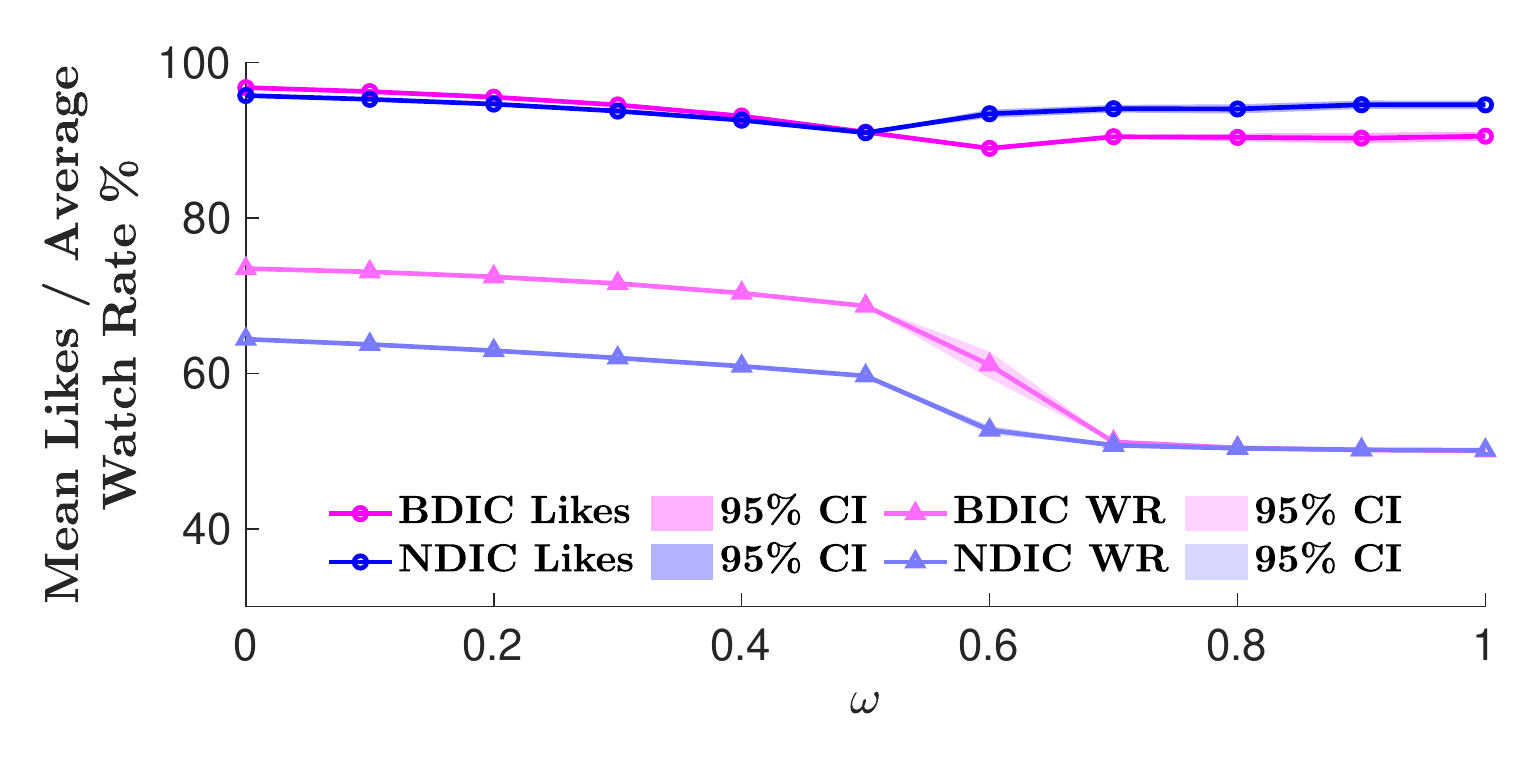}
        \caption{Full investigation of $k$ values spanning 2 to 101.}
        \label{fig:RQ3_performance_02}
    \end{subfigure}
    \hfill
    \begin{subfigure}[b]{0.49\textwidth}
        \centering
        \includegraphics[width=\textwidth]{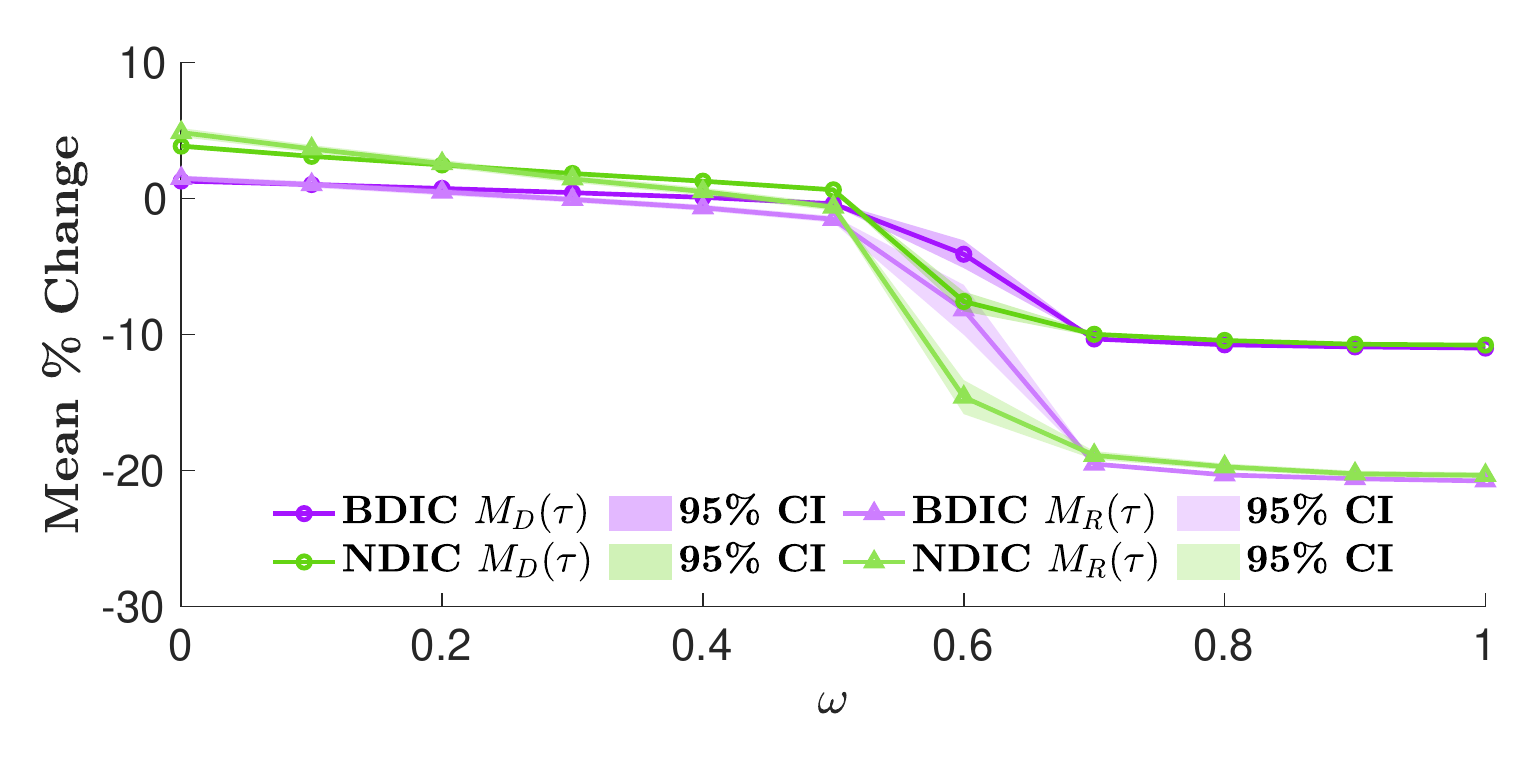}
        \caption{Zooming in on $k$ values spanning 2 to 11.}
        \label{fig:RQ3_polarisation_02}
    \end{subfigure}
    \caption{\textit{Research Question Three}: In (a) the performance metrics of the RS; $\mathrm{Likes\,\%}$, Eq.(11), and $\mathrm{Average\,Watch\,Rate\,\%}$ (WR), Eq.(12), are shown for different values of $\omega$, the virality-based filtering weighting parameter. In (b), the final percentage changes of the polarisation metric $M_D(\tau)$, Eq.(13), and radicalisation metric $M_R(\tau)$, Eq.(14), are plotted for different values of~$\omega$. Note that $\omega = 0$ and $\omega = 1$ correspond to entirely content-based filtering and entirely virality-based filtering, respectively.}
    \label{RQ3_together_02}
\end{figure}

\begin{figure}[ht]
    \centering
    \begin{subfigure}[b]{0.49\textwidth}
        \centering
        \includegraphics[width=\textwidth]{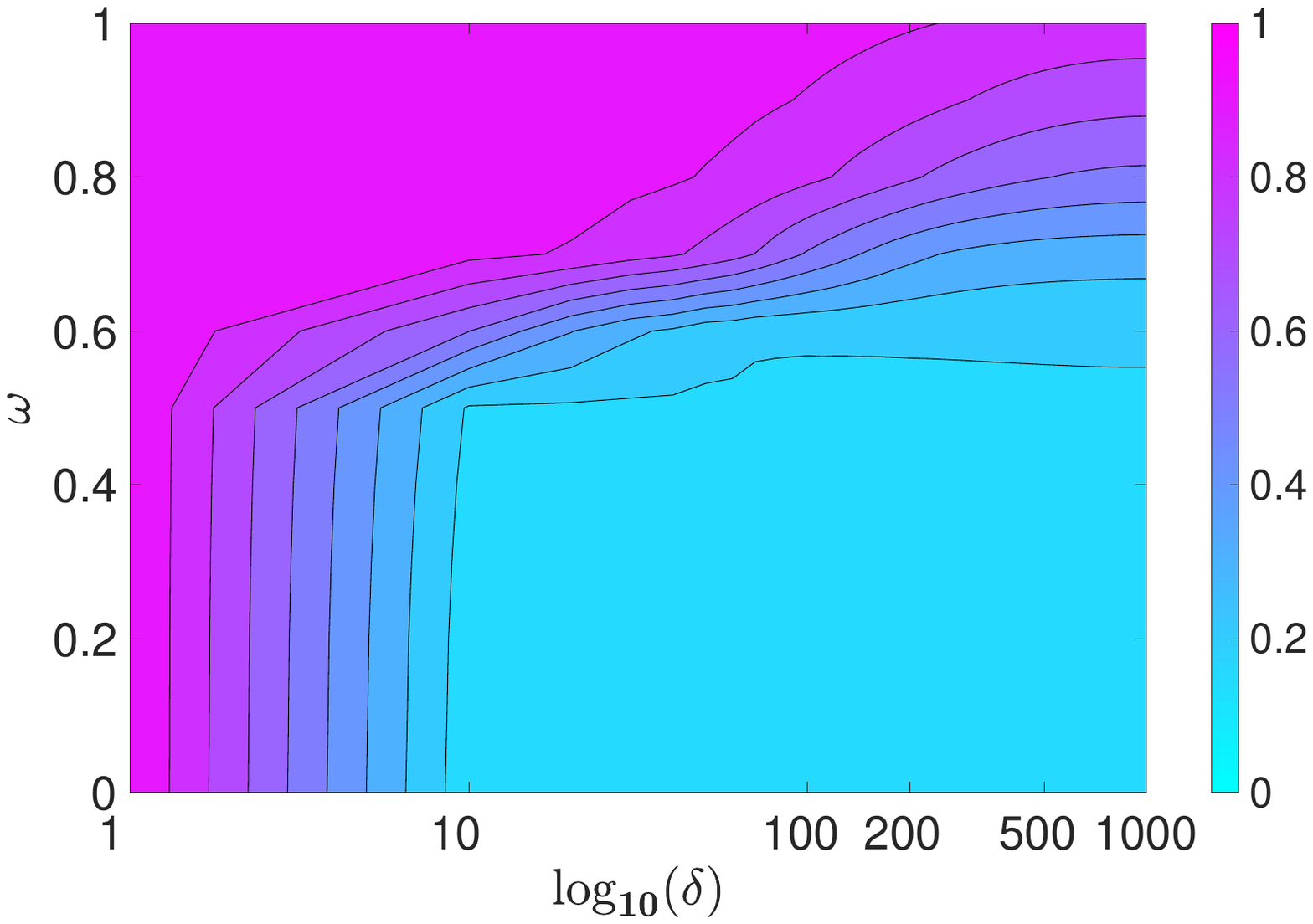}
        \caption{NDIC.}
    \end{subfigure}
    \begin{subfigure}[b]{0.49\textwidth}
        \centering
        \includegraphics[width=\textwidth]{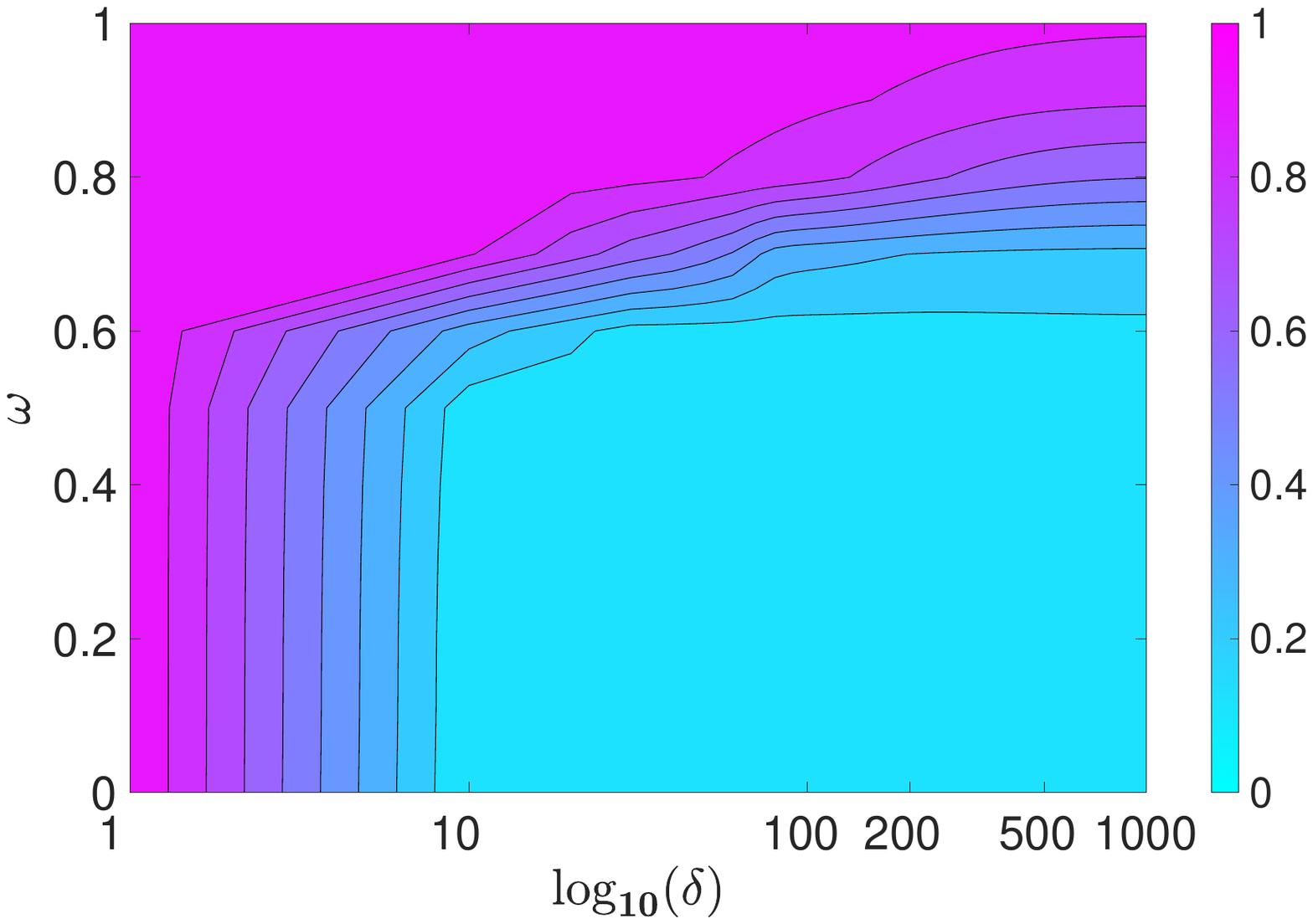}
        \caption{BDIC}
    \end{subfigure}
    \caption{\textit{Research Question Three}: Contour plots exploring the relationship between $\omega$, $\delta$ and when content goes viral within simulations. Here, $\omega$ is varied from 0 to 1, $\delta$ is varied from 1 to 1000 (note the log scale), and the colour bar refers to the proportion of the amount of likes the most liked piece of content received out of the total amount of likes recorded in the system for that simulation.}
    \label{fig:RQ3_viral_02}
\end{figure}

Here we provided results for Monte Carlo simulations as in Sections IV, V and VI, however Fig.~\ref{RQ1_together_02} to Fig.~\ref{fig:RQ3_viral_02} show the results of a homogenous population of $\lambda_i =0.2$ rather than $\lambda_i =0.9$. The results here do not substantially differ from those presented in the main paper hence they are reported here.

For the results of a homogenous population of $\lambda_i=0.2$ we see that the performance metrics in Fig.~\ref{fig:RQ1_performance_02} are not vastly different from those in Fig. 4a, just that when $\alpha>16$, the average watch rate of the NDIC does not begin to converge with that of the BDIC's, this is to be expected given the content is having less influence of the evolution of users' opinions. The results in Fig.~\ref{fig:RQ1_polarisation_02} are again to be expected given $\lambda_i=0.2$; however, both the NDIC and BDIC are consistently below zero, meaning polarisation and radicalisation were always decreased within the populations across the different $\alpha$ values. Interestingly, the dip for the NDIC when $\alpha=14$ is more pronounced than that of the one in Fig. 4b, suggesting \textit{viral} content was also experienced at these settings. 

For Fig.~\ref{fig:RQ2_performance_21}, we note no major differences from Fig. 5, just the similarity that when $k>9$ and $k$ is odd there are notable decreases in the opinion distribution metrics due to a neutral piece of content going \textit{viral.} 

For Fig.~\ref{fig:RQ2_polarisation_101}, we note that the change in metrics when $k \in [11,101]$ has asymptotically diminished, as per Fig.~5.

\newpage

The results shown in Fig.~\ref{RQ3_together_02} do not largely differ from those in Fig. 7. One notable difference is that in Fig.~\ref{fig:RQ3_performance_02} when $\omega>0.6$ the average total likes of the NDIC and BDIC do not converge as they did in Fig. 7a, this is due to the \textit{viral} content having less influence of an influence on the opinion evolution of the BDIC.

There are no notable differences between Fig.~\ref{fig:RQ3_viral_02} and Fig. 8.

\end{document}